\documentclass[preprint,authoryear,12pt]{elsarticle}

\usepackage{graphicx}

\usepackage{amssymb}
\usepackage{url}

\usepackage[
a4paper=true,%
breaklinks=true,%
colorlinks=true,%
pdfauthor={S.P. Moschou et al.},%
pdftitle={Simulating coronal condensation dynamics in 3D}%
]{hyperref}

\journal{Advances in Space Research}

\begin{document}

\begin{frontmatter}

\title{Simulating coronal condensation dynamics in 3D}


\author[1,2]{S. P. Moschou}
\ead{sofiaparaskevi.moschou@wis.kuleuven.be}

\author[2]{R. Keppens}
\author[2]{C. Xia}
\author[2]{X. Fang}

\address[1]{Belgian Institute for Space Aeronomy, Ringlaan-3-Avenue Circulaire, B-1180 Brussels, Belgium}
\address[2]{Centre for mathematical Plasma Astrophysics, Department of Mathematics, KU Leuven, Celestijnenlaan 200B, B-3001 Leuven, Belgium}

\begin{abstract}
We present numerical simulations in 3D settings where coronal rain phenomena take place in a magnetic configuration of a quadrupolar arcade system. 
Our simulation is a magnetohydrodynamic simulation including anisotropic thermal conduction, optically thin radiative losses, and parametrised heating as main thermodynamical features to construct a realistic arcade configuration from chromospheric to coronal heights. 
The plasma evaporation from chromospheric and transition region heights eventually causes localised runaway condensation events and we witness the formation of plasma blobs due to thermal instability, that evolve dynamically in the heated arcade part and move gradually downwards due to interchange type dynamics. 
Unlike earlier 2.5D simulations, in this case there is no large scale prominence formation observed, but a continuous coronal rain develops which shows clear indications of Rayleigh-Taylor or interchange instability, that causes the denser plasma located above the transition region to fall down, as the system moves towards a more stable state. Linear stability analysis is used in the non-linear regime for gaining insight and giving a prediction of the system's evolution. 
After the plasma blobs descend through interchange, they follow the magnetic field topology more closely in the lower coronal regions, where they are guided by the magnetic dips.

\end{abstract}


\end{frontmatter}

\parindent=0.5 cm

\section{Introduction}

Coronal rain blobs and prominences consist of cool and dense plasma and can be formed due to thermal instability, as it was described in \citet{Parker53} and \citet{Field65}.
First, chromospheric and coronal matter gets heated due to a physical trigger (e.g. a solar flare) and gets redistributed to coronal heights, where it starts loosing energy and cools down through radiative losses and thermal conduction \citep{Colgan08,Goldsmith71,Levine77,Townsend09,Xia11}. Thermal instability magnifies temperature perturbations and causes condensation in both hydro and magnetohydrodynamic regimes, isotropically and anisotropically, respectively \citep{Hildner74}. 
Observational studies provided indications that prominences are composed of embedded cool plasma inside a flux rope with a sheared arcade on top \citep{Chae01}.
Catastrophing cooling can take place within preformed flux ropes~\citep{Xia14b} or at the top of dynamic loop systems reducing the temperature of the plasma to transition region and chromospheric temperatures causing runaway events and the evacuation of the loop, as presented in \citet{Antolin10,Mendoza05,Muller04,OShea07,Schrijver01}. 

These loops show strong emission in lines with characteristic temperatures $T\leq10^5$K \citep{Muller03,Schrijver01} and weak emission at lines corresponding to higher temperatures. The cool material is observable in EUV lines, and can either be a large scale prominence structure \citep{Keppens14,Xia14b} or form smaller blobs with sizes of the order of a few hundreds of kilometres \citep{Antolin12a,Fang13,Muller03,Schrijver01,Tripathi09}. These blobs appear bright in 304\,\AA (EIT), while intensity variations appear to travel from the loop top towards the footpoints, suggesting cool temperatures and providing supporting evidence for heating-evaporation-condensation cycles \citep{DeGroof05,DeGroof04,Mok08,Xia12}.

The importance of the magnetic field geometry in coronal rain and prominence events was highlighted in \citet{Kawaguchi70}, as it guides the cool plasma falling from the corona back to the chromosphere \citep{Leroy72}.
Coronal rain can provide information about the magnetic fields supporting it, as condensed blobs, with lifetimes of a few tens of minutes \citep{Fang13}, slide and deform along the magnetic field lines revealing their small-scale topology. They slide down with a wide range of speeds and with accelerations different from the one predicted by gravity alone, suggesting the important influence of other (M)HD forces \citep{Antolin11}.
Both hot plasma upflows from chromospheric heights and multi-temperature downflows along magnetic field lines were observed to take place mainly in EUV wavebands \citep{Beckers62,Kamio11}, with hot downflows only close to the loop top \citep{Tripathi09}. 

Magnetic field topologies used for the simulations of loop systems extending to the solar corona have e.g. used force-free field approximations to explore how heating affects EUV and soft X-ray views \citep{Mok05}. In a full flux rope configuration, \citet{Xia14b} also demonstrated prominence formation and demonstrated its SDO/AIA appearance, in accord with actual observations.
Computational efforts have already successfully produced prominence condensations resulting from different heating functions \citep{Antiochos91,Xia12}. 
Numerical studies \citep{Antiochos99,Dahlburg98} have suggested that for the prominence to be formed through catastrophic cooling \citep{Xia12} the heating function of the loop needs to be localised close to the footpoints of the loop at chromospheric heights, whereas the magnetic dip topology is not a necessity \citep{Karpen01}. Downflows of condensed matter were found to accelerate in the dipped part of the configuration in numerical simulations \citep{Antiochos99}, while observational studies verify the accelerated sliding of the cool plasma towards both footpoints \citep{Oliver14,Schrijver01,Wang99} with propagation speeds in the range of 10 up to 200$\mathrm{km/s}$ \citep{Kleint14} and accelerations smaller than the effective gravitational one \citep{Antolin12a} in MHD setups.

\label{Motivation}
High resolution observations with instruments such as Hinode/Solar Optical Telescope, CRISP/Swedish Solar Telescope, SDO and IRIS are available and provide insight in the dynamic fine-structures \citep{Scullion14} of the solar corona.
Numerical studies have also started to explore the physical conditions that play a key role in dynamic loop systems, such as the detailed magnetic topologies and the heating mechanisms, since they determine the temperatures, densities, and speeds of the condensed material~\citep{Fang13}, and influence relevant waves and instabilities driving the phenomena.
With the already obtained knowledge about these physical conditions we want to further investigate the driving mechanism(s) for the creation \citep{Xia11} and the evolution of the condensed blobs.

\section{Setup: physical and computational aspects}
\subsection{Governing equations and initial configuration}
\begin{figure}[htbp]
\begin{center}
\includegraphics[width=0.8\linewidth]{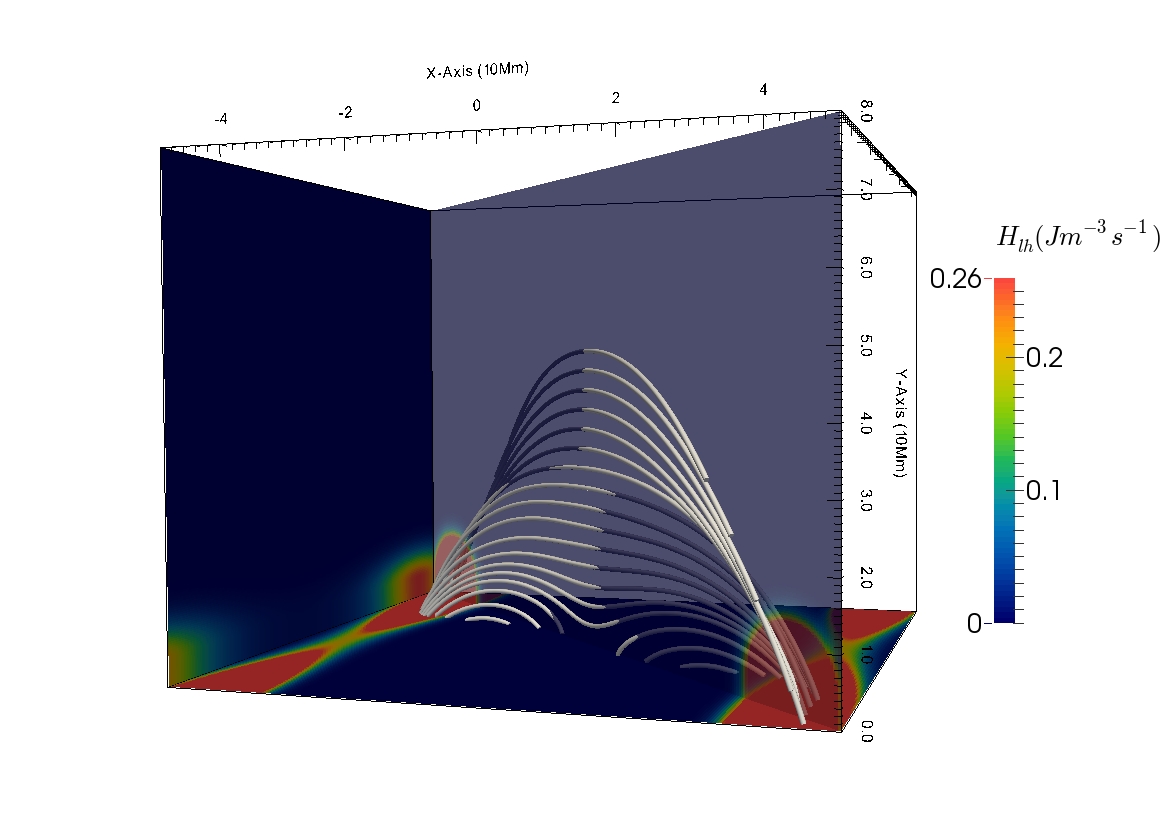}
\caption{Here we present in a 3D fashion the localised heating function and the magnetic field topology, to emphasise that the heating is applied at the footpoints of the enveloping arcade. Three different cross-sections are shown, namely the diagonal and the left and bottom boundary planes of our simulation domain. The image corresponds to a physical time of about $205 \mathrm{min}$.}
\label{heating}
\end{center}
\end{figure}

We present numerical simulations in 3D settings where coronal rain phenomena occur in a quadrupolar magnetic environment. 
Simulations are done with the grid adaptive MPI-AMRVAC \footnote{\url{https://gitorious.org/amrvac}} code as described in \citet{Keppens12} and updated in \citet{Porth14}.
As initial conditions we take a 2.5D, potential magnetic field, forming a quadrupolar arcade configuration with a dip, as demonstrated in figure~\ref{heating}, augmented with a 1D stratified equilibrium, such that in ideal MHD we realise a force-balanced state.
The velocity is set to zero throughout, and the magnetic field is taken as
\begin{eqnarray}
\label{eq:magfield1}
B_x&=& B_{p0}\cos\left(\frac{\pi x}{2L_0}\right)e^{-\frac{\pi y}{2L_0}}-B_{p0}\cos\left(\frac{3\pi x}{2L_0}\right)e^{-\frac{3\pi y}{2L_0}}{,} \\
\label{eq:magfield2}
B_y&=& -B_{p0}\sin\left(\frac{\pi x}{2L_0}\right)e^{-\frac{\pi y}{2L_0}}+B_{p0}\sin\left(\frac{3\pi x}{2L_0}\right)e^{-\frac{3\pi y}{2L_0}}{,} \\
\label{eq:magfield3}
B_z&=&B_{z0} {.}
\end{eqnarray}
where the $y-$direction is the height in our simulation, $B_0=4\times 10^{-4}\mathrm{T}$ and the local angle between the $(x,y)$-plane and the field lines is set to $\pi/4$ at $x=0$, $y=2L_0/\pi$, determining $B_{p0}$ and $B_{z0}$ uniquely, making $B_{p0}=\frac{B_0}{(e^{-1}-e^{-3})\sqrt{2}}$ and $B_{z0}=B_{p0}(e^{-1}-e^{-3})$.
The density and pressure are calculated by the gravitational stratification. We initialise the 1D arrays of density and pressure in the bottom boundary from hydrostatic equilibrium as follows:  
\begin{equation}
\frac{p_j-p_{j-1}}{\Delta y}=\frac{1}{4}(g_j+g_{j-1})\left(\frac{p_j}{T_j}+\rho_{j-1}\right) \/.
\end{equation}

The stratification gets a transition region temperature $T_{tr}=1.6\times 10^5\mathrm{K}$ at a chosen height $h_{tr}=0.27\times 10^7 \mathrm{m}$ and adopts a constant vertical thermal conduction flux upwards, taken at $\kappa(T)\frac{dT}{dy}=200 \mathrm{Jm^{-2}s^{-1}}$. The temperature in the top boundary is later on restricted to be $T_{top}=2\times 10^6\mathrm{K}$. 

Nevertheless, we extend physically the ideal MHD equations by including non-ideal terms and study non-ideal effects including (a) optically thin radiative losses $Q$, (b) anisotropic thermal conduction and (c) a parametrised heating function $H$, such that the evolution equation of the total energy becomes
\begin{equation}
\frac{\partial E}{\partial t}+\nabla\cdot (E\vec{v}+p_{tot}\vec{v}-\vec{B}\vec{B}\cdot\vec{v})=\rho \vec{g}\cdot\vec{v}+\nabla\cdot (\vec{\kappa}\cdot\nabla T)-Q+H
\end{equation}
where $E=p/(\gamma-1)+\rho v^2/2+B^2/2$ is the total energy density including internal, kinetic and magnetic energy densities accordingly for each term. We take $\gamma=5/3$ and set magnetic permeability $\mu_0=1$, making the total pressure given by $p_{tot}=p+B^2/2$. The thermal conduction is field-aligned and it has the following form: $\kappa_{||}=10^{-11}T^{5/2}\mathrm{Jm^{-1}s^{-1}K^{-3.5}}$. The optically thin cooling uses a tabulated temperature dependence $\Lambda(T)$ and as the equation $Q\propto n_H^2\Lambda(T)$ suggests it is proportional to the squared hydrogen number density. We adopt the came cooling table $\Lambda(T)$ as described and used in \citet{Xia11}, section 2.1 and figure 1 therein (solid line).

The parametrized heating term has the following prescription 
\begin{eqnarray}
H&=&H_{bg}+H_{lh} {,} \\
H_{bg}(y)&=&H_0 e^{-y/L_{bg}}{,} \\
H_{lh}(x,y,z,t)&=&H_1 R(t) C(y) F(x,z) {,}
\end{eqnarray}

\begin{eqnarray}
\label{eq:fxz}
F(x,z)&= \left[exp\left(-\frac{(x-x_l)^2}{\sigma_2}\right) exp\left(-\frac{sin\left(\frac{\pi(z-x_l)}{\Delta z}\right)^2}{\sigma_3}\right)\right.\nonumber\\
 &\qquad \left. {}+exp\left(-\frac{(x-x_r)^2}{\sigma_2}\right) exp\left(-\frac{sin\left(\frac{\pi(z-x_r)}{\Delta z}\right)^2}{\sigma_3}\right)\right]{,}
\overfullrule 5pt
\end{eqnarray}

\begin{equation}
    C(y)=\left\{
                \begin{array}{ll}
                  1, & \mathrm{if }\; y<y_h\/,\\
                 e^{-\frac{(y-y_h)^2}{\lambda_h}}, & \mathrm{if}\;  y\geq y_h \,.
                \end{array}
              \right.
\end{equation}

We have two different heating phases, as is evident by the heating function. First, we relax our system from begin time $t=t_0$ to $t=t_{relax}$ ($\approx 51$min) using only the background heating term $H_{bg}$ till a quasi-equilibrium state is reached. Afterwards, we add an extra localised heating term $H_{lh}$, as visualised in the figure~\ref{heating}.
For the background heating we have an amplitude of $H_0=3\times 10^{-5} \mathrm{Jm^{-3}s^{-1}}$, while for the localised heating we have $H_1=2\times 10^{-3} \mathrm{Jm^{-3}s^{-1}}$, i.e. $H_1\approx 10^2 H_0$. The length and pressure units are taken as $L_{unit}=10^7\mathrm{m}$ and $p_{unit}=0.03175\mathrm{Jm^{-3}}$, while the scale height of the background heating is $L_{bg}=5L_{unit}$ and the localised heating is prescribed by a ramp function $R(t)$ with a linear variation from zero to one, for a specific initial time and duration. For the rest of the heating parameters we take $\sigma_2=0.2L^2_{unit}$ and $\sigma_3=0.3$, $x_l=-x_r=-4.2L_{unit}$, $y_h=0.4L_{unit}$, $\lambda_h=0.25L^2_{unit}$. We assume fully ionised plasma with 10:1 H:He abundance. 

\begin{figure}
 \begin{center}
 \includegraphics[trim=2cm 2cm 2cm 2cm,width=\textwidth]{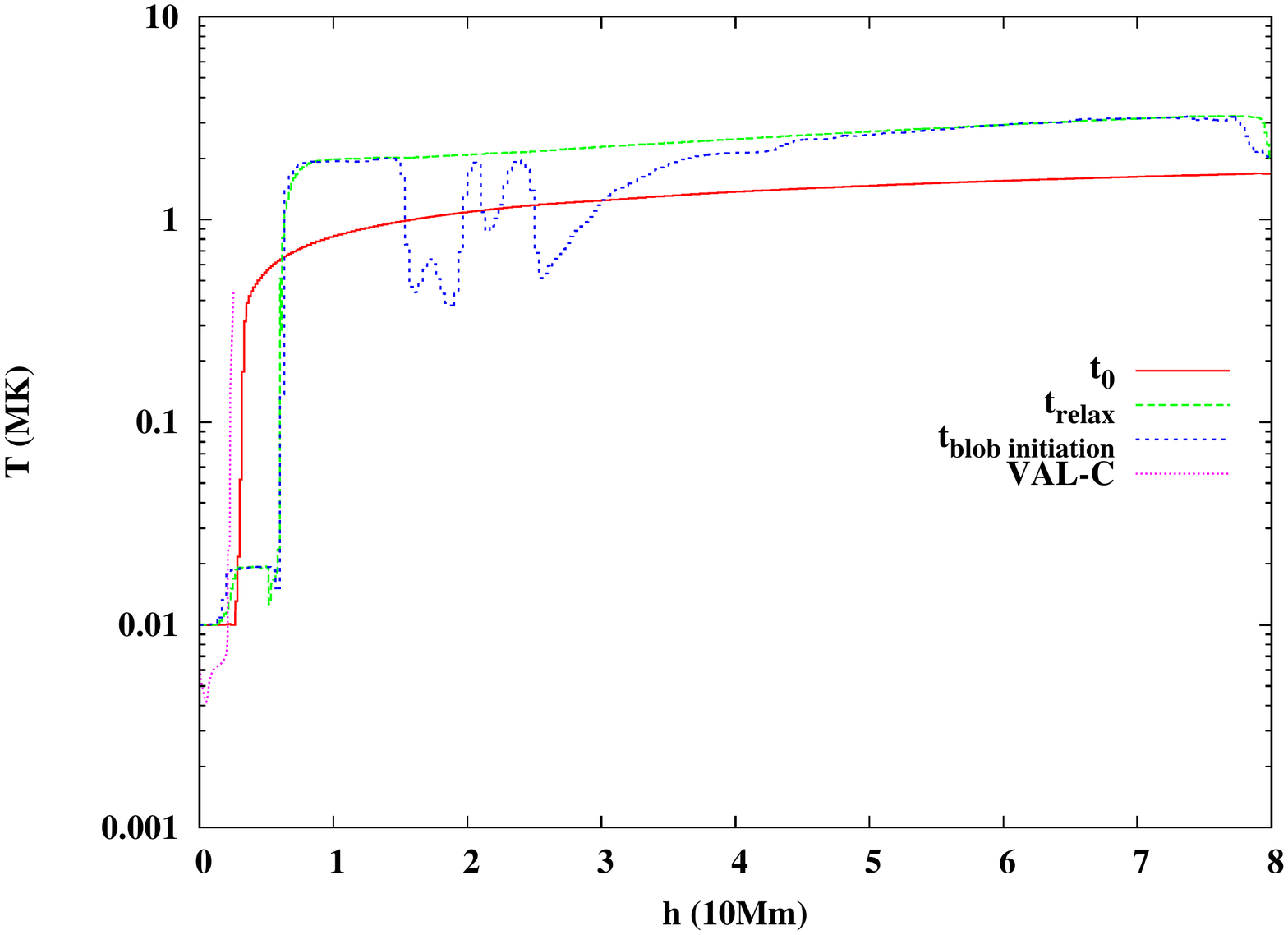}
 \end{center} 
 \caption{ Temperature profiles are shown at times a) $t=t_0$, b) $t=t_{relax}$, c) $t=t_{blob\ initiation}$, together with d) the static averaged temperature profile according to semi-empirical VAL-C model for comparison.}
 \label{Tprofile}
\end{figure}

In figure~\ref{Tprofile} we present the temperature profile along a line parallel to the $y-$axis passing through the middle of the simulation domain, i.e. $(0,y,0)$, in three different times throughout our simulation. More specifically, we show a) $t=t_0$, b) $t=t_{relax}$, c) $t=t_{blob\ initiation}$ and d) the static averaged temperature profile according to semi-empirical VAL-C model \citep{Vernazza81}, which only extends to low heights ($<2.6\mathrm{Mm}$) and is a plane-parallel model for quiet sun. We start at  $t=t_0$ from a temperature profile that includes chromosphere, transition region and corona and we apply a background heating with a scale length of 50 $\mathrm{Mm}$ till we reach a quasi-equilibrium state at $t=t_{relax}$. The transition region is shifted to larger heights and heated. In physical units, the relaxation phase follows the 3D heated arcade for $\approx 51$ min. Also, $t=t_{blob\ initiation}$ is actually $\approx$ 205 min after $t=t_{relax}$.

\subsection{Boundary conditions and numerical aspects}
As boundary conditions we use ghost cells of 2 grid layers exterior to the physical domain everywhere in accord with the discretisation scheme.
For numerical advancing of the partial differential equations we use a three-step Runge-Kutta scheme. 
As evaluation scheme to move from cell-centered to edge values we use a third-order-accurate limited reconstruction, as described in \citet{Cada09}.
For fluxes we adopt an HLLCD scheme, which is a suitably mixed contact-resolving HLLC with fallback to TVDLF, hence D(iffused) hybrid scheme introduced in~\citet{Zak}.
As Courant parameter we use 0.9 of the CFL computed timestep. A diffusive approach is used for magnetic monopole control.

For our simulation we typically impose boundary conditions on the primitive variables: ($\rho$, $v_x$, $v_y$, $v_z$, $p$, $B_x$, $B_y$, $B_z$).
\begin{itemize}
\item Side boundaries: $(x_{min}, x_{max})=(-5L_{unit},5 L_{unit})$ and $(z_{min}, z_{max})=(-5L_{unit},5 L_{unit})$\\
For each physical quantity on the side surfaces in $x-$direction of the simulation box we choose the symmetric boundary conditions for ($\rho$, $v_y$, $v_z$, $p$, $B_y$, $B_z$) and asymmetric boundary conditions for ($v_x$, $B_x$), while taking periodic boundary conditions for the $z-$direction for all physical quantities. 

\item Bottom boundary at: $y_{min}=0$ \\
At the bottom, we use asymmetric boundary conditions for the three components of the velocity to set zero flow there. We fix the density and pressure as indicated by the gravitational stratification of the initial state and the magnetic field is fixed as prescribed by the analytic relations given in Eq.~(\ref{eq:magfield1}), (\ref{eq:magfield2}), (\ref{eq:magfield3}).

\item Top boundary at: $y_{max}=8L_{unit}$\\
First, the pressure and density in the top layers is calculated giving a top temperature.
This local temperature, but limited to $T_{top}$ is then used together with a discrete extrapolation of the stratification.
The velocity is set asymmetric. 
For the magnetic field, we fix $B_z$ to its initial value, adopt a one-sided difference expression to set $\frac{\partial B_x}{\partial y}=0$.
Finally, in order to set the normal component $B_y$ we apply the central differenced condition $\nabla\cdot\vec{B}=0$.

\end{itemize}

The simulation uses a base grid of $120\times120\times120$ grid points, activating a 3-level mesh refinement based on mixed evaluation of weighted discrete second derivates involving density and the $x-$ and $y-$ magnetic field components $\rho:B_x:B_y$ in a $0.6:0.2:0.2$ ratio.
   The effective mesh size is $480\times480\times480$ corresponding to physical distances of $208\mathrm{km}\times167\mathrm{km}\times208\mathrm{km}$ in a single highest resolution grid cell.
   
\subsection{Filter for analysis of thermodynamic evolution}

We use filters to analyse the presence of blobs in our simulations, which only start appearing after more than 205 minutes of physical time elapsed after extra heating is switched on.
We define coronal rain blobs as cool and dense plasma suspended in the corona, after it condensed in-situ there. 
Condensation happens after a long period of gradual mass transport to the corona due to evaporation from transition region and chromospheric heights, as a consequence of the localised heating function that heats the footpoints of the large arcade, as shown in figure~\ref{heating}.  

Blobs are found as dense, cool matter well above transition region heights. Practically, the criteria used for the classification of a plasma element, according to our numerical resolution and grid size, as a blob element, are $y>h_{TR}(x,z,t)$ and $\rho>7\rho_{unit}$ and $T<0.1\mathrm{MK}$. More specifically, we use criteria that take into account the number density, the temperature and the height of each grid cell to characterise a cell volume as containing blob material. First of all, we need to quantify the transition region height $h_{TR}(x,z,t)$. We do so in 3D by searching our simulation grid starting from the bottom looking for the height where the temperature and density gradient become maximum for the first time. If this technique fails, we set $9\mathrm{Mm}$ as the transition region height locally, which is typically an overestimate of the true transition region height found. 
After having quantified the transition region surface, we are able to calculate the total mass of the cool material and the number of the blobs, the latter by clustering of neighbouring plasma cells into larger parts.
Furthermore, we define a spatial cutoff for blobs that are in agreement with the blob-filter criteria, but have a total volume smaller than eight individual grid cells, i.e. we only take into account blobs with a volume of at least eight grid cells in total. 

More specifically, the grid cells, that satisfy the blob-filter conditions and are above the set cutoff, are labelled as cool matter and kept in a list.
With the cool material thereby mapped in our 3D grid, we can group the different grid cells into individual blobs at each saved snapshot and calculate their number.
After this calculation is completed, we drop out groups with less than eight grid cells, remap the cool material and then recalculate the blob number.
When all individual blobs are defined and mapped, we can collect their physical properties (such as the total mass of the blob, the average density) and calculate the centroid location of each blob with the temperature as weight.
If the maximum density grid cell is close to the centroid and if their separation distance is larger than the resolution, we adopt as the centroid the grid cell with the maximum density value for that blob. 

\begin{figure}
 \begin{center}
 \begin{tabular}{cc}
 \includegraphics[trim=4cm 4cm 4cm 4cm,width=0.45\textwidth]{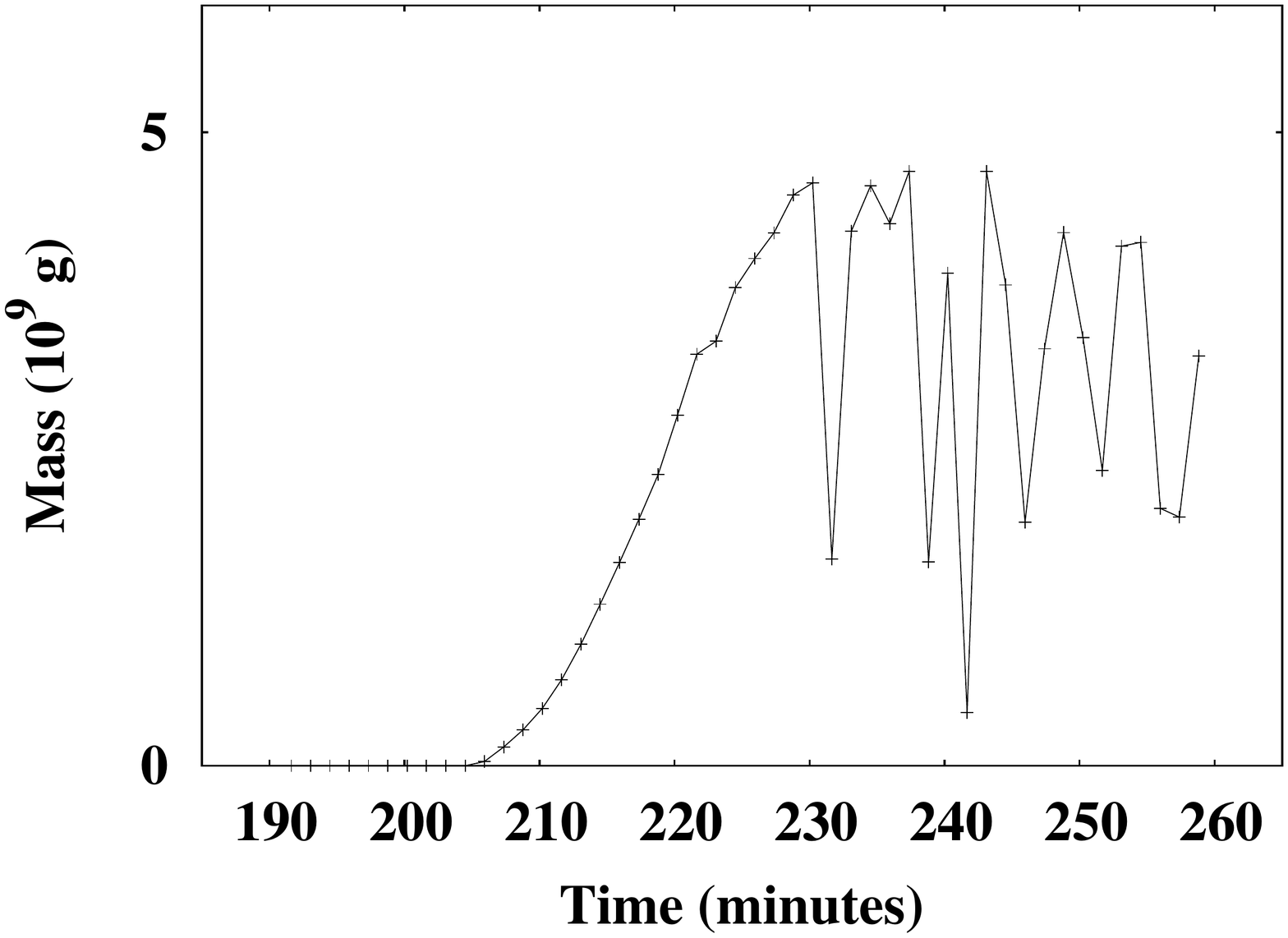}&
  \includegraphics[trim=4cm 4cm 4cm 4cm,width=0.45\textwidth]{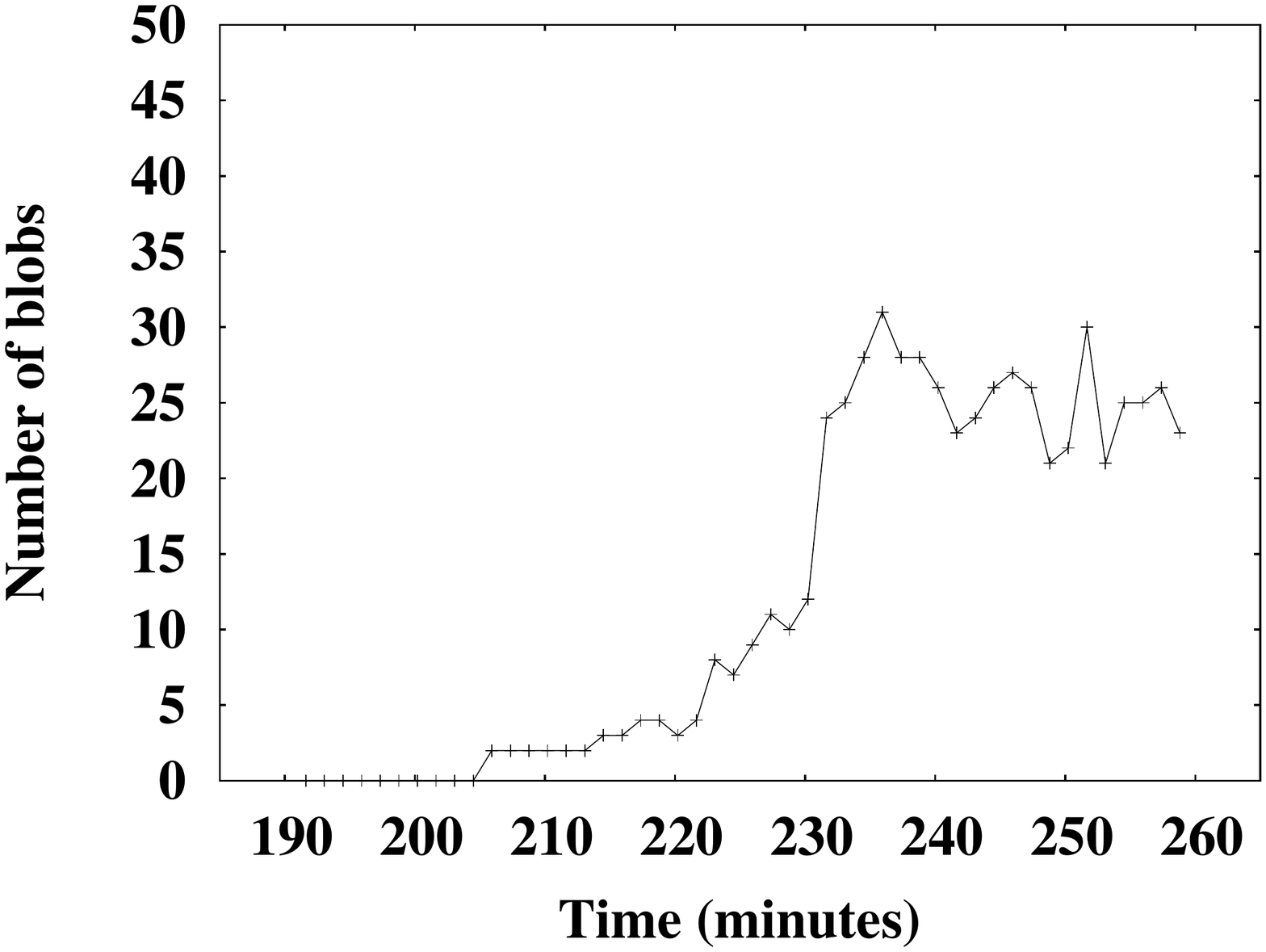}
 \end{tabular}
 \end{center} 
 \caption{The left plot shows the evolution of the mass contained in the blobs with time and the right one shows how the number of the blobs evolves with time.}
 \label{massnumber}
\end{figure}

\begin{figure}
\begin{center}
 \begin{tabular}{ccc}
 \includegraphics[trim=4cm 4cm 4cm 4cm,width=0.45\textwidth]{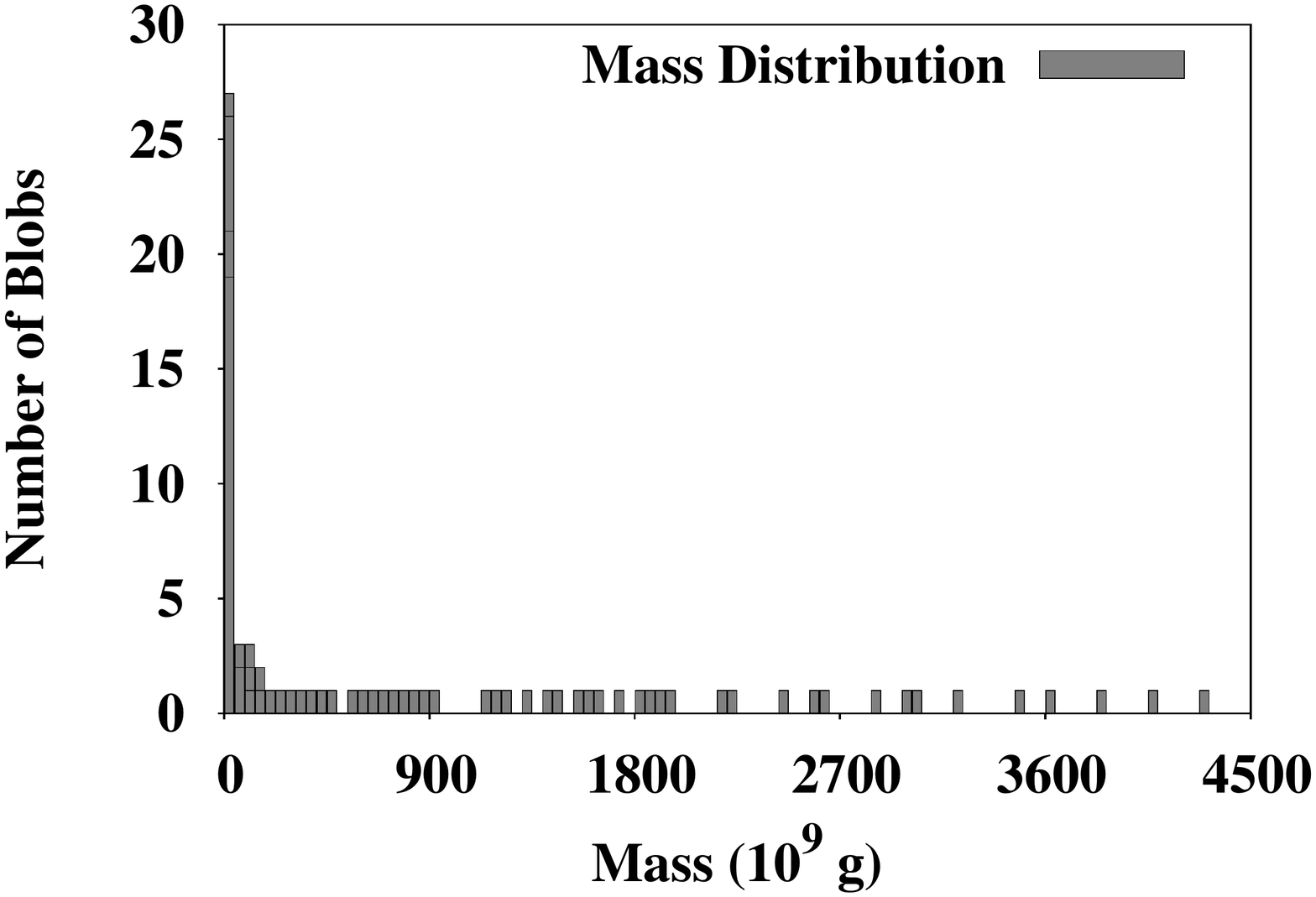}&
  \includegraphics[trim=4cm 4cm 4cm 4cm,width=0.45\textwidth]{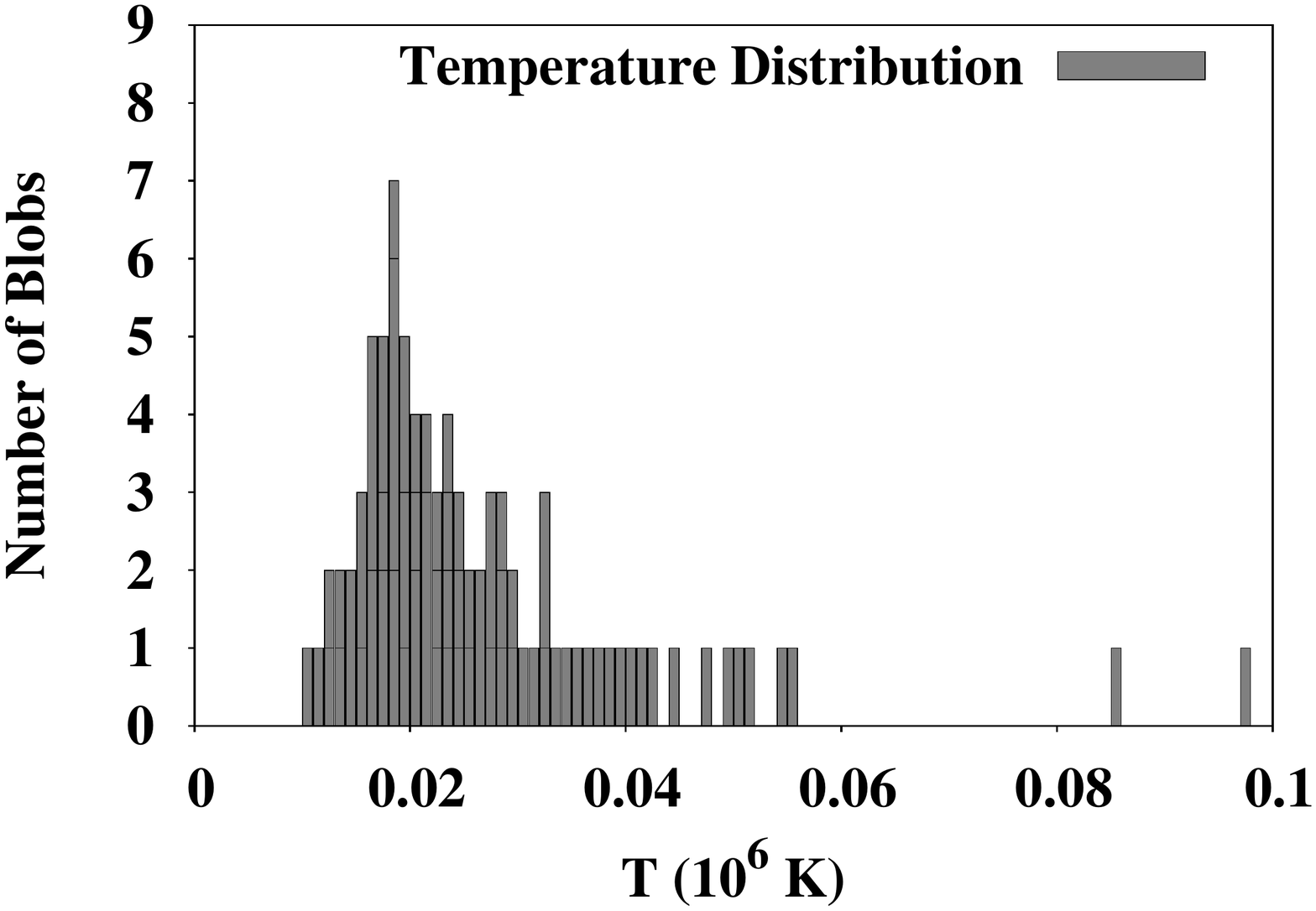}
  \end{tabular}
  \end{center}
\caption{The quantities shown on this figure are the temperature (right) and mass (left) distribution of all the blobs throughout the entire simulation. The blob material is rather cool with most of them having a temperature of 20,000K, while the range is from 10,000K to 100,000K. The vast majority of the blobs has a very small mass, close to the lower limit for their definition, while the distribution range is rather large.}
\label{Tmass}
\end{figure}

In figures~\ref{massnumber} and \ref{Tmass} we demonstrate the results of the analysis using the blob filter.
Specifically, in figure~\ref{massnumber} we show the total mass accumulated in the blobs (left panel) as well as the number of the blobs as the simulation evolves (right panel),
while in figure~\ref{Tmass} we present the distribution of the mass in blobs (left panel), as well as the distribution of the average temperature inside each blob (right panel) throughout the entire simulation.
The minimum mass that can be contained in the smallest grid cell corresponding to the cutoff density criterion of the blob filter is about $\Delta m_{min}=1.1845\times 10^5\mathrm{kg}$.
This means that the peak in the blob mass distribution is about three orders of magnitude bigger than this minimum mass, i.e. $m_{peak}\propto 10^3 \Delta m_{min}$, and is thus not influenced by resolution effects (also aided by the way we introduced a lower cut-off in blob size).
In figure~\ref{massnumber}, we notice that there are two phases, namely, a build-up phase and a saturation phase. 
The build-up phase corresponds to mass accumulation at the middle of the domain, due to evaporation and blob formation.
At some point, blobs obtain enough mass and are subject to the gravitational instability, so they start moving downwards.
When they fall back into the transition region, our filter stops taking them into account in the calculation of the total mass and the number of blobs.
Thus, the episodic behaviour that we witness to take place in the left panel, after $t=230 \mathrm{min}$ and the saturation that we observe in the right panel of figure~\ref{massnumber}, after $t=235 \mathrm{min}$, reflect the time variation of the blob creation and the blob loss, which sets up a mass recycling process. 

\section{Blob formation and evolution}

\begin{figure}[htbp]
\begin{center}
\begin{tabular}{cc}
\includegraphics[trim=1cm 0cm 3cm 0cm,clip=true,width=0.5\linewidth]{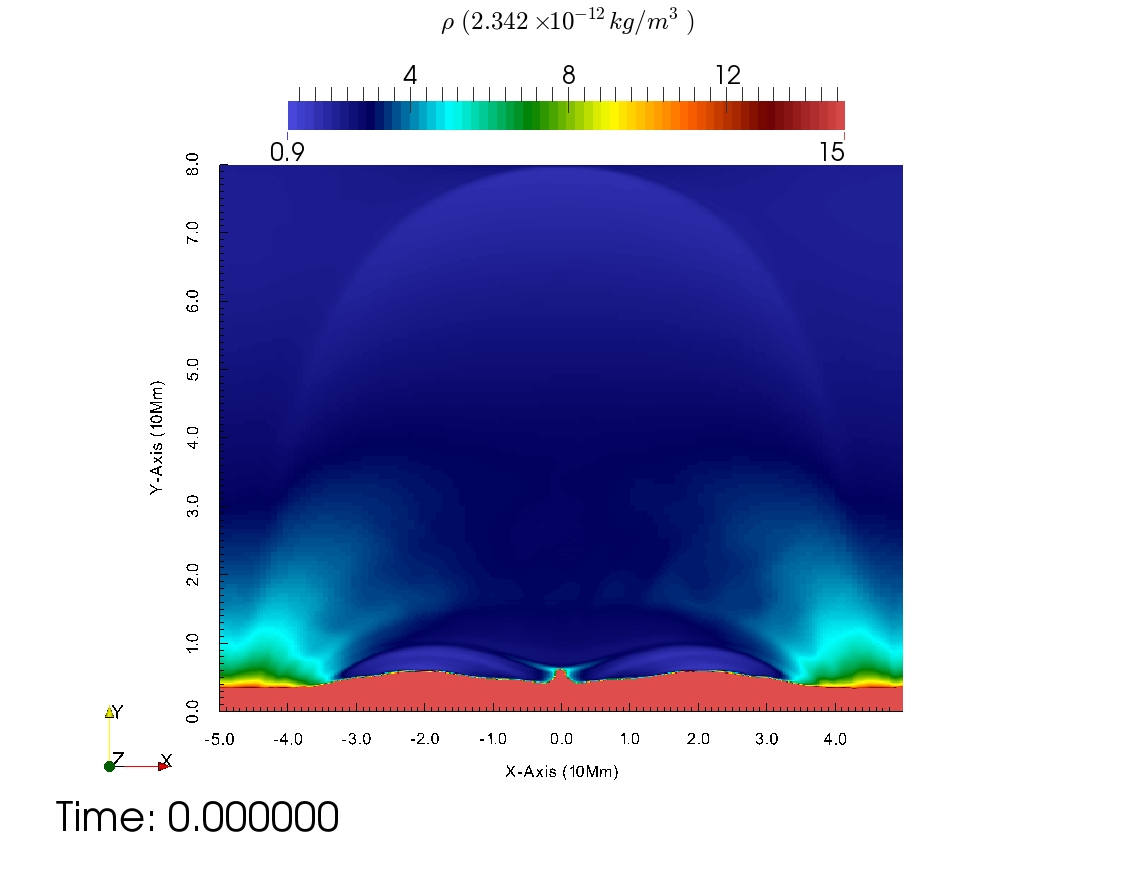}&
\includegraphics[trim=1cm 0cm 3cm 0cm,clip=true,width=0.5\linewidth]{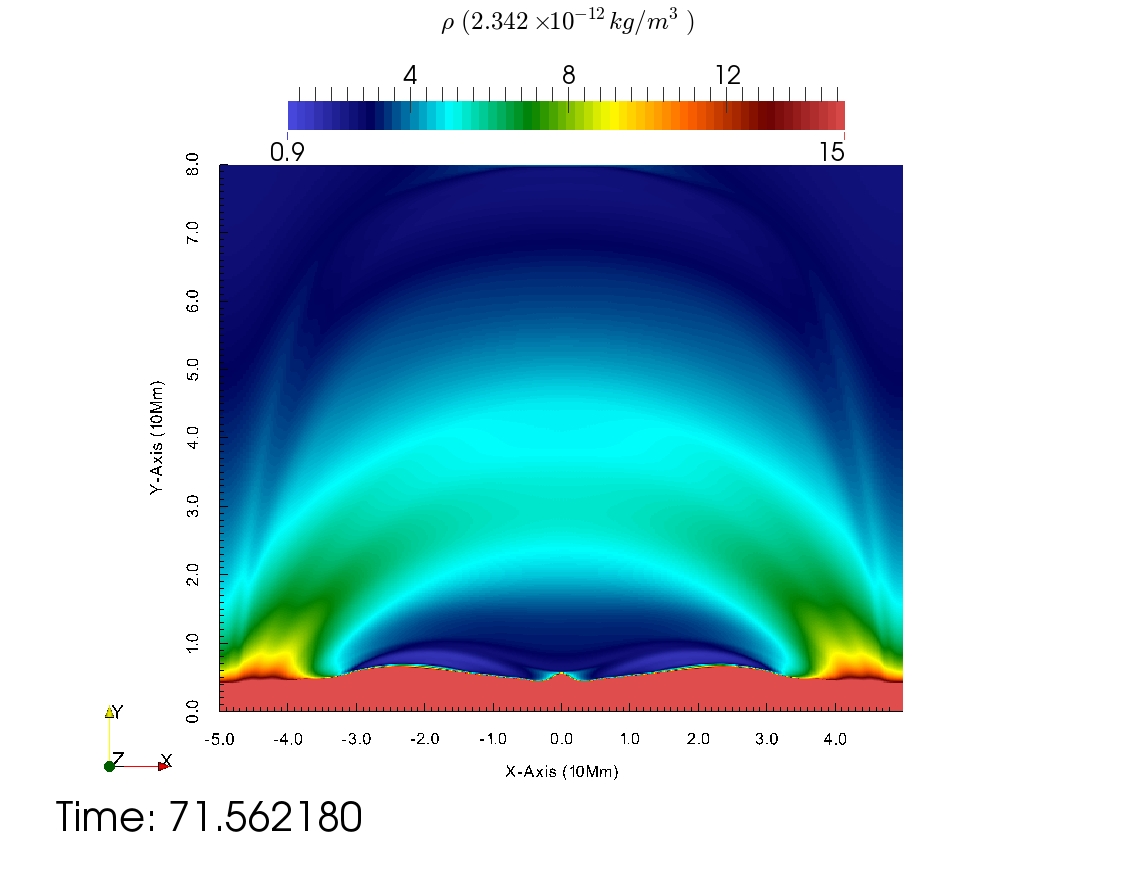}\\
\includegraphics[trim=1cm 0cm 3cm 0cm,clip=true,width=0.5\linewidth]{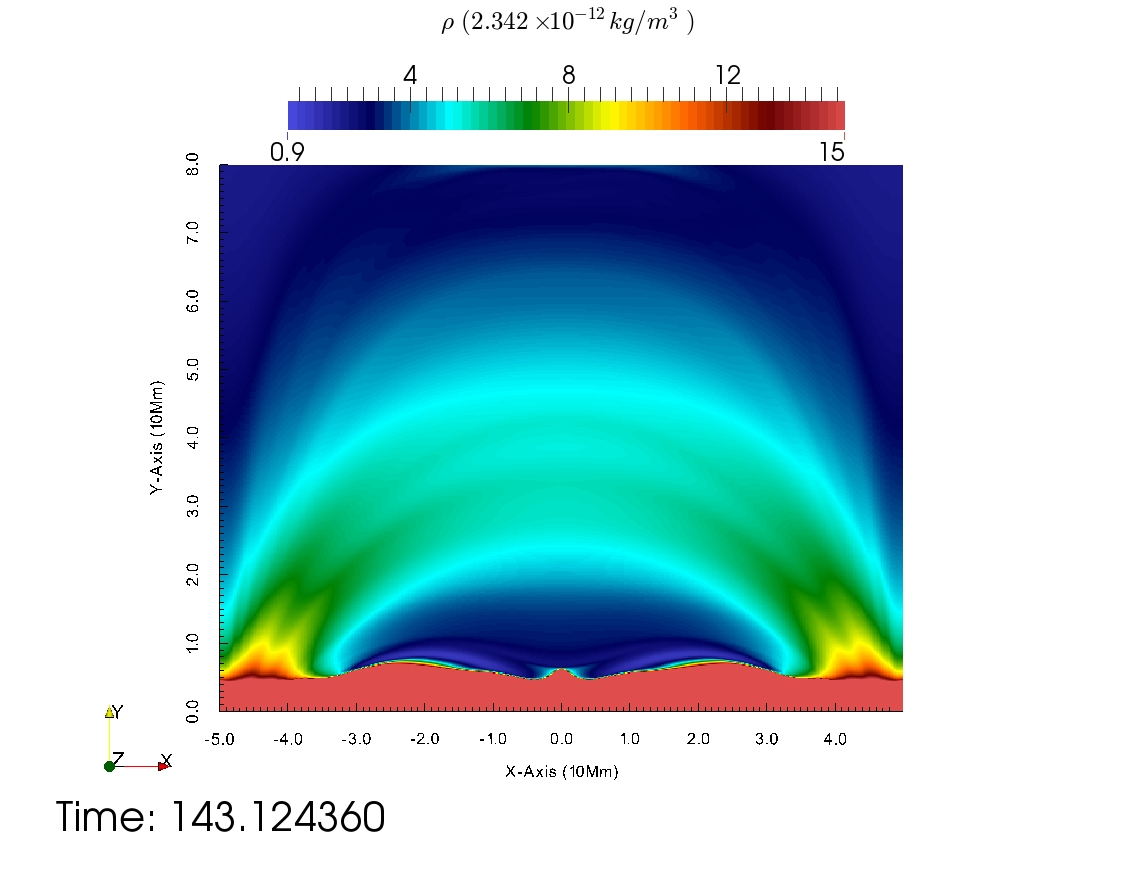}&
\includegraphics[trim=1cm 0cm 3cm 0cm,clip=true,width=0.5\linewidth]{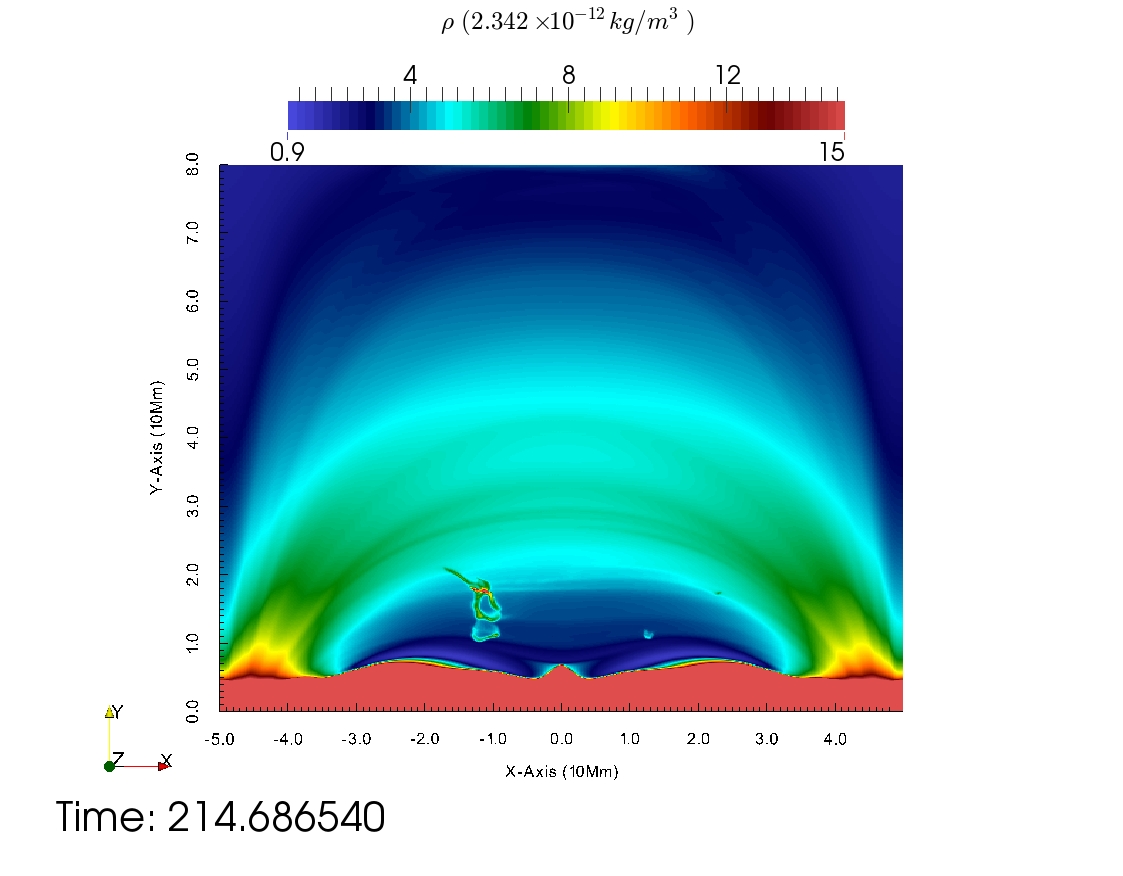}
\end{tabular}
\caption{Here we show the integrated view along the $z-$ direction of the density at four different snapshots with a corresponding time interval of about $72\mathrm{min}$, to demonstrate the evolution and to show where the mass is accumulated due to the heating of the footpoints of the enveloping arcade. The time is annotated in minutes and the density is shown with the desaturated rainbow colorbar. In the top left panel, we show the $t=0 \mathrm{min}$ snapshot, when we activate the localised heating function $H_{lh}$ ($t=t_{relax}$). Later on, (top right and bottom left panel) the mass gets accumulated at heights around $40\mathrm{Mm}$. In the bottom right panel, the phenomenon of the blob runaway events has already started and condensations have made their appearance (at about 205 $\mathrm{min}$ blobs appear for the first time).}
\label{density}
\end{center}
\end{figure}

Due to evaporation from chromospheric and transition region heights, cool material starts concentrating at the top part of the large arcade preferentially within the heated loop part. 
As time passes, and more mass is accumulated at that specific region, at a height of about $4\mathrm{Mm}$, as shown in figure~\ref{density}, the density starts rising and condensation centres make their appearance in our simulation. 
This formation of blobs containing cool material has been found and analysed in 2.5D simulations for a force-free arcade by \citet{Fang13} and causes runaway events. 
In figure~\ref{density}, the evolution of the mass transport and accumulation is presented in four different snapshots with a time interval of about $72\mathrm{min}$.
The mass transport starts from the heated loop footpoints and moves upwards depositing mass, due to evaporation, on the loops above the magnetic dip of the arcade configuration, at a height of about $40\mathrm{Mm}$. The top left panel shows the moment when we activate the localised heating $H_{lh}$ ($t=t_{relax}$) and reset the time to $t=0$.
The first blobs are formed at the time $t\approx 205 \mathrm{min}$ and the last snapshot for our simulation corresponds to a time of about $259 \mathrm{min}$.
We have studied the phenomenon of the blobs for a time duration of about $54 \mathrm{min}$, as we present further on.

\begin{figure}
 \begin{center}
\begin{tabular}{c}
  \includegraphics[trim=0cm 2cm 0cm 4cm,width=0.8\textwidth]{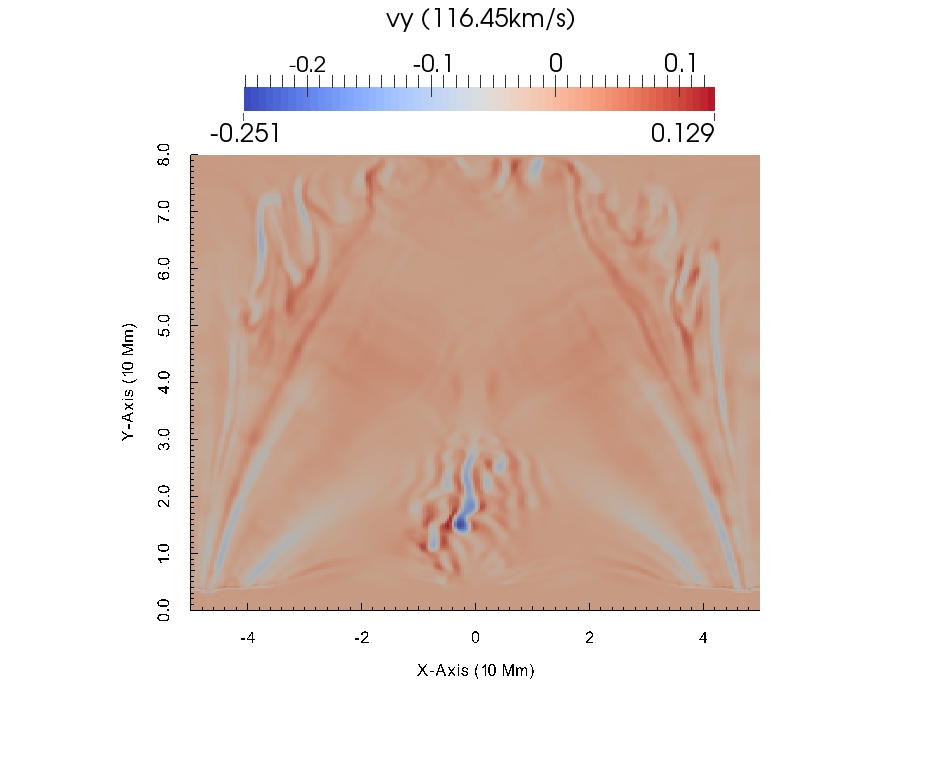}\\
  \includegraphics[trim=0cm 2cm 0cm 0cm,width=0.8\textwidth]{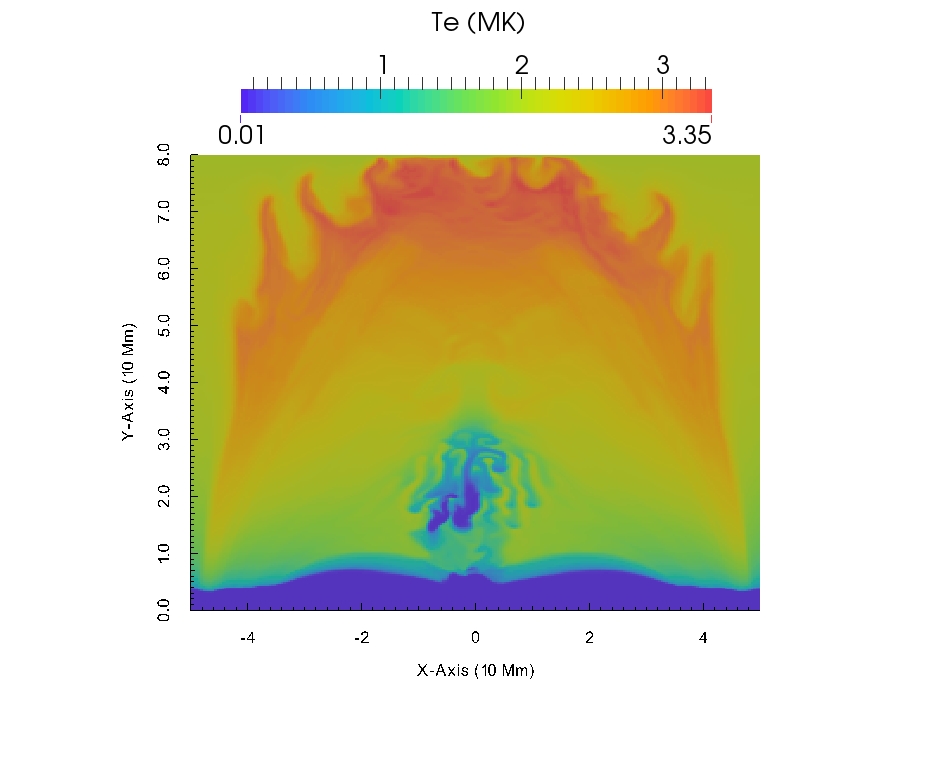}
  \end{tabular}
  \end{center}
 \caption{We present here the vertical component of the plasma velocity (top panel) and the temperature profile (bottom panel) at the time corresponding to 235$\mathrm{min}$ of physical time. Both $\mathrm{v_y}$ and T are shown in cross-section on the $x-y$ plane corresponding to $z=0$. We observe clear indications of Rayleigh-Taylor instability, seen as finger-like structures at heights $y\approx 20-30\mathrm{Mm}$.}
 \label{temp}
\end{figure}

A continuous coronal rain develops, but here in a fully 3D fashion and the blob dynamics show clear indications of Rayleigh-Taylor or interchange instability.
More specifically, as can be seen in figure~\ref{temp}, we observe vertical finger-like structures in the middle part of the velocity map (top panel), with opposite direction velocities appearing next to each other. 
Interesting features appear also close to the top of the simulation domain, with opposite velocities coexisting next to each other, indicative of a kind of stratification driven mixing occurring there. 
The magnetic field strength decreases exponentially with height, according to its analytic prescription that was presented in the previous section. 
In the second interesting region, at the top, the dominant component of the magnetic field is the third horizontal component, $B_z$.

The temperature profile helps to locate the cooler blob as demonstrated in the bottom panel of figure~\ref{temp}, corresponding to the plane $z=0$. 
Finger-like structures appear again at the same region as in the velocity profile. 
Interestingly, the first blobs make their appearance above the magnetic dip of our arcade configuration and this is the region from where on we start seeing the finger-like formations. 
From this perspective the condensed blobs seem to fall through the almost horizontal magnetic field lines, i.e the blobs look like they pass through the field lines, a scenario which would mean that the frozen-in condition gets violated. 
We will further analyse this phenomenon from different angles, to clarify the real blob motion later in this paper.

\section{SDO synthetic views}

\begin{figure} 
\begin{center}
\begin{tabular}{cc}
\includegraphics[width=0.5\linewidth]{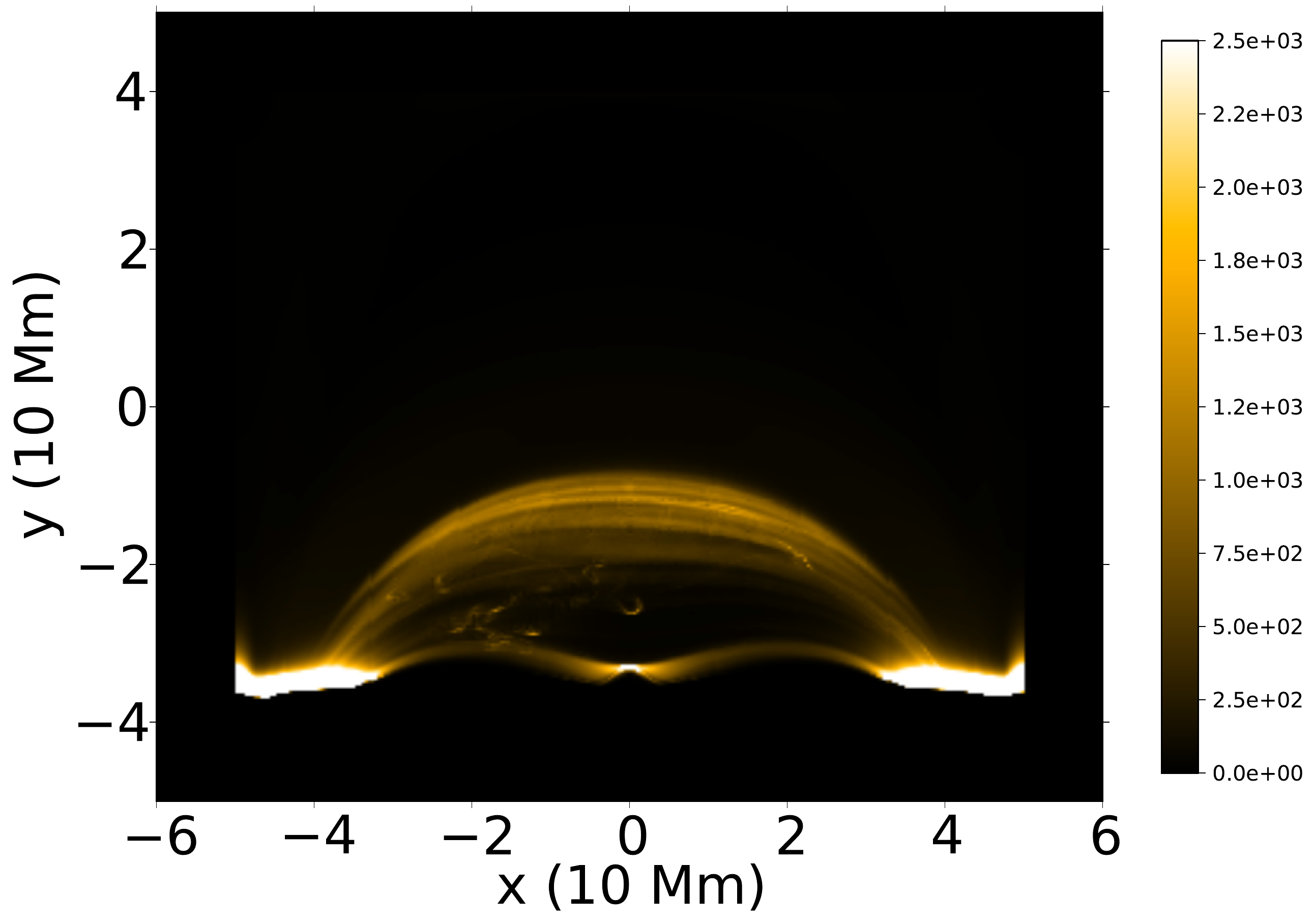}&
\includegraphics[width=0.5\linewidth]{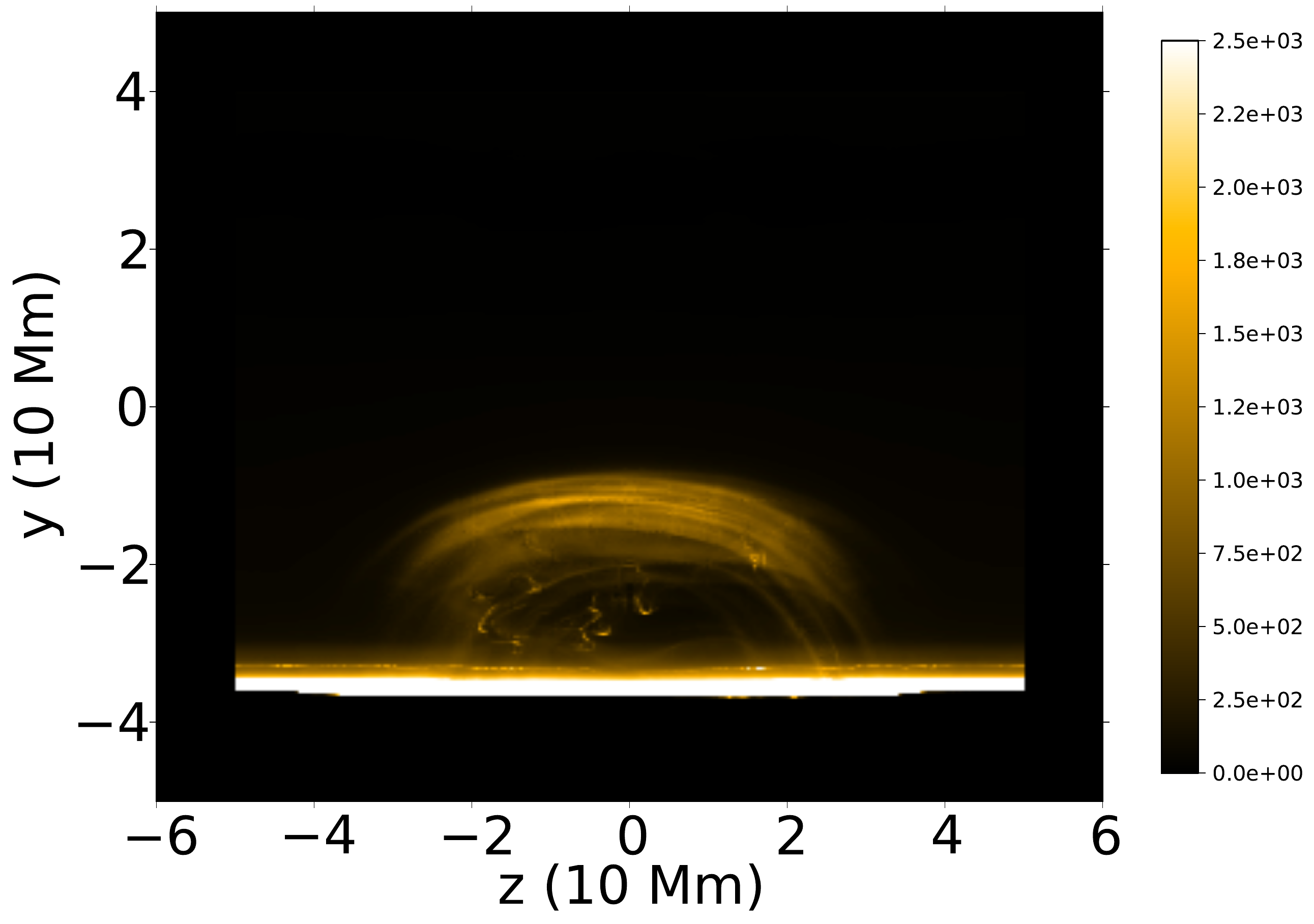}\\
\includegraphics[width=0.5\linewidth]{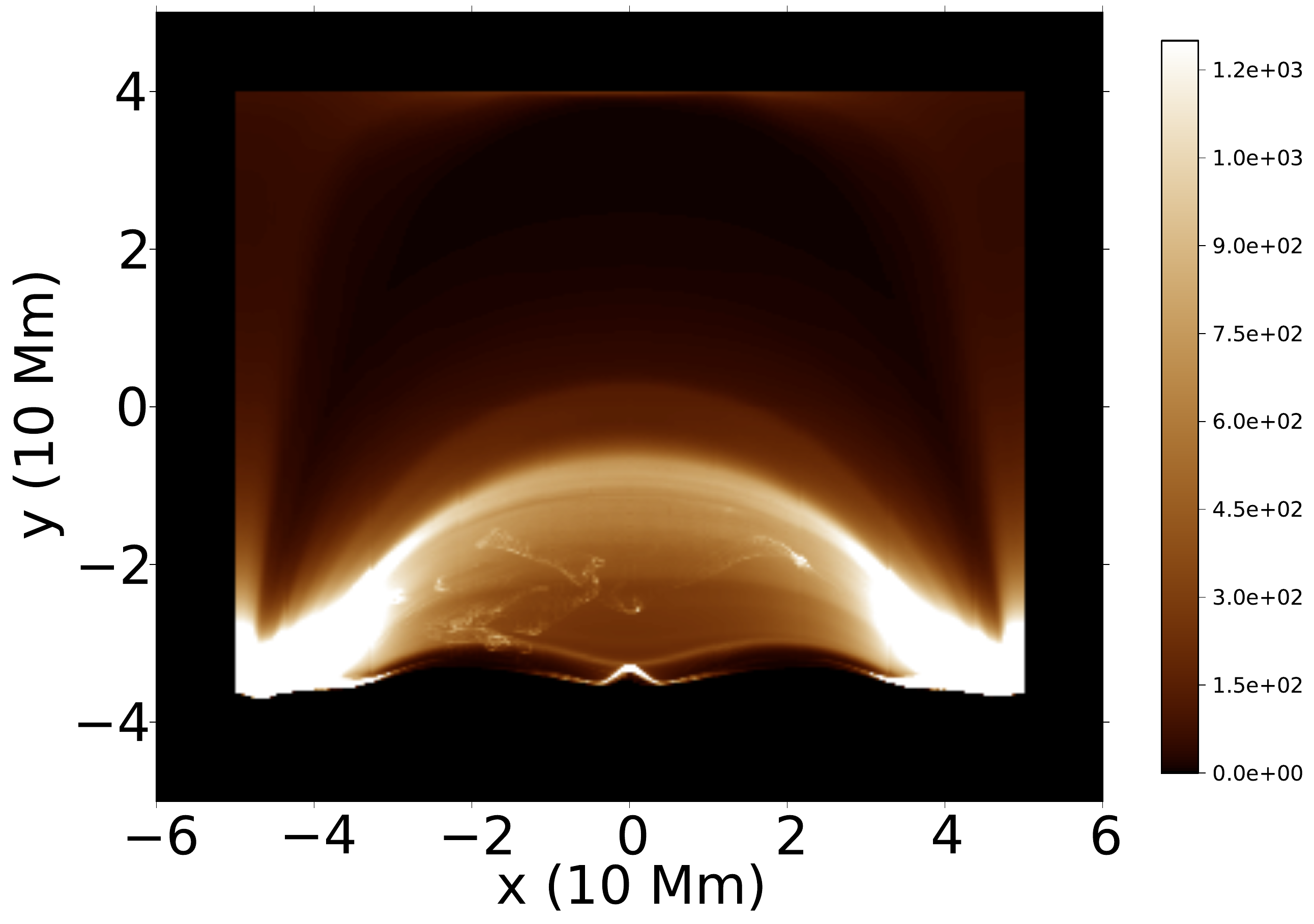}&
\includegraphics[width=0.5\linewidth]{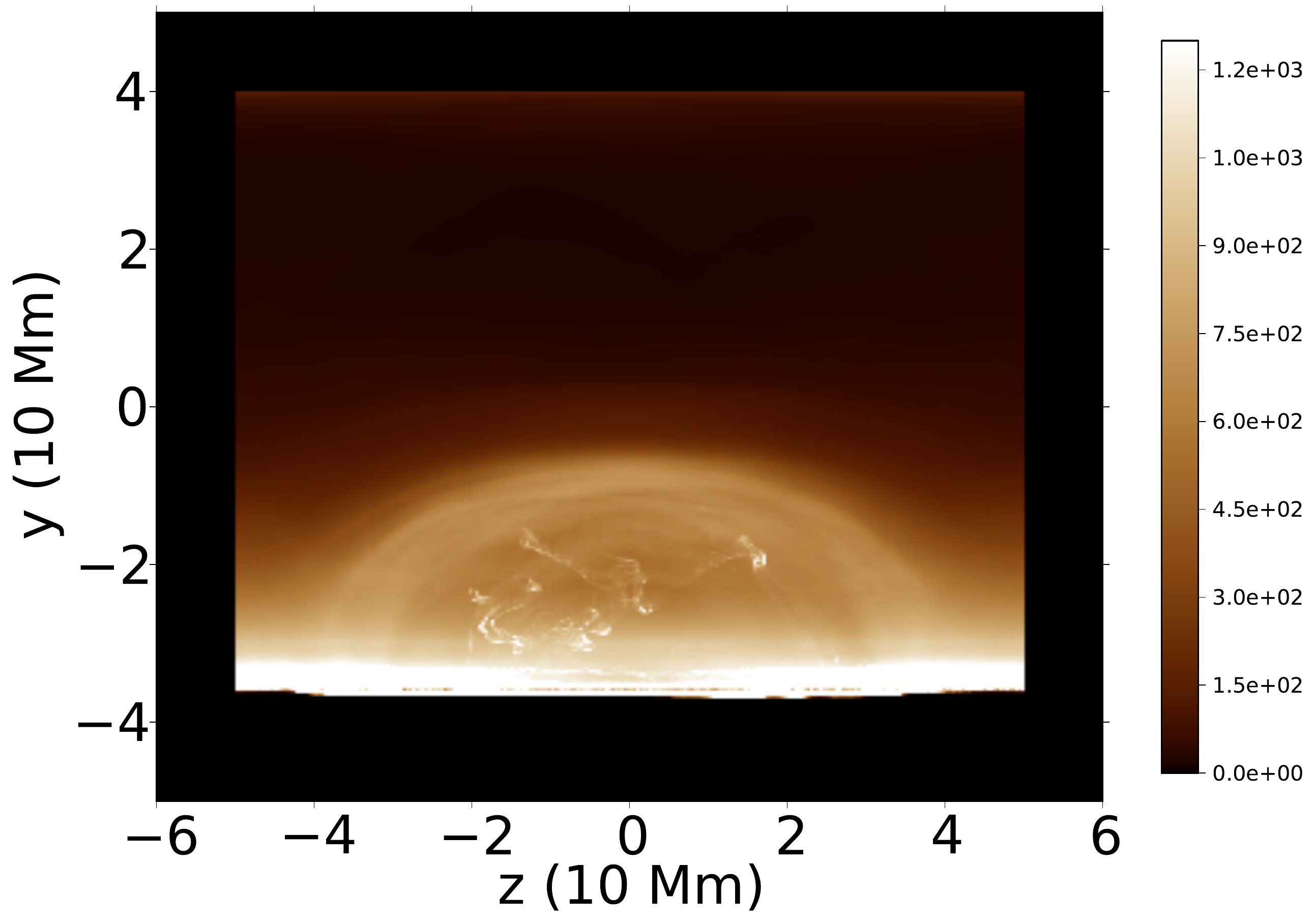}
\end{tabular}
 \end{center}
   \caption{SDO/AIA synoptic views in two different wavelengths showing the $z-$integrated side view on the left and the $x-$integrated side view on the right of the whole simulation box. This shows a specific snapshot corresponding to the 235$\mathrm{min}$ of physical time in 171 \AA (top panels) and 193 \AA (bottom panels) EUV channel.}
   \label{sdo171}
\end{figure}

\begin{figure} 
\begin{center}
\begin{tabular}{cc}
\includegraphics[width=0.5\linewidth]{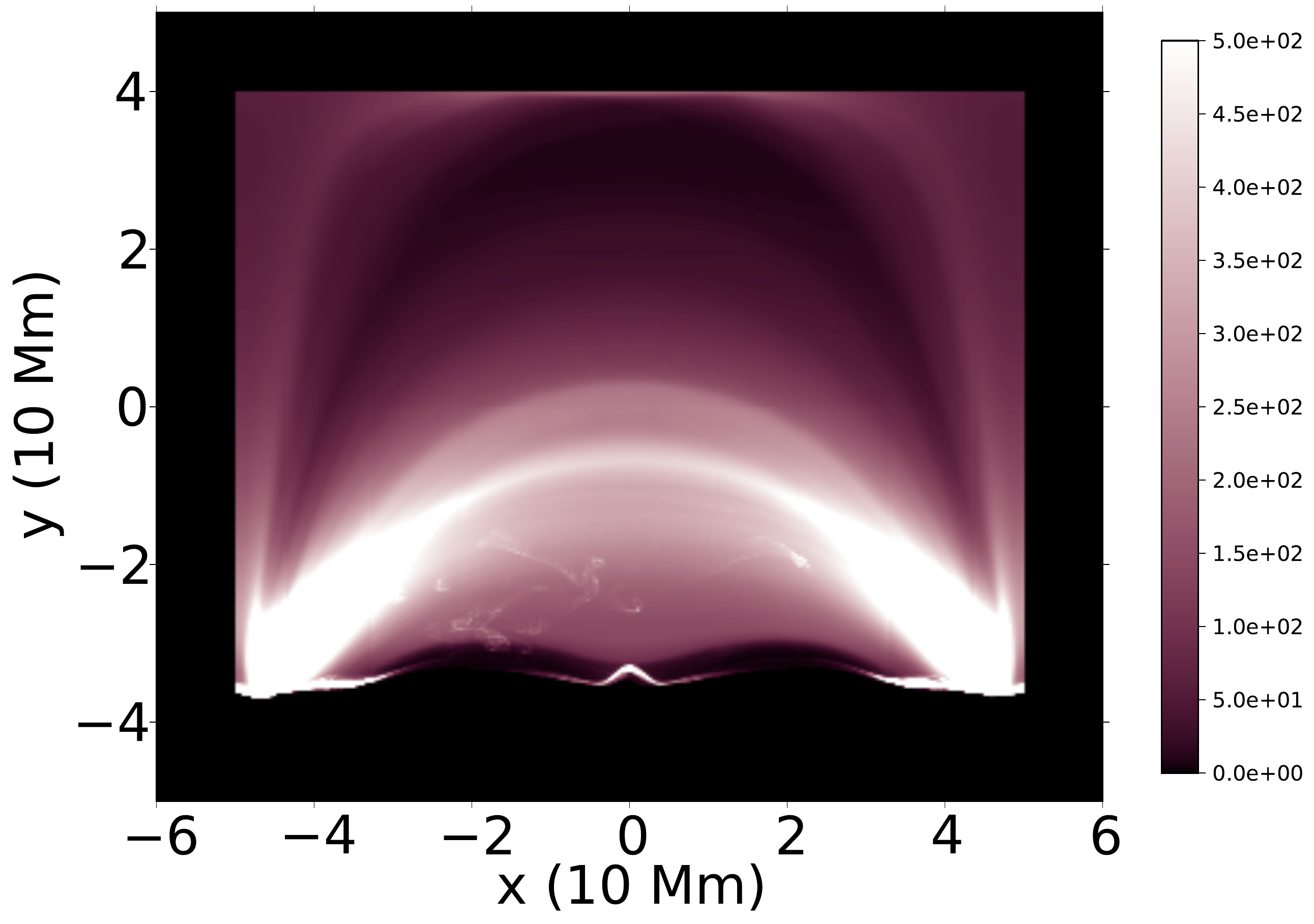}&
\includegraphics[width=0.5\linewidth]{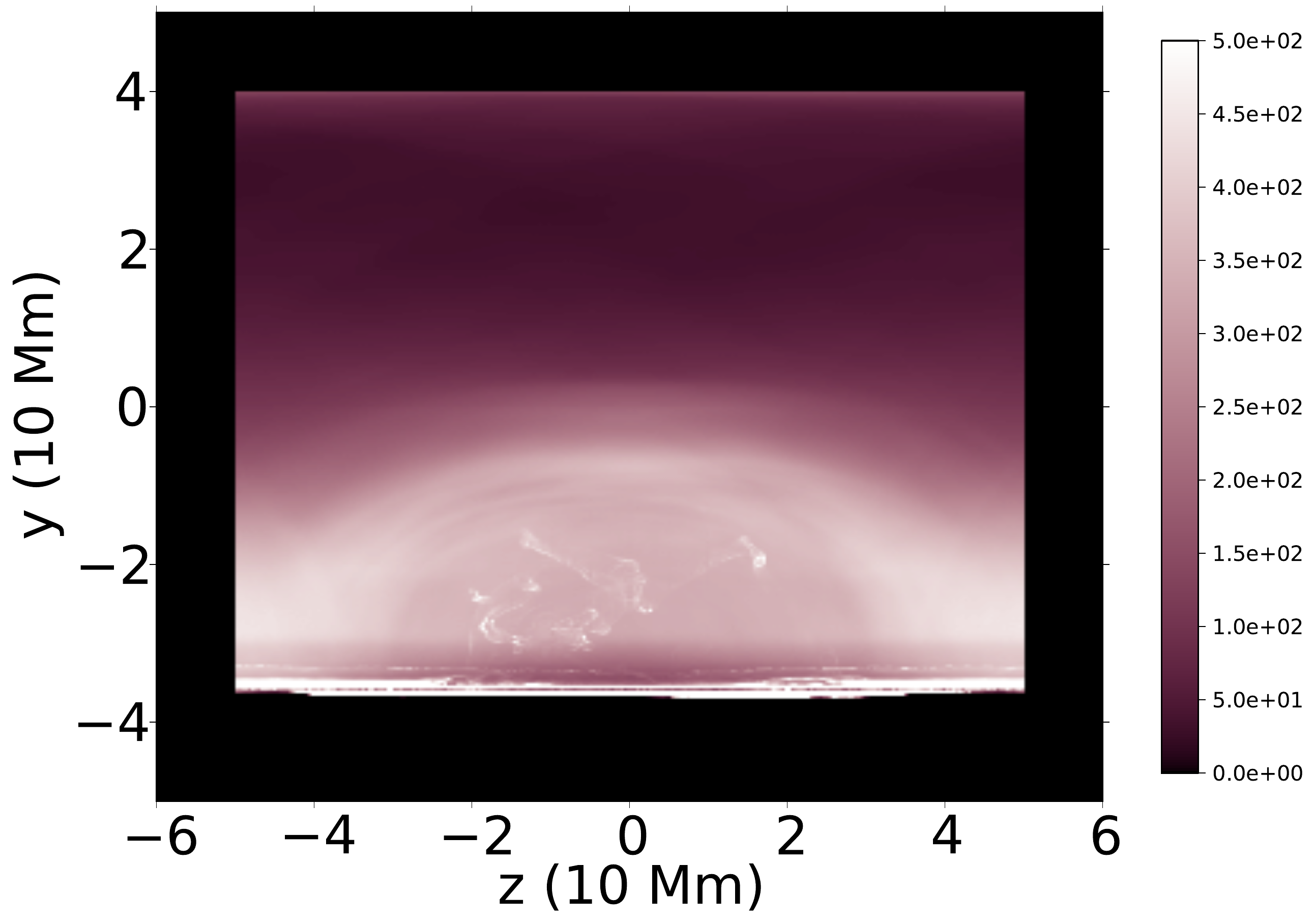}\\
\includegraphics[width=0.5\linewidth]{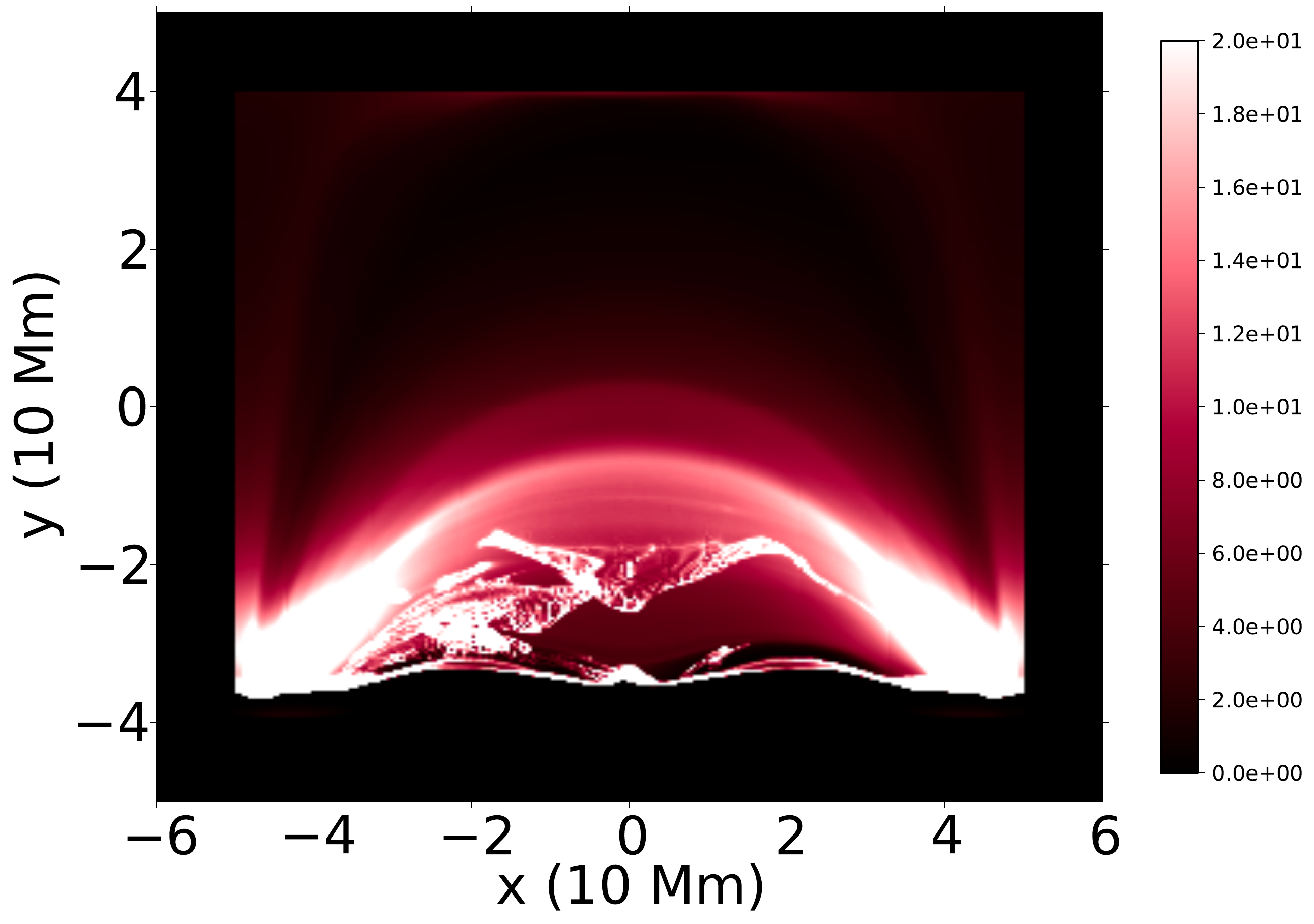}&
\includegraphics[width=0.5\linewidth]{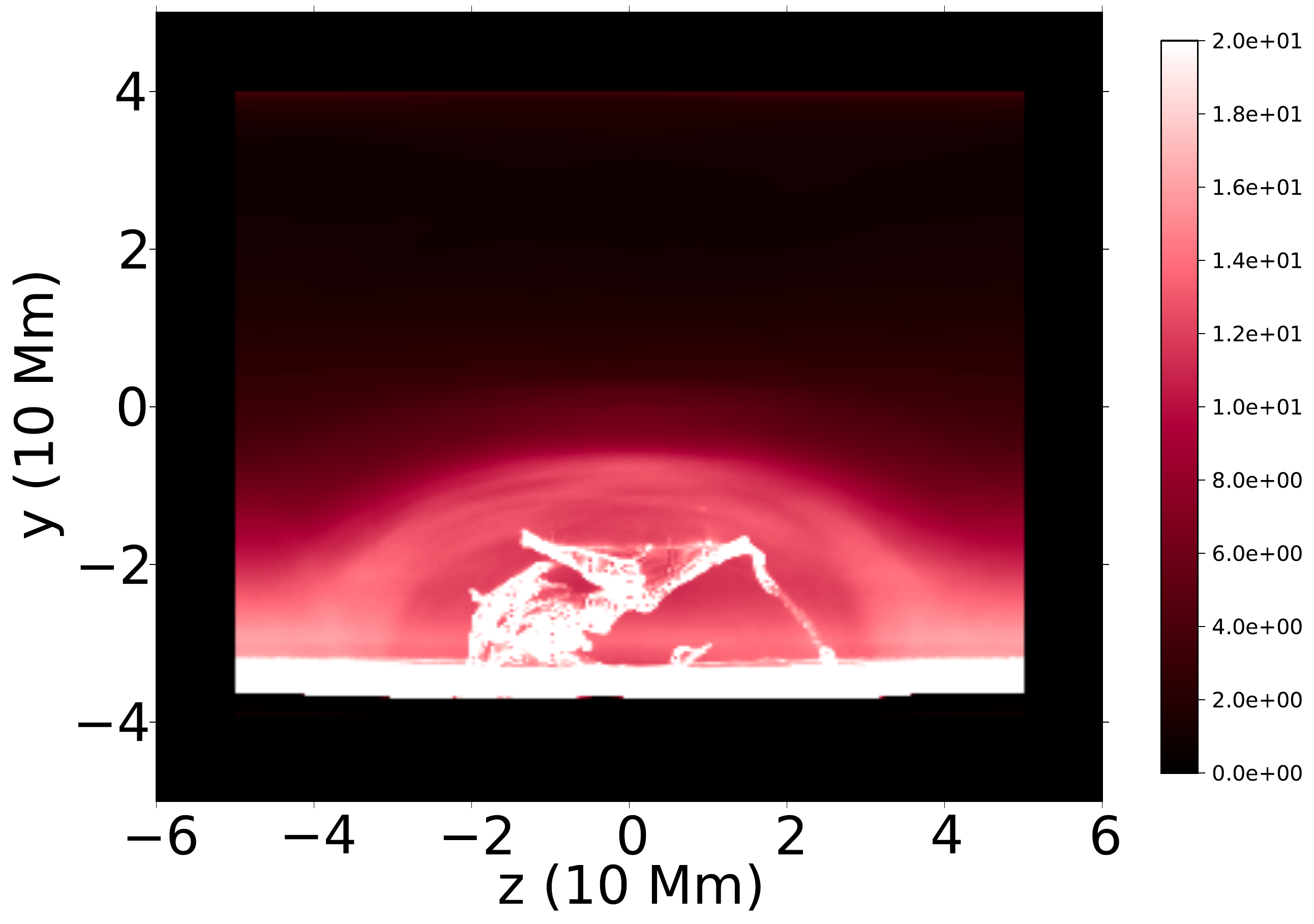}
\end{tabular}
 \end{center}
   \caption{SDO/AIA synoptic views in 211 \AA (top panels) and 304 \AA (bottom panels) wavelengths showing the $z-$integrated side view on the left and the $x-$integrated side view on the right of the whole simulation box. This shows a specific snapshot corresponding to the 235$\mathrm{min}$ of physical time, and demonstrates that the cool blobs are mostly seen in the 304\AA EUV channel.}
   \label{sdo211}
\end{figure}

We construct and present synthetic views from SDO/AIA in four different wavelengths 171 \AA,193 \AA, 211 \AA, 304 \AA, that have their main emission contributions for plasma at temperatures of 0.8MK, 1.5 MK, 1.8MK, 0.08MK respectively. For the same time $t=235\mathrm{min}$ as shown in figure~\ref{temp} we present these four EUV views in figures~\ref{sdo171} and \ref{sdo211}, with the two first wavelengths (171 \AA,193 \AA) corresponding to the first and second row of the figure ~\ref{sdo171} and the two last ones (211 \AA, 304 \AA) corresponding to the first and second row of figure~\ref{sdo211}. We show two different sideway views corresponding to the $z-$integrated and the $x-$integrated view point for the left and right column of both figures accordingly, for one specific snapshot corresponding to $235\mathrm{min}$ of physical time. 
The most informative view is the one that corresponds to the cool filter, i.e. the bottom row of figure~\ref{sdo211}. 
The blob dynamics and their structure becomes evident in this 304 \AA\ channel, allowing us to follow more closely the evolution of the position of the condensation centres with time. 
In animated views, we observe that mostly blobs form and move on the left part of the $z-$integrated view and on the right part of the $x-$integrated view. 
This means that (numerically) accumulated asymmetries have rendered one half of the heated arcade loops close to thermal instability.
The cool material overall seems to follow the magnetic configuration during its evolution. 
In all filters, the footpoints of the enveloping heated large loop system appear enhanced in brightness, as the heating process is captured. 
At the 171\AA\ and 193\AA\ wavelengths (figure~\ref{sdo171}) the blob centres are enhanced and the phenomenon resembles coronal rain on these views (movie provided).

\section{Mass circulation}

As we will quantify further on, blobs form due to thermal instability,  creating condensation centres. 
After the blobs are formed, we will show further on that they are cospatial with regions liable to interchange instability.
Its nonlinear evolution dominates the local dynamics and the blobs obtain an overall downward velocity. 

\begin{figure}[htbp]
\begin{center}
\includegraphics[width=0.8\linewidth]{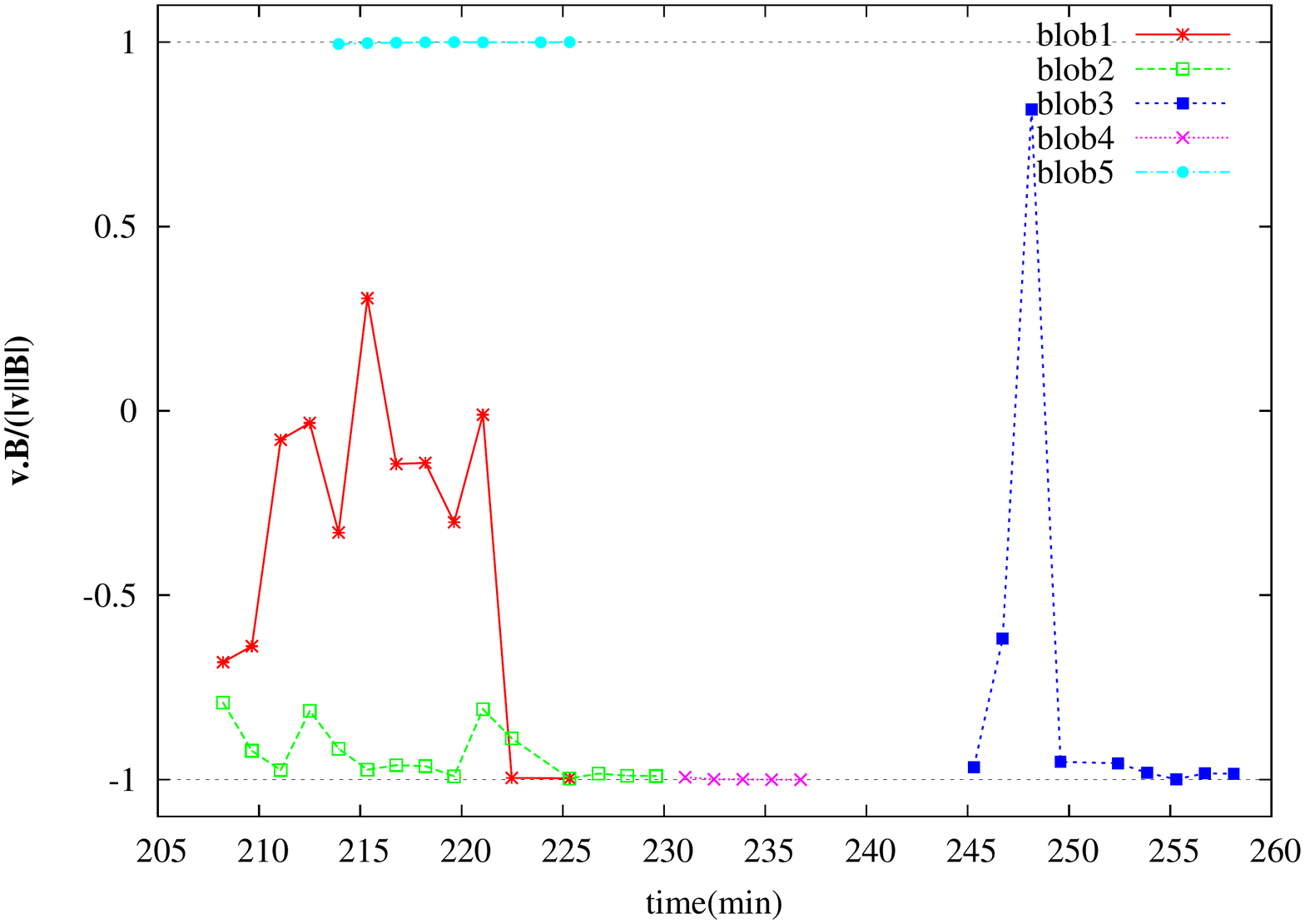}
\caption{The velocity component parallel to the local vector magnetic field is quantified here for five of the heaviest blobs in our experiment. We observe that there is a tendency for each blob in our sample to parallelise to the local magnetic field direction (parallel or antiparallel) later in their circulation through the corona as they approach the loop footpoints.}
\label{blobs}
\end{center}
\end{figure}

\begin{figure}[htbp]
\begin{center}
\begin{tabular}{cc}
\includegraphics[width=0.46\textwidth]{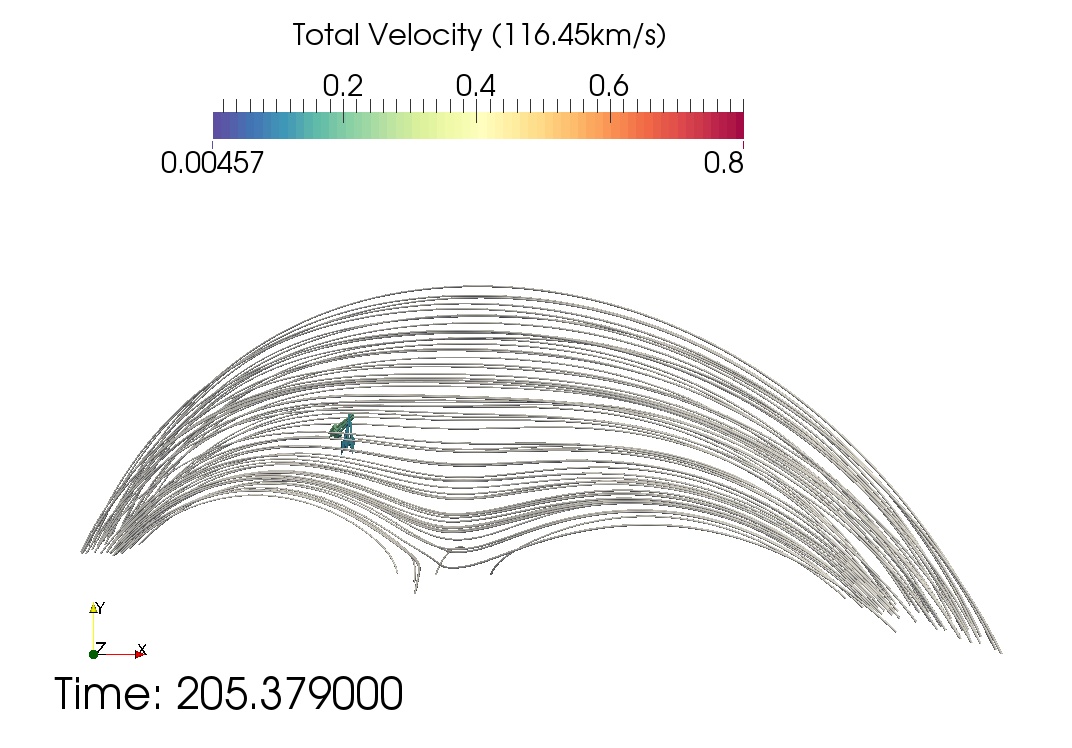}&
\includegraphics[width=0.46\textwidth]{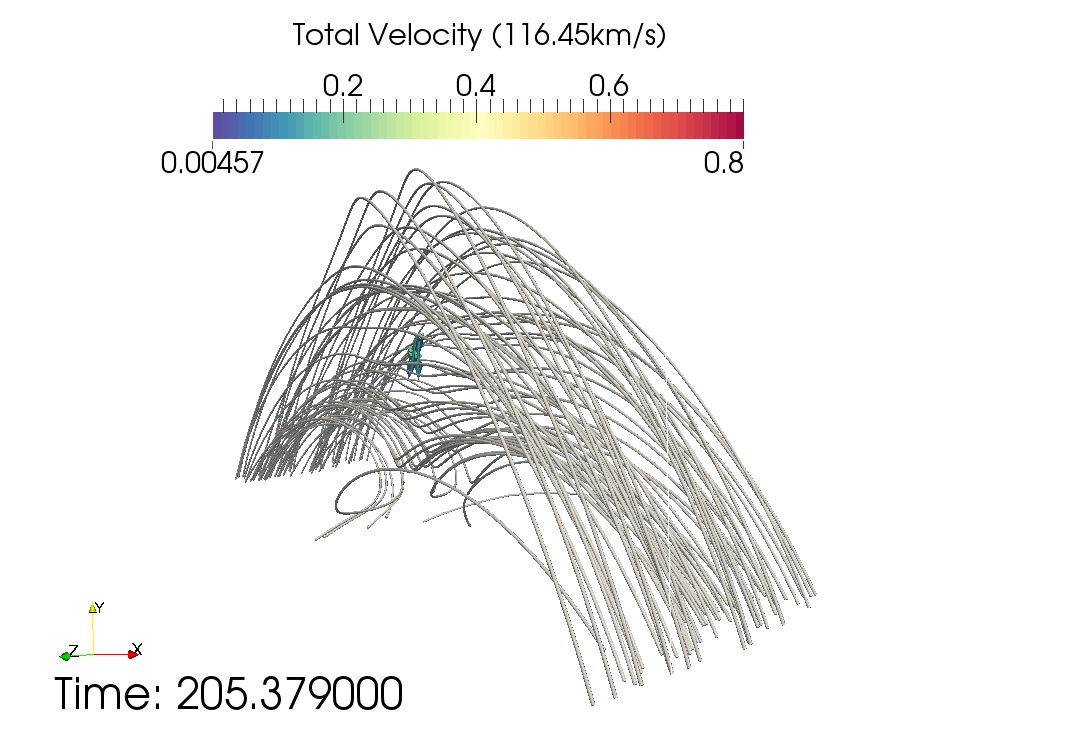}\\
\includegraphics[width=0.46\textwidth]{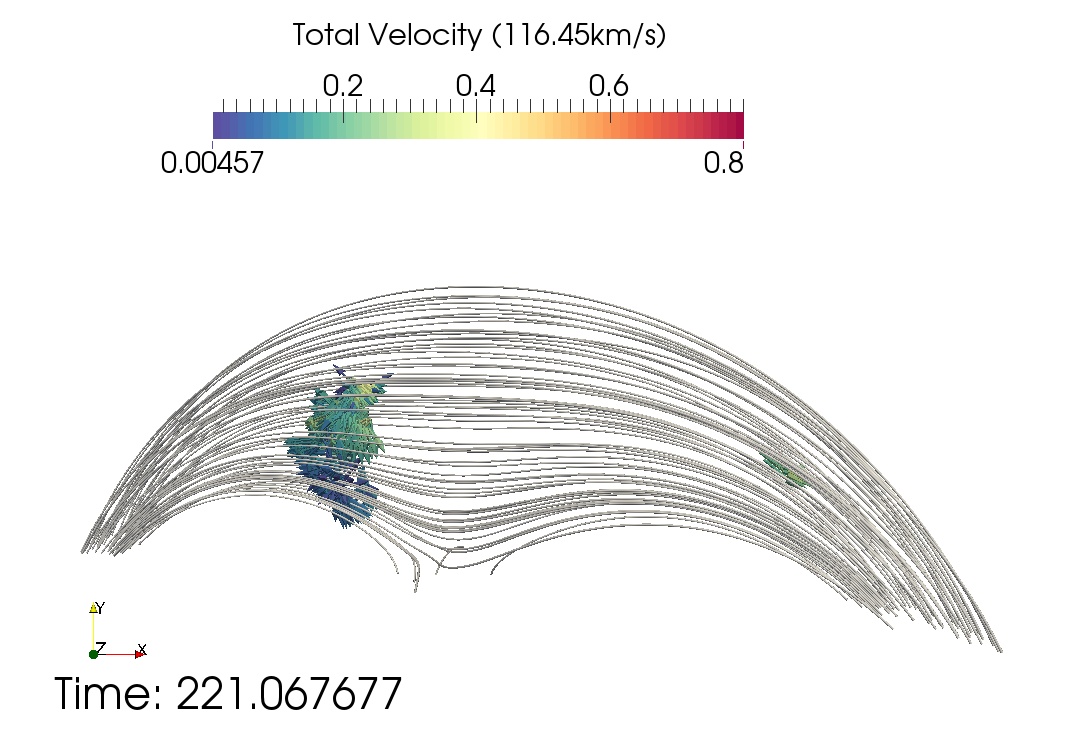}&
\includegraphics[width=0.46\textwidth]{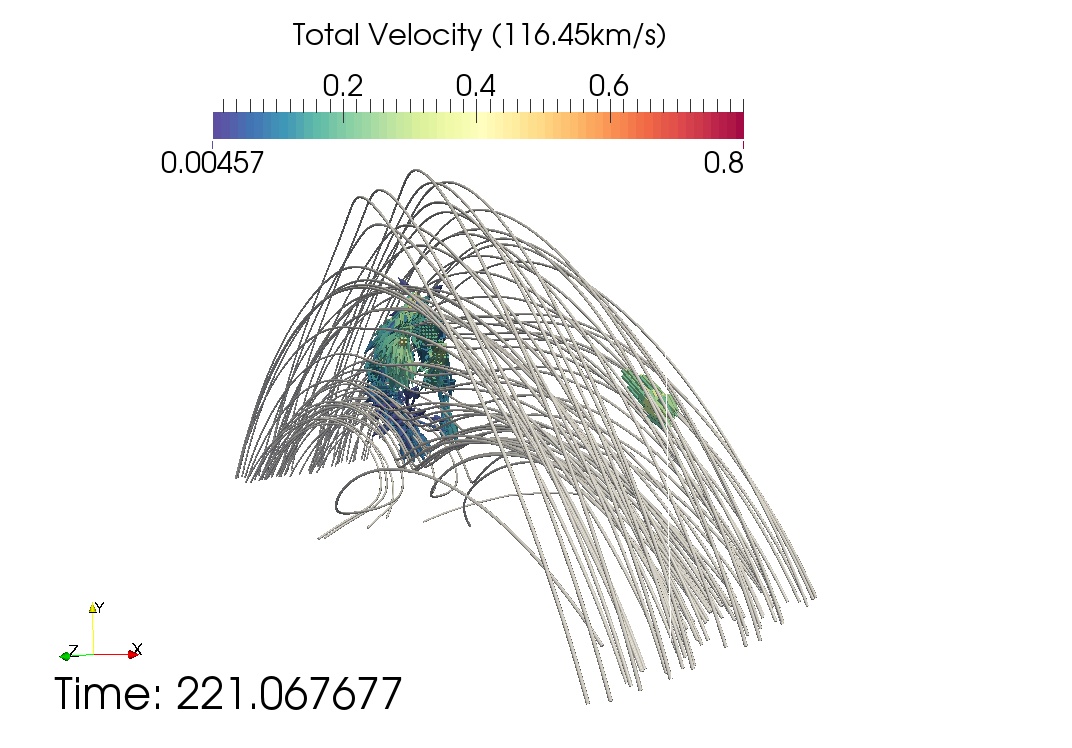}\\
\includegraphics[width=0.46\textwidth]{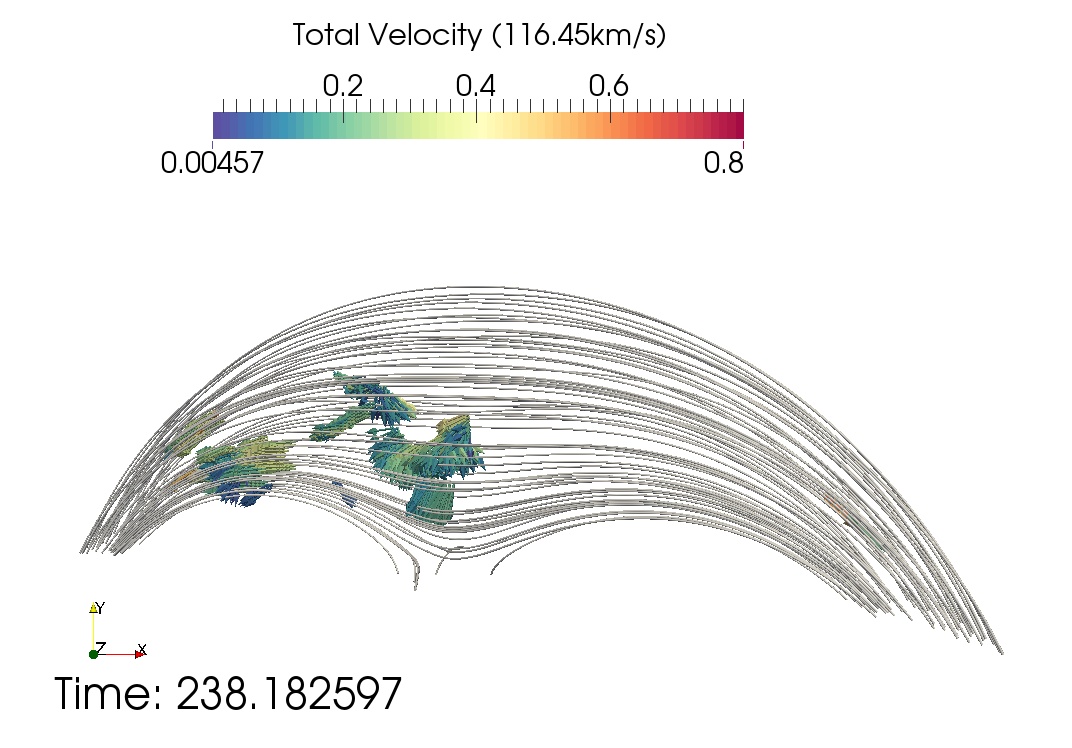}&
\includegraphics[width=0.46\textwidth]{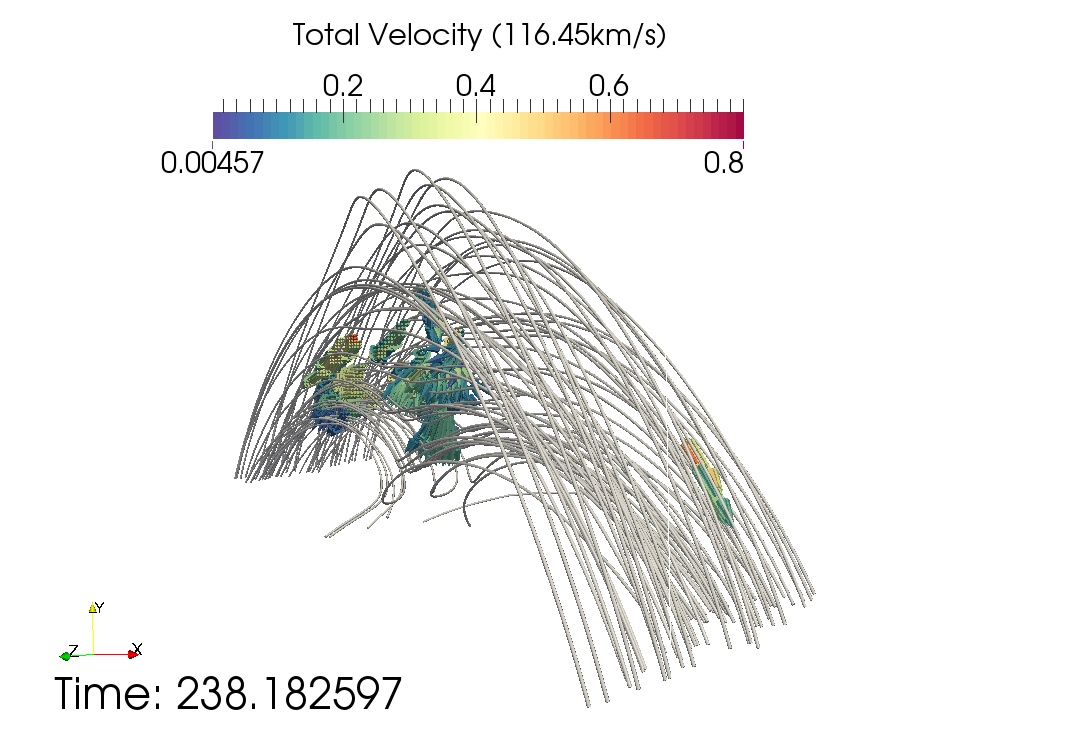}\\
\includegraphics[width=0.46\textwidth]{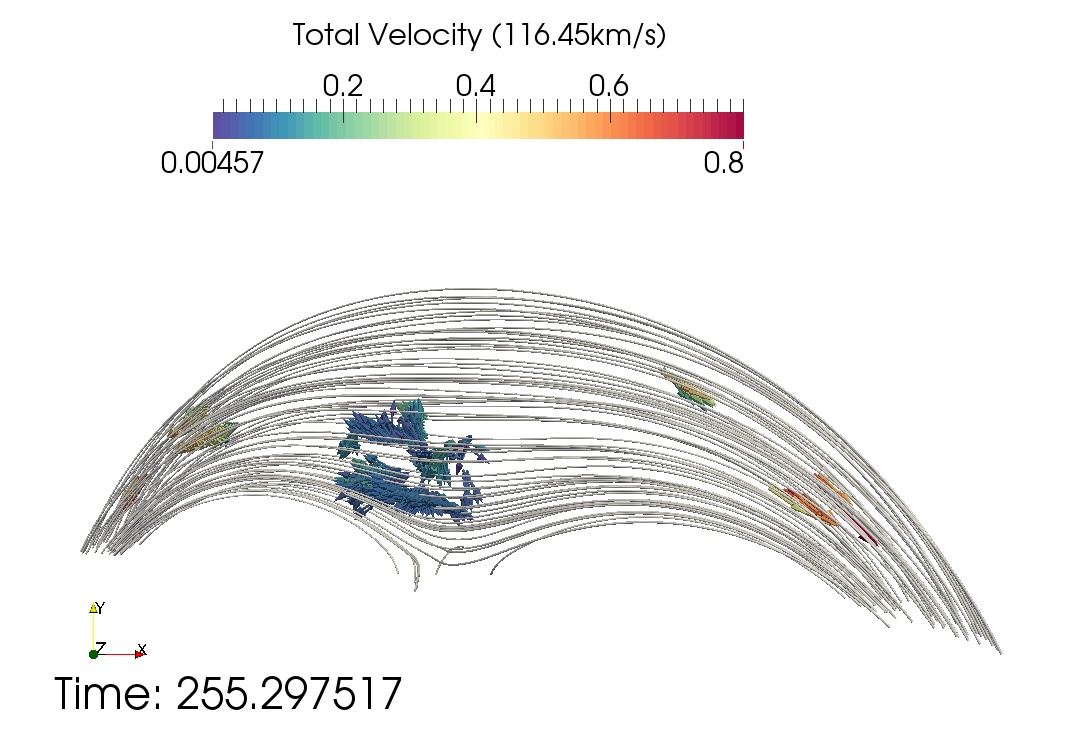}&
\includegraphics[width=0.46\textwidth]{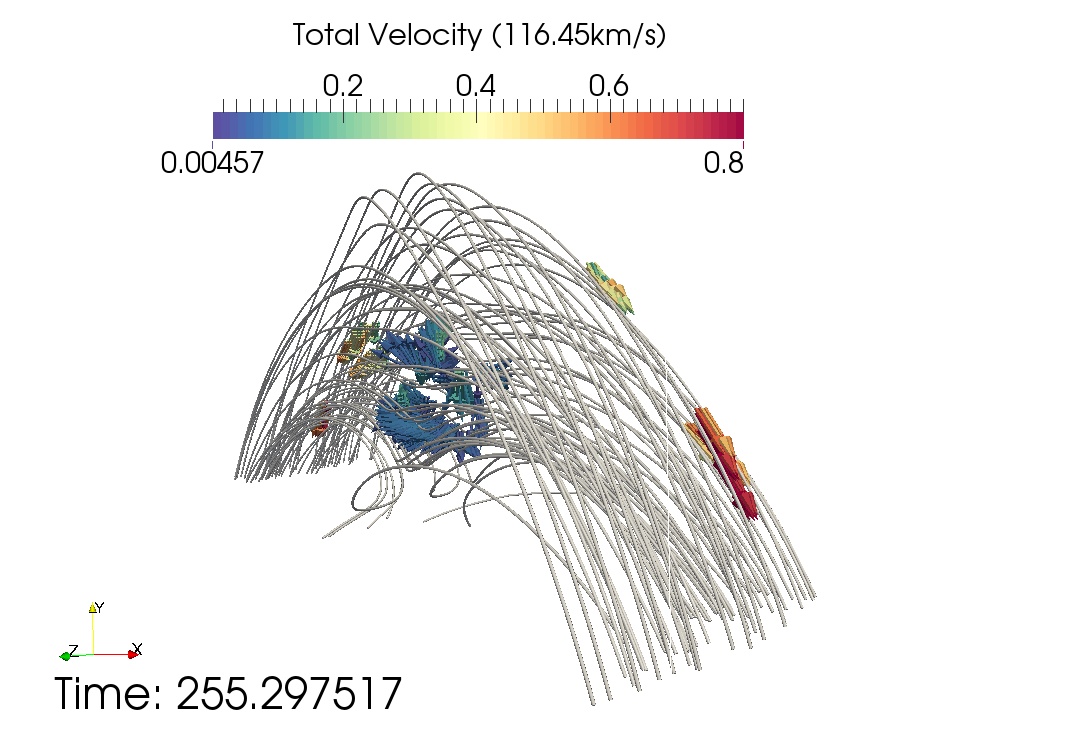}\
\end{tabular}
\caption{Here we present from two different lines of sight the magnetic field topology as well as the velocity vector field of the condensed material in four different snapshots.}
\label{circulation}
\end{center}
\end{figure}

More specifically, the blobs are very dynamic and as such they evolve changing their number, size and physical characteristics, splitting into smaller pieces or colliding forming larger cool and dense plasma elements. At the beginning, from the moment blobs start appearing in the simulation, they increase in mass and number, figure~\ref{massnumber}. The vast majority of the blobs have small masses, as shown in the left panel of figure~\ref{Tmass}, but overall the blobs have masses that spread out to a quite wide range of values varying from 45 to 4500 $10^9$g. As the time passes, they start splitting and small blobs start following more closely the magnetic field topology on the lower part of the configuration and specifically near the footpoints. Figure~\ref{blobs} provides evidence of this gradual alignment of the plasma clump motion with the local magnetic field direction, regardless their earlier directionality. At blob birth, blobs can have velocity components perpendicular to the magnetic field acquired when condensing (due to e.g. pressure difference). Note that we just quantify the centroid motion here. The blob growth during formation also influences motion of the blobs perpendicular to the magnetic field. Then they ultimately descend into the transition region with accelerating speeds. At time $t\approx225\mathrm{min}$, one part of the cool plasma follows the left footpoint of the large arcade to reach the transition region with increasing speed downwards while another part moves towards the centre of the configuration, just on top of the magnetic dip of the quadrupolar system, where a more complicated motion takes place. Specifically, overall we observe how blob features develop a rather circular motion on the top of the magnetic dip, that is accompanied by further blob splitting with some blobs falling along the field line topology towards both sides of the footpoints of the large arcade, while some parts have an apparent falling motion in the middle of the magnetic structure. This angular motion accelerates slightly. While this circular motion takes place new small blobs merge with the existing circulating plasma in that region. Another part of the cool plasma in the blobs that was found above the centre of the quadrupolar arcade system follows the topology on top of the small arcade and falls back to the transition region following their footpoints. The plasma that falls through the central part of the configuration, where the neighbouring footpoints with opposite polarities of the two small arcades come close, doesn't show any noticeable acceleration.
An impression of the blob dynamics is shown in Figure~\ref{circulation}, which shows, from two perspectives, the field topology, and where each blob element has a fixed length vector representation with its direction defined by the flow, and colored by the flow magnitude.

\begin{figure}[htbp]
\begin{center}
\begin{tabular}{cc}
\includegraphics[width=.45\linewidth]{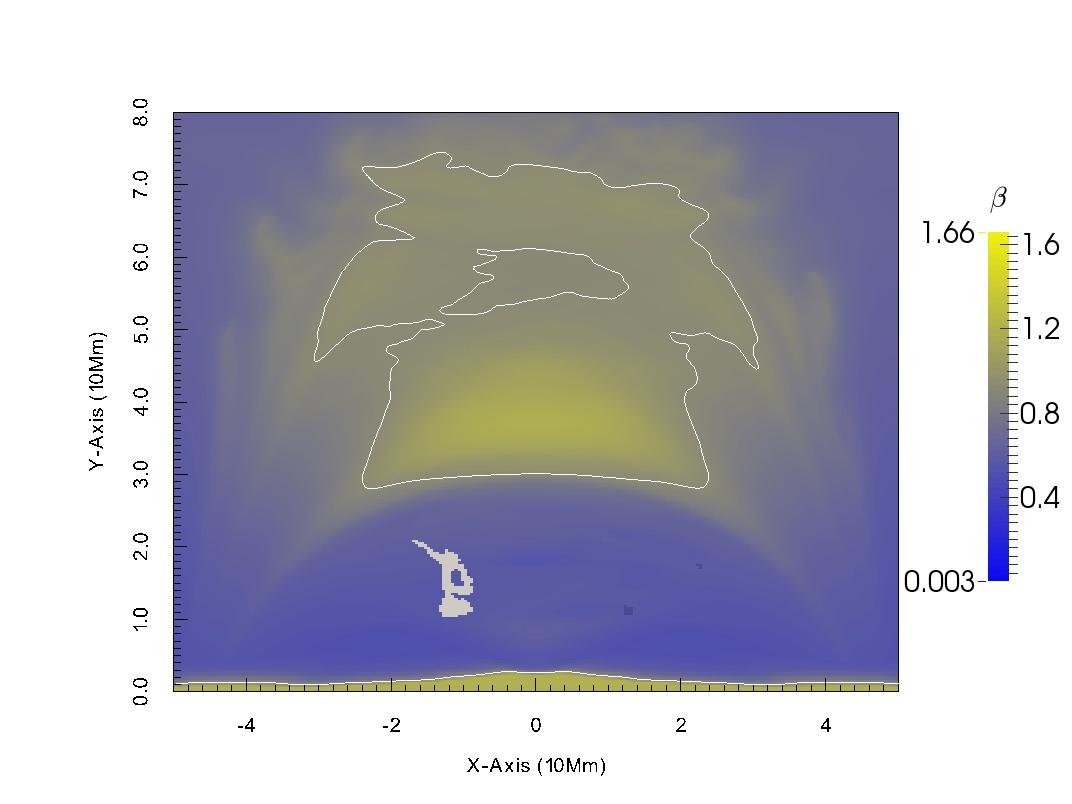}&
\includegraphics[width=.45\linewidth]{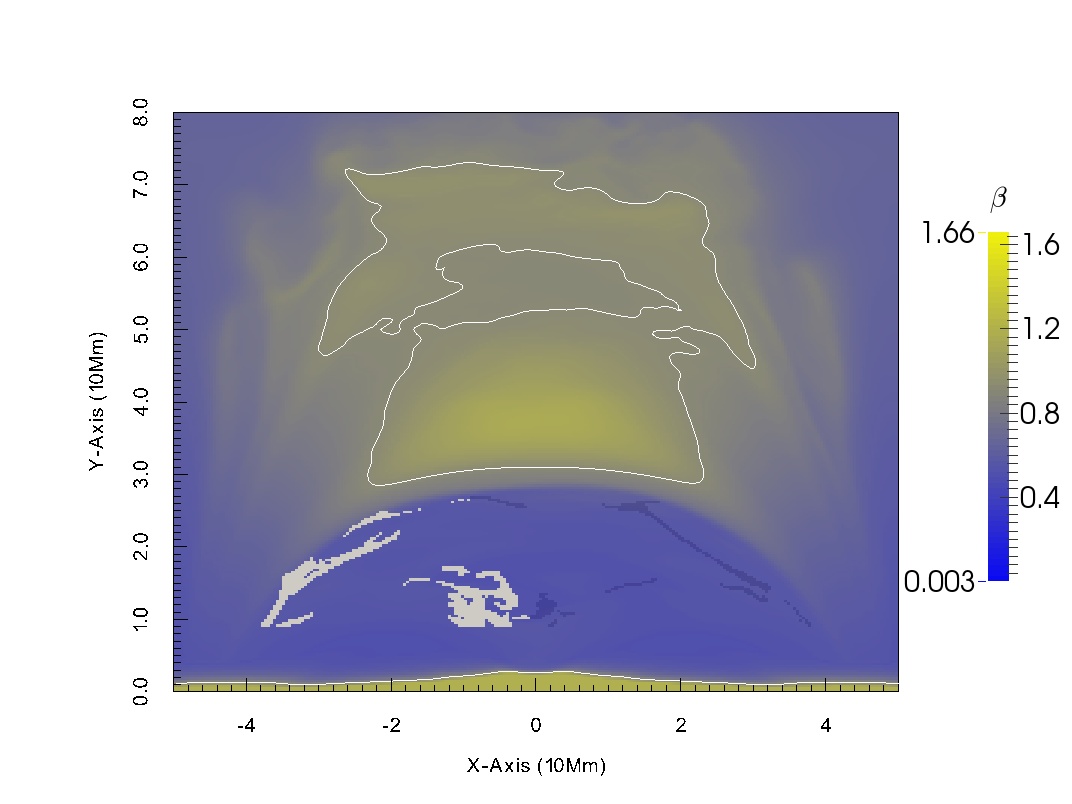}
\end{tabular}
\caption{Plasma-$\beta$ profile on the slice that corresponds to the diagonal plane $x=-z$. The white curve indicates the region where $\beta=1$. Yellow represents high $\beta$ values and blue low $\beta$ values. We depict (left panel) one of the first snapshots that show the blob formation ($\approx$214min), as well as, (right panel) the final snapshot of our simulation ($\approx$259min), when there are still blobs circulating inside the coronal volume. Blobs are represented with white colour and they are projected on the 2D view. The plane $x=-z$ has opacity 0.7 so that it allows for blobs that are behind it to be seen in dark blue shades.}
\label{beta}
\end{center}
\end{figure}

The plasma-$\beta$, representing the ratio of plasma pressure to magnetic pressure, in the lower part of the simulated region is smaller than unity, as indicated in figure~\ref{beta}. This is the region where the blobs get formed and evolve moving overall downwards. As the phenomenon progresses and the blobs move towards lower latitudes the plasma-$\beta$ decreases further in that region. More specifically the plasma-$\beta$ range of values is a) $8.4\times 10^{-3}<\beta < 0.557$ for the blobs at all times b) $3.4\times 10^{-3}<\beta < 1.7$ for the physical domain above 10Mm and c) $2.4\times 10^{-3}<\beta < 121$ for the whole physical domain. So we conclude that the blobs have low-beta values throughout the simulation. This observation is an indication that the magnetic field dominates. In order to accurately resolve and explain the plasma displacements and then link with the magnetic topology of our system, we need to develop particle tracing techniques on top of our MHD simulation to see how fluid elements obey frozen-in conditions. In this paper we only present more indirect means to analyse the Lagrangian property and we leave this computationally demanding particle tracing for future analysis.

\section{MHD stability analysis}
In the remainder of this paper, we will now make direct contact with linear MHD theory governing especially the instability criteria that may drive further nonlinear evolution.
In general, the linearised MHD equations (about time independent physical quantities in equilibrium states $p(\vec{r}),\vec{B}(\vec{r}),\rho(\vec{r})$) accept solutions in normal mode fashion, according to \citet{Goedbloed04}:
\begin{equation}
\vec{\xi}(\vec{r},t)=\hat{\xi}(\vec{r})e^{-i\omega t} \ ,
\end{equation}
where $\vec{\xi}$ is the plasma displacement vector. In ideal MHD, this leads to the eigenvalue problem:
\begin{equation}
\vec{F}(\hat{\xi})=-\rho \omega^2 \hat{\xi}
\end{equation}
where $\vec{F}$ is the ideal MHD force operator.
Both discrete and continuous eigenvalues are accepted and their set of \{$\omega^2$\} form collectively the spectrum. For true ideal MHD conditions, the operator $\rho^{-1} \vec{F}$
is self-adjoint so that the eigenvalues $\omega^2$ are real, which means the eigenvalues $\omega$ can be either real or imaginary, which respectively lead to waves and instabilities.

For instabilities we have an exponential growth of the initial perturbation. However, our current setup is characterised by inclusion of non-ideal effects like thermal conduction and radiative losses. This makes that exponentially growing entropy modes can exist, or that magnetosonic wave modes can become overstable. Moreover, bulk flow also complicates the actual linear spectrum quantification \citep{Goedbloed10}.

Still, we will now use earlier linear instability criteria for idealised setups, to quantify and study the first stages of the evolution of each perturbation.
This will demonstrate that our simulation has thermal and interchange instabilities happening simultaneously.

\subsection{Instabilities and frequency cutoff}
The physical mechanism that creates the blobs and the complex dynamics that drives them around in the lower corona is related to two kinds of instabilities. When unstable, the system has a natural tendency to move towards more stable states through the mixing resulting from the developing instabilities. 

Since we have prominent radiative losses, as well as a strong magnetic field in the presence of gravity, we will consider: 
\begin{enumerate}
\item the thermal instability and
\item the interchange instability.
\end{enumerate}

The first one likely creates the condensation centres and thus the blobs and the second one can drive their evolution and circulation in the corona.
More specifically, due to localised cooling and mass accumulation, the criteria for thermal instability can be fulfilled for condensation of cool material to take place in situ. This creates the observed blobs at first. After they are formed, they possess a higher density compared to the layers underneath them, so this configuration can become gravitationally unstable and the system needs to adjust to a new more stable state. In search of this new energetically economic configuration, the blobs start moving even in the region of strong magnetic field at the lower part of the simulation domain.
The characteristic of the interchange instability is to do this with none to minimal field line bending $\vec{k}\cdot\vec{B}$, 
where $\vec{k}$ denotes the wave vector when a plane wave type description is used.

The spatial resolution of our simulation limits the wavelengths detectable in our data to (a multiple of) the grid cell size, $\Delta x$. Furthermore, data storage limits introduced a time cadence $\Delta t_s\approx 85s$, between successively stored full 3D data, so the instabilities we can detect most easily have growth rates $N$ in $e^{Nt}$, that grow sizeably in about 85s. Specifically for the Lagrangian displacement:
\begin{equation}
\xi=\xi_0 e^{-i\omega t}=\xi_0 e^{Nt} \,,
 \end{equation}
we observe noticeable dynamic changes evolving from one saved frame to the next one when their displacement gets larger than our spatial resolution or when:
\begin{equation}
e^{N \Delta t_s}-1\geq \frac{\Delta x}{\xi_0} \,.
\end{equation}
Initially, since blobs are created in situ, their Lagrangian displacement is small, so:
\begin{equation}
\frac{\Delta x}{\xi_0}\geq 1 \,.
\end{equation}
So, finally we get: 
\begin{equation}
e^{N\Delta t_s} > 2\Rightarrow N > \frac{\ln{2}}{\Delta t_s} \Rightarrow N> 0.008\mathrm{Hz} \,.
\end{equation}

We use this criterion further on to cut off lower growth rates artificially as they have not enough time to be determined by the system's evolution between two snapshots. Specifically, we are only interested in frequencies higher than 0.008$\mathrm{Hz}$, which means that we are looking for negative squared eigenvalues in the region: $\omega^2<-6.4\times 10^{-5}\mathrm{Hz^2}$. These eigenvalues are the most important ones for determining the blobs' formation and dynamical evolution at the temporal and spatial resolution that we have in this simulation. 

\subsection{Thermal Instability: isochoric and isobaric criteria}
The prescribed localised heating function that heats up the footpoints of the external magnetic loop causes evaporation that gradually adds cool plasma from the transition region into the large loop of our main configuration. So the evaporated cool plasma increases the density in that region from about $\rho=0.1\rho_{unit}$ ($\rho_{unit}=2.342\times10^{-12}\mathrm{kg/m^3}$) at $t=0$ to $\rho=\rho_{unit}$ at $t\approx205\mathrm{min}$. 
The density increase enters gradually the radiative loss and condensation centres are formed and blobs are created.

Two thermal instability criteria were introduced, the isochoric by \citet{Parker53} and the isobaric one by \citet{Field65}, that control thermodynamic evolution in astrophysical systems. From these criteria, we should be able to predict the condensation regions, i.e. regions where catastrophic cooling takes place. Although the criteria are only valid for gaseous uniform media, we will now use them to locate where blobs will be formed due to thermal instability.

According to \citet{Parker53} and also used for prominence onset in \citet{Xia11} the isochoric criterion is:
\begin{equation}
C=k^2-\frac{1}{\kappa}\left(\frac{\partial H}{\partial T} -\frac{\partial R}{\partial T} \right) < 0 \/,
\end{equation}

where $k$ is the wavenumber of the perturbation that undergoes thermal instability, $\kappa$ is the heat conduction coefficient, and $R=n_H n_e \Lambda(T)=Q (n_e/n_H)$ is the radiative loss.

Similarly, (see equation (8) in \citet{Xia11}) the criterion for an isobaric thermal instability is given by:
\begin{equation}
C_{isobaric}=\rho \left(\frac{\partial \mathcal{L}}{\partial T} \right)_{\rho}-\frac{\rho^2}{T}\left(\frac{\partial \mathcal{L}}{\partial \rho}\right)_T+k^2\kappa<0 \/,
\end{equation}
where $\mathcal{L}=(n_Hn_e\Lambda(T)-H)/\rho=(R-H)/\rho=(Q(n_e/n_H)-H)/\rho$ is the generalised heat-loss function. We use the parallel conduction coefficient value to quantify $\kappa$ (isotropic in a gaseous medium as analysed by \citet{Parker53}).

As perturbation lengths (i.e. to set the wavenumber $k$) we use two different length scales that play an important role in the dynamic processes.
\begin{enumerate}
\item On one hand, we adopt as perturbation wavelength the smallest detectable blob size $\lambda=208 \,\mathrm{km}$, which corresponds actually to our spatial resolution.
\item On the other hand, we can estimate the whole heated region size to be about $\lambda=40 \mathrm{Mm}$. This is done as follows. 
The temperature distribution indicates that most blobs have an average temperature of about 20,000K, but in general blobs with temperatures from 10,000K to 100,000K are observed, with the vast majority of them concentrated in the range between 10,000K and 60,000K, as demonstrated in the second panel of figure~\ref{Tmass}.
In order to estimate the size of the thermally unstable region, we use an isotemperature contour that corresponds to 40,000K in our physical domain at the moment that we start observing for the first time the condensation phenomenon ($205\mathrm{min}$). We estimate the size of the cool material region from the volume enveloping this isotemperature contour.  
The isosurface of 20,000K is present on the snapshot just before the first blob is formed, but vanishes on the snapshot that first captures the blob formation. This is in agreement with previous studies and is explained by the fact that cooling takes place in order for the condensations to happen, up to a point where cooling stops and inflows start converging on it, which results to a pressure increase that results to a localised increase in the temperature. In the same region the density is increasing strongly, forming condensation centres that gradually evolve into blobs.
\end{enumerate}

\begin{figure}[htbp]
\begin{center}
\begin{tabular}{cc}
\includegraphics[trim=0cm 1cm 0cm 3cm,clip=true,width=0.5\linewidth]{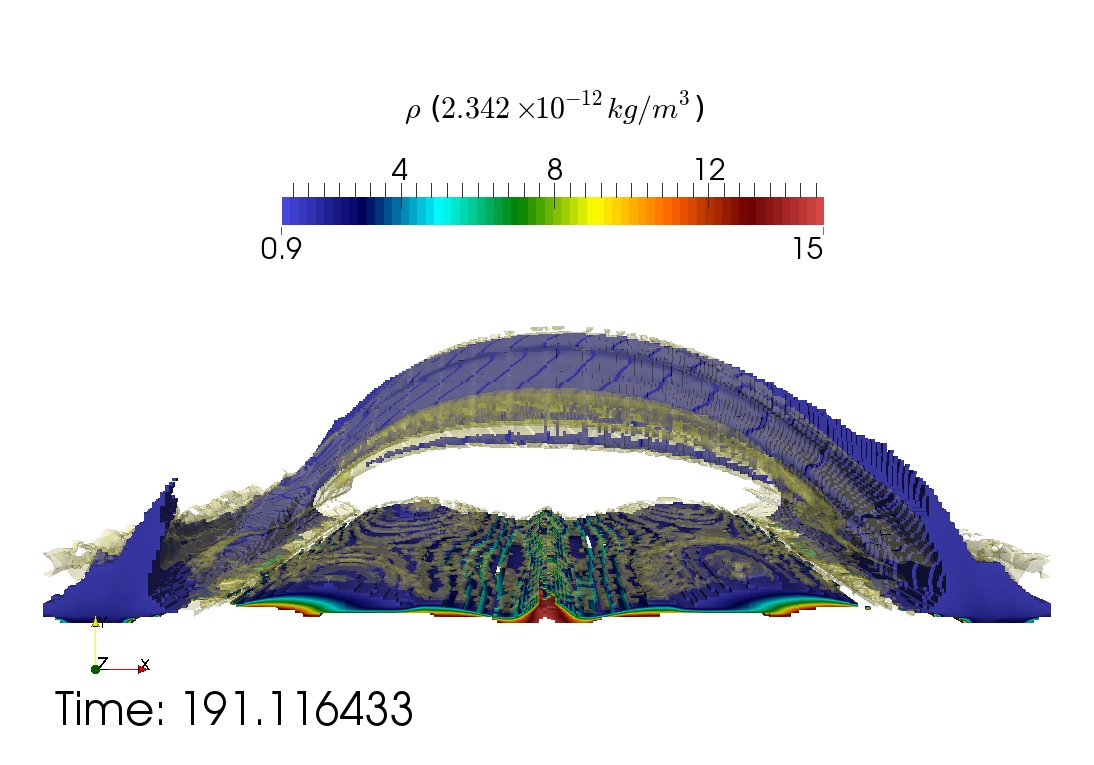}&
\includegraphics[trim=0cm 1cm 0cm 3cm,clip=true,width=0.5\linewidth]{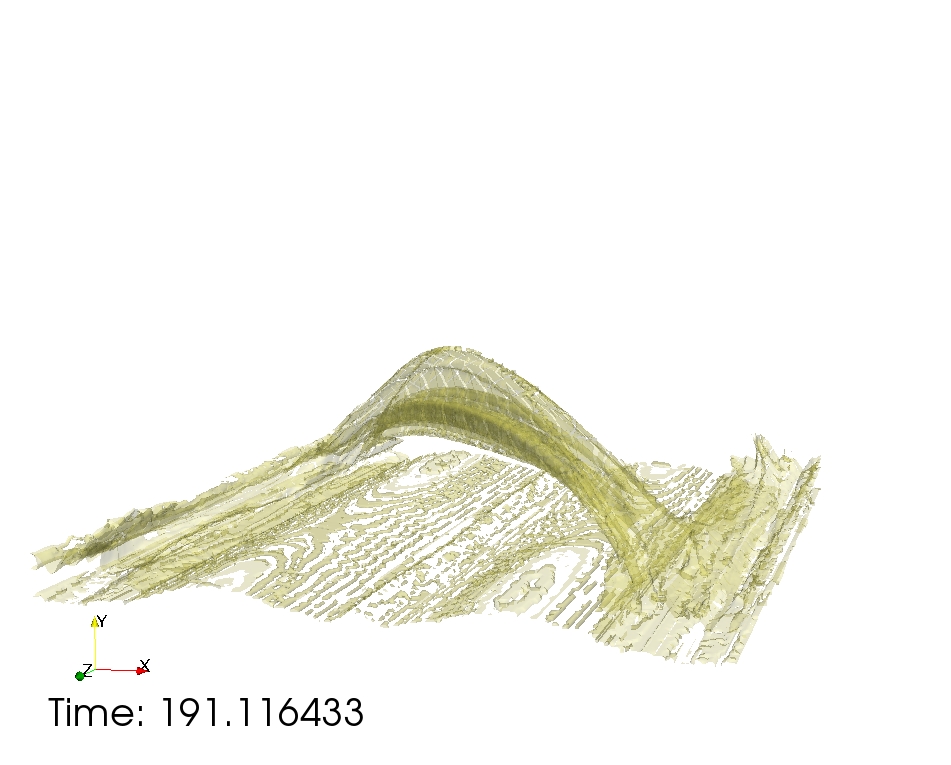}\\
\includegraphics[trim=0cm 1cm 0cm 3cm,clip=true,width=0.5\linewidth]{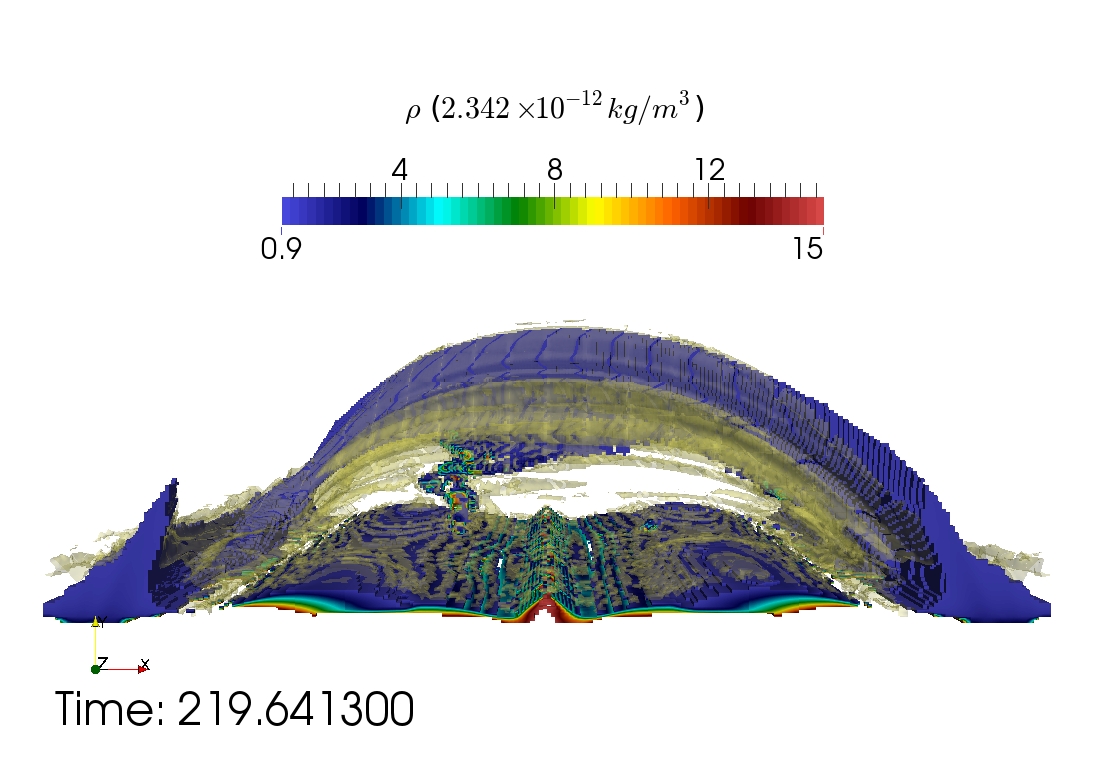}&
\includegraphics[trim=0cm 1cm 0cm 3cm,clip=true,width=0.5\linewidth]{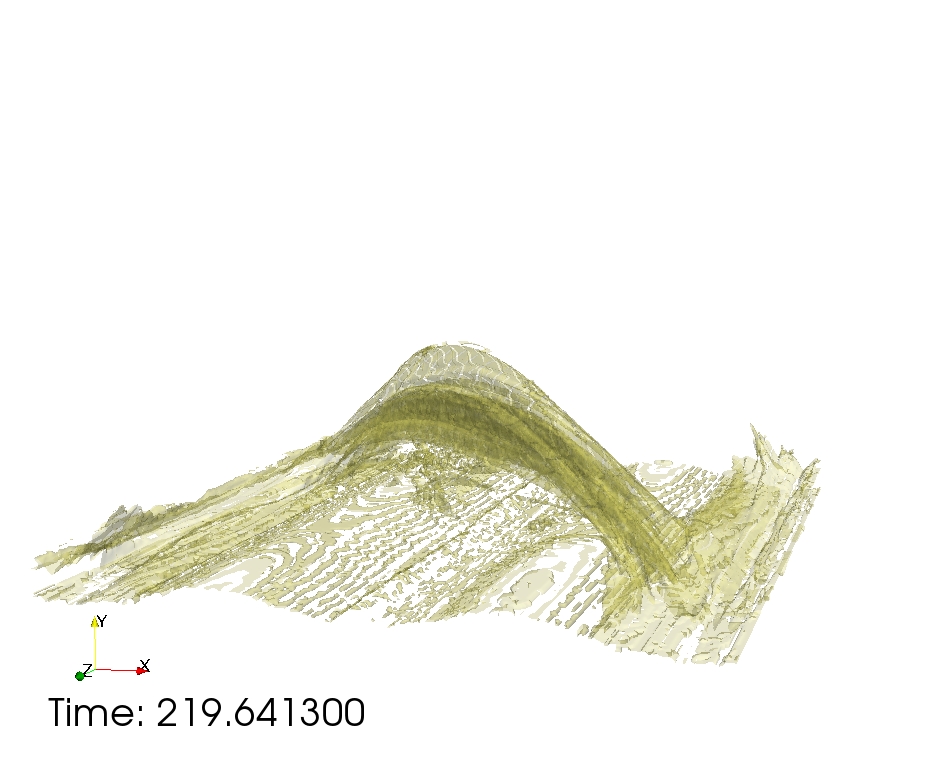}\\
\includegraphics[trim=0cm 1cm 0cm 3cm,clip=true,width=0.5\linewidth]{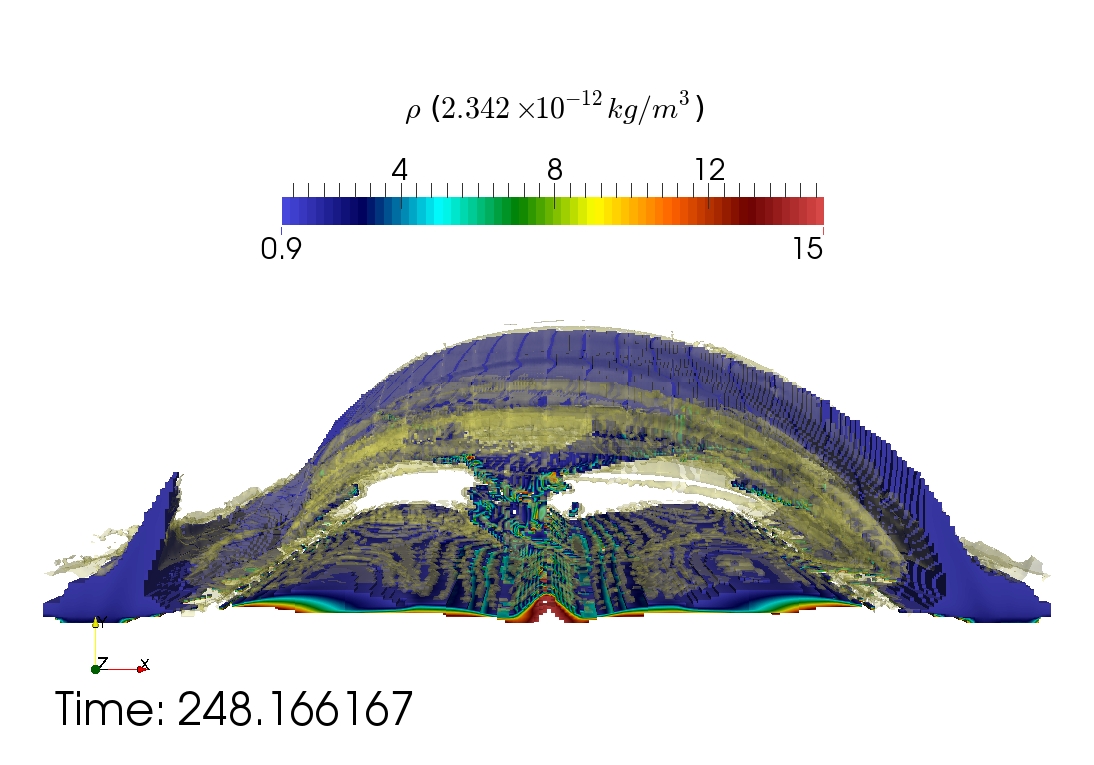}&
\includegraphics[width=0.5\linewidth]{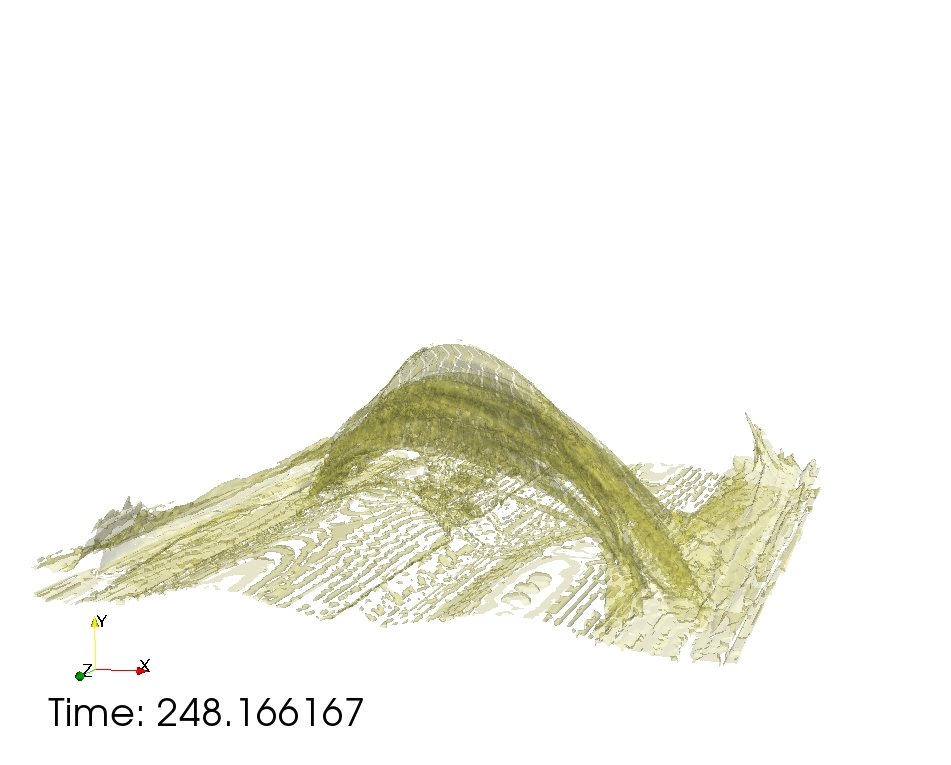}
\end{tabular}
\end{center}
\caption{Estimation of the unstable region from an isochoric thermal instability criterion shown at three times (top to bottom) and for two visualising angles (left-right column). 
In the panels of the left column we demonstrate the plasma densities with coloured isodensity surfaces as indicated by the colourbar.
The yellow isosurface visualised here in both left (with the density isosurfaces) and right columns (on its own) corresponds to growth rates above $0.008 \mathrm{Hz}$ and captures well the unstable region, as it follows the changing location of the condensed material in the corona with time in our simulation and highlights the blobs' birth places.}
\label{thermalinsta}
\end{figure}

Using the isochoric thermal instability criterion, we conclude that the thermally unstable region is very well identified, as we can see on figure~\ref{thermalinsta}. Indeed, for the wavelength $\lambda=40\mathrm{Mm}$, the condensation centres are captured and followed throughout the evolution of the phenomenon, as new high density blobs keep emerging till the end of our simulation. 

The isochoric criterion is rather independent of the length scale of the perturbation in our analysis, as after using the two wavelengths that were estimated above, that differ about two orders of magnitude from each other, we do not observe large differences in the estimated unstable region with the isosurfaces' method. This fact is in agreement with \citet{Xia11}, who first showed the agreement with Parker's results in a 1D setup. So the value of the perturbation wavelength $\lambda$ and thus the perturbation wavenumber $k=2\pi/\lambda$, doesn't affect the results of the thermally unstable analysis in highlighting the regions where the blob formation will take place.

For the isobaric case though, we noticed a clear dependence on the length scale of the perturbation.
Specifically, for the wavelength of $\lambda=208\mathrm{km}$, the isobaric criterion suggests fully stable states throughout the simulation, as it gives $C_{isobaric}>0$.
For the wavelength of $\lambda=40\mathrm{Mm}$ we find $C_{isobaric}\approx -8\times10^{-6}$ in individual cells for certain snapshots, without clear correspondence to the evolution of the blobs.

Concluding, our results agree with previous studies on the fact that the isobaric thermal instability criterion depends strongly on perturbation wavelength and seems less appropriate to explain catastrophic cooling during condensation formation in solar corona. The isochoric instability threshold gives a rather nice correspondence with blob formation regions. Hence for our case, the isochoric thermal instability criterion is the most appropriate to examine blob formation.

\subsection{Hydrodynamic buoyancy and Atwood numbers}
If for whatever reason we arrive at a configuration where heavier fluid is positioned above lighter fluid in a gravitational field, this configuration can be gravitationally unstable leading to Rayleigh-Taylor (RT) development in pure hydro setups. The system will search for a more stable state by forcing the heavy fluid to change position with the lighter one and fall down. The classical Rayleigh-Taylor instability results when small perturbations are applied at the interface between the two fluids and prevent the interface from keeping a perfectly flat shape.
A small perturbation at the interface then grows exponentially with a growth rate of:\\
\begin{center}
$\exp(N t)$, with $N=\sqrt{\mathcal{A}gk}$ and $\mathcal{A}=\frac{\rho_{heavy}-\rho_{light}}{\rho_{heavy}+\rho_{light}}$\\
\end{center}
where $N$, is the growth rate, $k$ is the spatial wavenumber and $\mathcal{A}$ is the Atwood number indicating the density disparity in a gravity field quantified by $g$ \citep{Chandrasekhar61,Glimm01}.
In pure Eulerian fluids, small wavelengths grow first.

This simple criterium is drastically modified in MHD, but it is known that the density disparity $\mathcal{A}$ between the fluids, which takes values $0\leq \mathcal{A} \leq 1$, determines the morphology and the evolution of the plasma undergoing non-linear Rayleigh-Taylor instability, since for $\mathcal{A}$ close to 0, Rayleigh-Taylor instability shows a symmetric mixing for both finger-like plasma structures for the heavy down falling fluid and bubble-like ones for the light fluid that rises;
while for $\mathcal{A}$ close to 1, falling spikes have larger growth rates and penetrate deeper in the opposite region than rising bubbles \citep{Mikaelian14}.

\begin{figure}[htbp]
\begin{center}
\includegraphics[width=0.8\linewidth]{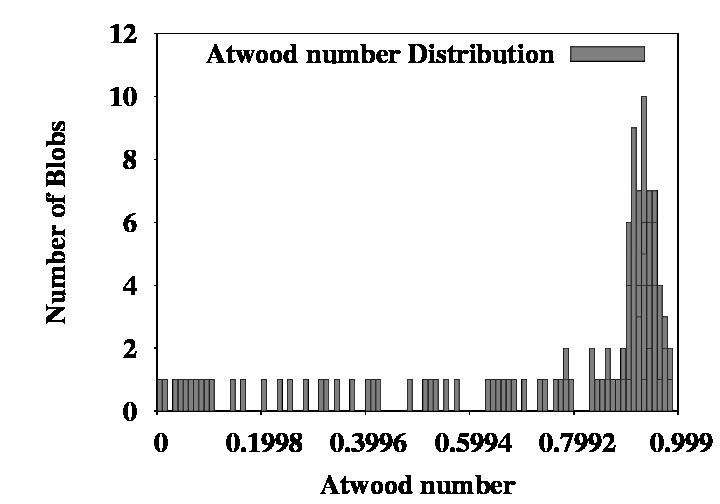}
\caption{Atwood number distribution for all the blobs throughout the entire simulation. The vast majority of the blobs appears to have Atwood numbers close to unity, suggesting a blob density of about 20 times higher than the plasma elements underneath the blob.}
\label{atwood}
\end{center}
\end{figure}

In order to quantify the Atwood number for each individual blob in our simulation, we set as $\rho_{heavy}$ the maximum density of the blob and search underneath the blob on the same column ($y-$direction) starting from the first grid cell just underneath the blob and moving downwards, for a grid cell with density $\rho_{light}$ smaller than the density of the centroid. There can be grid cells underneath the blob with similar density, but hotter material.
In our simulation the vast majority of the blobs take on Atwood numbers of the second limiting case, i.e. values close to unity, as is evident from the distribution in figure~\ref{atwood}. For the extreme cases we have the following:
\begin{itemize}
\item for $A=0.00145\Rightarrow \rho_{blob}=1.0029\cdot\rho_{corona}$,
\item for $A=0.99\Rightarrow \rho_{blob}\approx 20\cdot\rho_{corona}$.
\end{itemize}

As already mentioned, the Atwood number examines Rayleigh-Taylor instability from the purely hydrodynamic perspective, quantifying the density disparity between the plasma elements of interest and their surroundings and particularly the layer just underneath them, ignoring any influence the magnetic field might have on the event. The distribution of the Atwood number for the blobs throughout the entire simulation clearly shows that most blobs get Atwood numbers very close to unity, which suggests that the condensation centres have about 20 times higher density than their surroundings. On the other side of the distribution range we conclude that those blobs with small Atwood numbers close to 0 have a density only slightly higher than their surroundings. We now address their mixing tendency according to more relevant magnetohydrodynamic criteria.

\subsection{Interchange instability and Brunt-V\"{a}is\"{a}l\"{a} frequencies}
After the blobs are created, they are situated in low $\beta$ regions, so gravitational stratification can cause interchange instability to take place and drive their evolution. Thermal instability may still control the delicate growth process of individual blobs and sympathetic runaway cooling. Hence, new blobs continue appearing at new regions due to catastrophic cooling. After their creation, blobs can get born in a gravitationally unstable situation and then they start moving to help the configuration find a new more stable state. 
When gravity projected along field lines wins, they fall back to the transition region.

To examine the role played by interchange instability, likely causing the blob motion after they are formed, we quantify the relevant Brunt-V\"{a}is\"{a}l\"{a} frequencies. 
These are known to appear in standard instability criteria for gravity driven modes in up to strong magnetic fields.

We will use again various instability criteria to trace and follow the evolution of the blobs and their circulation in the corona.
We will find that buoyancy frequencies trace the unstable regions very well and follow the plasma condensations as they move around the corona from the moment they are formed onwards. We use different quantification strategies for two relevant Brunt-V\"{a}is\"{a}l\"{a} frequencies. 

The first method in a sense ignores the magnetic topology of the configuration and only takes into consideration the vertical components of the gradients (due to gravitational stratification) of density and pressure. This is appropriate in true 1D plane-parallel atmospheres, such as an exponentially stratified plasma with constant sound and Alfv\'{e}n speed. We then obtain

\begin{eqnarray}
N^2_B&=&\frac{1}{\rho^2}\frac{dp}{dy}\left(\frac{d\rho}{dy} -\frac{\rho}{\gamma p}\frac{dp}{dy} \right) \/, \\
N^2_m&=&\frac{1}{\rho^2}\frac{dp}{dy}\left(\frac{d\rho}{dy} -\frac{\rho}{\gamma p+B^2}\frac{dp}{dy} \right) \/, 
\end{eqnarray}
where $N^2_{B}$ is the classic Brunt-V\"{a}is\"{a}l\"{a} frequency and $N^2_{m}$ is the classic magnetically modified Brunt-V\"{a}is\"{a}l\"{a} frequency.
The latter is especially relevant for perpendicular ($\vec{k}\perp \vec{B}$) wave modes.

The second method tries to do justice to the 3D magnetic topology. In translationally symmetric (2.5D) Grad-Shafranov type equilibria between gravity, Lorentz free and pressure gradient, a stability criterium for convective continuum instability (CCI) modes uses projections on the field lines \citep{Blokland11}.
Since we do not have translational symmetry, and we want to avoid the details of a straight field line (SFL) representation, we arrive at estimations of the Brunt-V\"{a}is\"{a}l\"{a} frequencies, taking into account the three-dimensional configuration of the magnetic structure of our system by
\begin{eqnarray}
N^2_{B,p}&=&\left[\frac{\vec{B}_p \cdot \nabla p}{\rho B}\right]\left[\frac{\vec{B}_p}{\rho B}\cdot\left(\nabla \rho -\frac{\rho}{\gamma p}\nabla p \right) \right] \/, \\
N^2_{m,p}&=&\left[\frac{\vec{B}_p \cdot \nabla p}{\rho B}\right]\left[\frac{\vec{B}_p}{\rho B}\cdot\left(\nabla \rho -\frac{\rho}{\gamma p+B^2}\nabla p \right) \right] \/, 
\end{eqnarray}
where $N^2_{B,p}$ is the Brunt-V\"{a}is\"{a}l\"{a} frequencies and $N^2_{m,p}$ is the magnetically modified Brunt-V\"{a}is\"{a}l\"{a} frequency, both projected on the magnetic field lines. 
In the actual CCI criteria, one must use the (SFL) poloidal field and project along a flux surface.

We used three different approximations to projected Brunt-V\"{a}is\"{a}l\"{a} frequencies, that differ in the way we include the magnetic field influence in our calculation. The three different approaches take $\vec{B}_p$ in the above formulae as
\begin{enumerate}
\item $\vec{B_p}=(B_x,B_y,B_z)$
\item $\vec{B_p}=(B_x,B_y,0)$
\item $\vec{B_p}=(0,B_y,0)$
\end{enumerate}

\begin{figure}[htbp]
\begin{center}
\begin{tabular}{cc}
\includegraphics[trim=2cm 3cm 7cm 4cm, clip=true,width=0.5\linewidth]{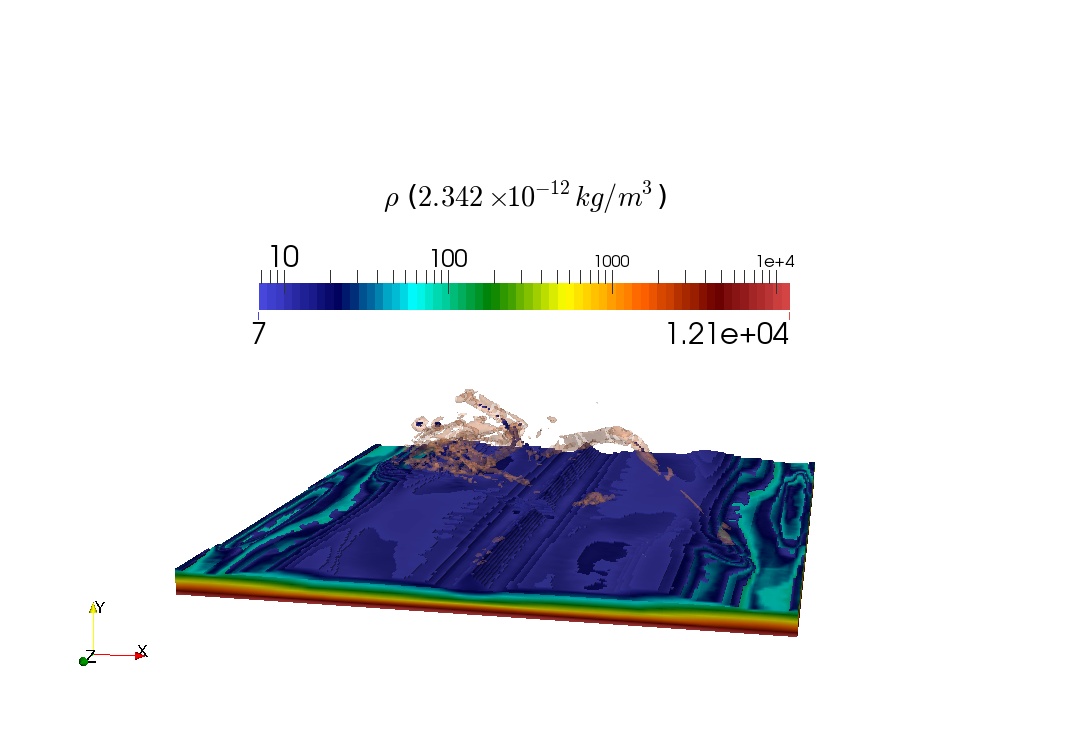}&
\includegraphics[trim=2cm 3cm 7cm 4cm, clip=true,width=0.5\linewidth]{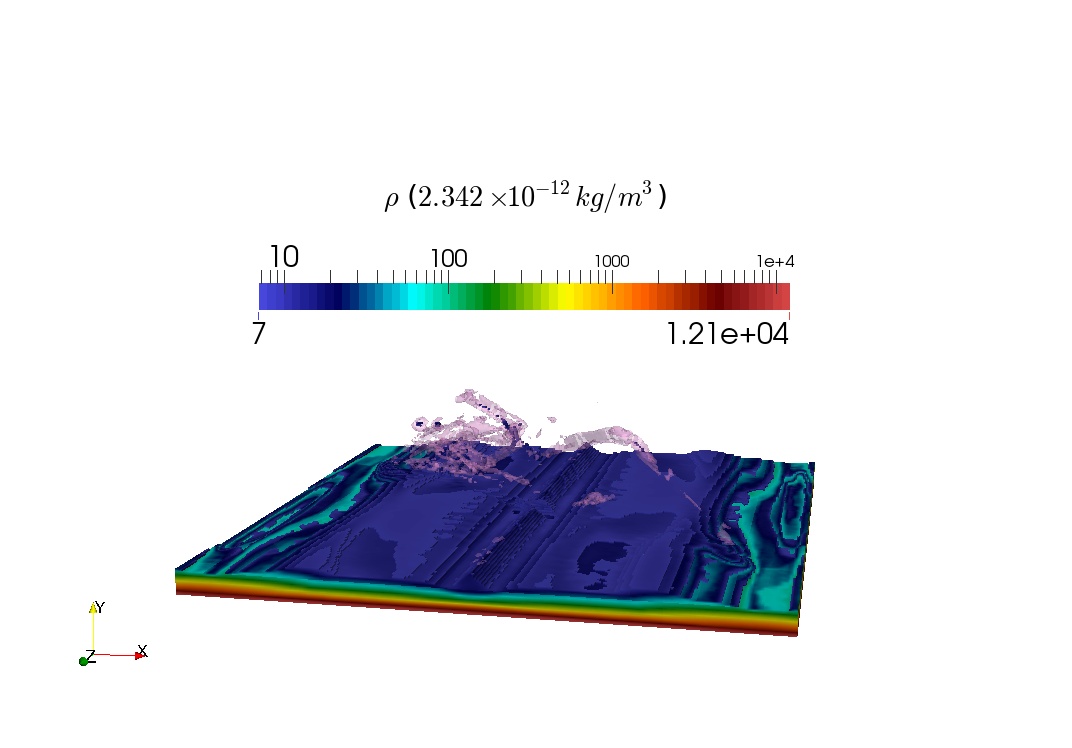}\\
\includegraphics[trim=2cm 3cm 7cm 4cm, clip=true,width=0.5\linewidth]{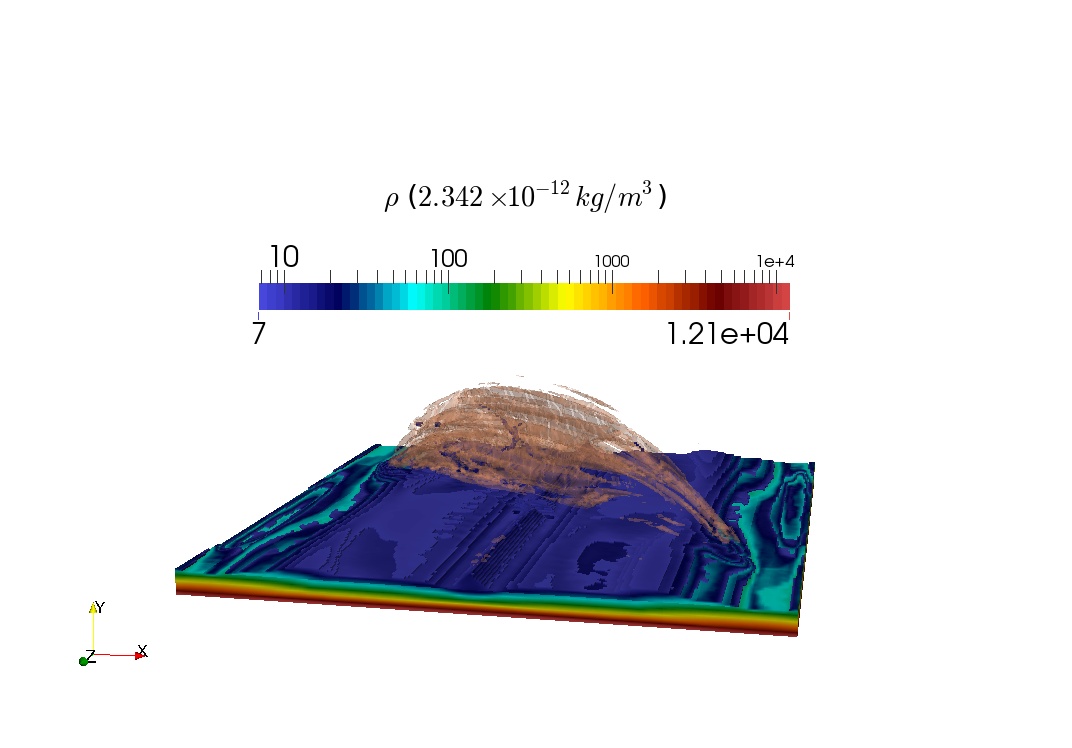}&
\includegraphics[trim=2cm 3cm 7cm 4cm, clip=true,width=0.5\linewidth]{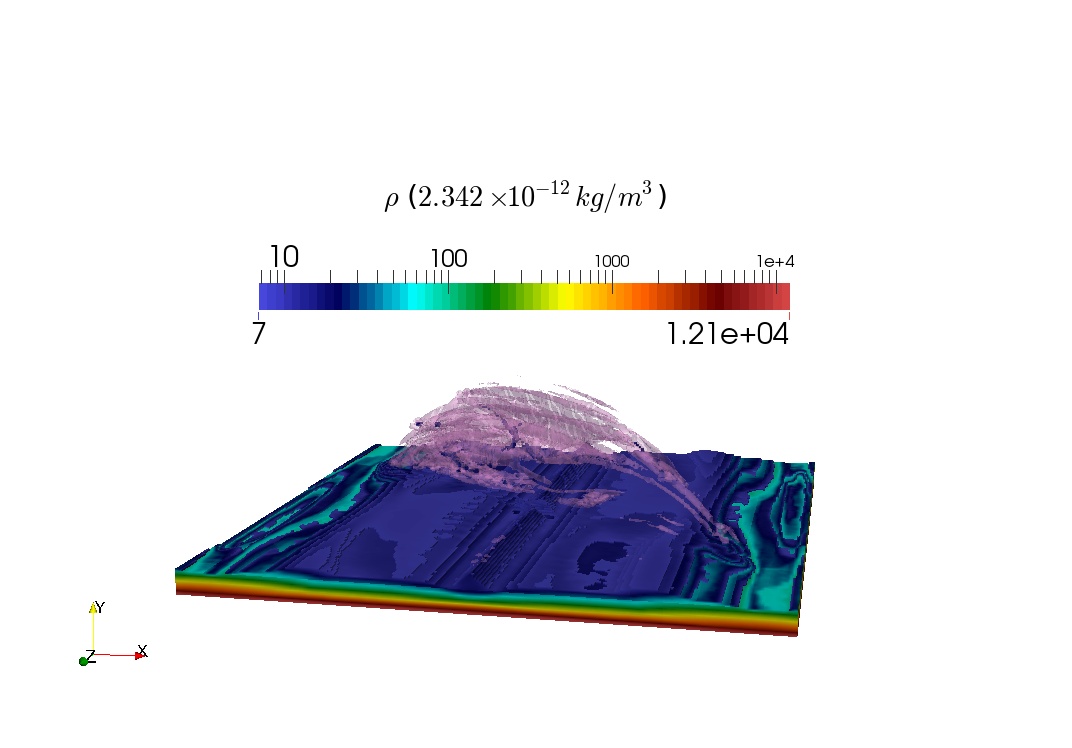}\\
\includegraphics[trim=2cm 3cm 7cm 4cm, clip=true,width=0.5\linewidth]{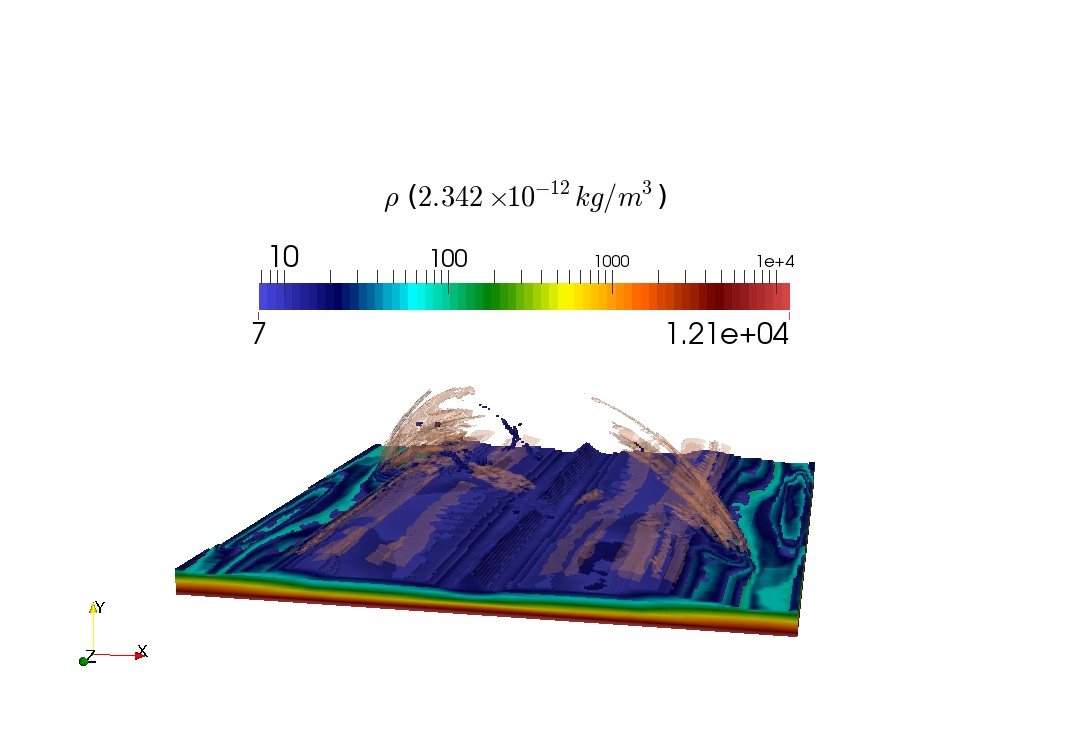}&
\includegraphics[trim=2cm 3cm 7cm 4cm, clip=true,width=0.5\linewidth]{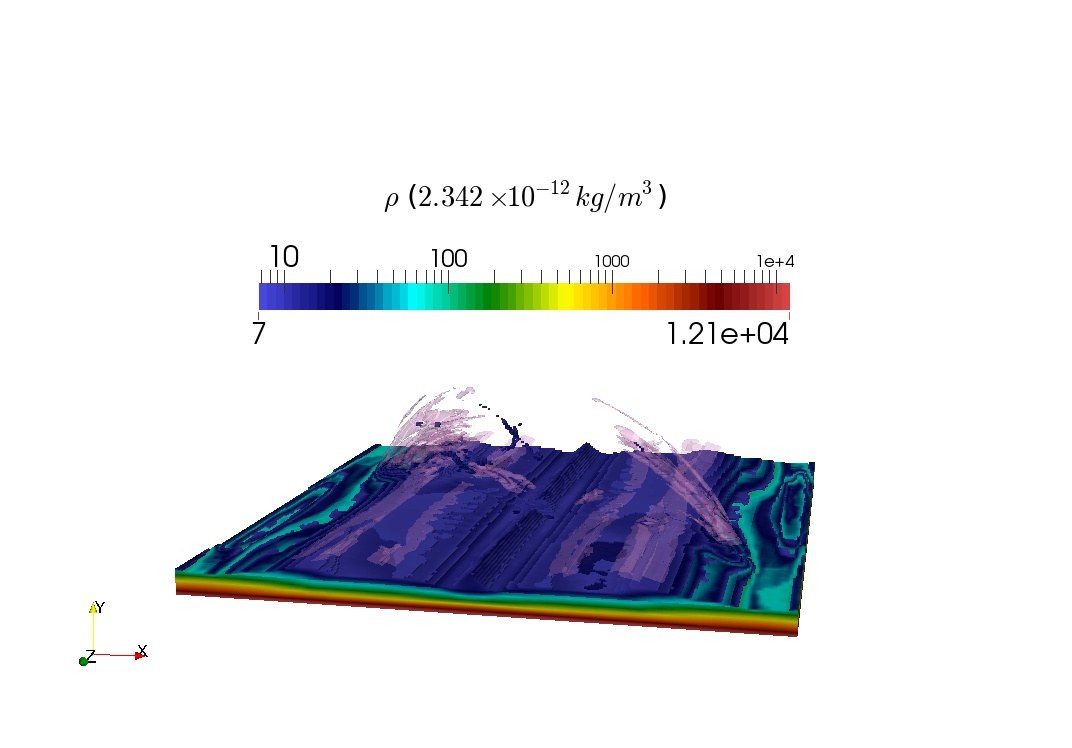}
\end{tabular}
\caption{Here, we demonstrate for the snapshot that corresponds to 235min of physical time the Brunt-V\"{a}is\"{a}l\"{a} frequencies, projected on field lines. We show frequencies $N^2_{B,p}$ (brown) and $N^2_{m,p}$ (purple) as isosurfaces in the left and right columns, respectively. Additionally, in all the panels we visualise in coloured isosurfaces the densities as indicated by the colourbar to capture all the regions of high density, i.e. the solar atmospheric regions underneath and the blobs inside the corona. We present quantifications done with three different ways to approximate field projections, with from top to bottom using a)  $\vec{B_p}=(B_x,B_y,B_z)$, b) $\vec{B_p}=(B_x,B_y,0)$, c) $\vec{B_p}=(0,B_y,0)$.}
\label{comparebxyz}
\end{center}
\end{figure}

Even slightly negative values of the squared Brunt-V\"{a}is\"{a}l\"{a} frequencies, as quantified according to the equations above, indicate unstable states. 
However, we explained earlier that due to the resolution of our simulation and our time resolution for data saves, we only care for instabilities with growth rates corresponding to time intervals of about 1.43min, i.e. the time interval between two sequential snapshots.

The comparison is presented in figure~\ref{comparebxyz} at time $t=235\mathrm{min}$. From that figure, we can conclude that the most appropriate method for tracing the unstable regions using projected Brunt-V\"{a}is\"{a}l\"{a} frequencies is the one that corresponds to the total three-dimensional magnetic field, namely the top panels in figure~\ref{comparebxyz}. This takes into account the three-dimensional variations of density and pressure. The unstable regions estimated in this way surround the blobs throughout their evolution (when we make animated views for all times) and appear to be a very consistent method to trace the important areas of the simulation domain, where the dynamical evolution is critical for the condensation centres. The second row, where we use only $x,y-$ variation (in line with the initial $z-$invariance of the setup) of density and pressure, overestimates the unstable regions, indicating the whole area underneath the loop where the blobs are located. The bottom row is least appropriate, as it only takes into account the vertical component of the density and pressure gradient and therefore does a bad job localising the perturbed regions. Instead, it indicates that the parts of the loops closer to the footpoints are most unstable, which is not true, as we observe no blobs or interchange dynamics there at all, at that time of the simulation. Hence, the full 3D prescription of all three quantities, i.e. the magnetic field and the 3D gradient of density and pressure, is the most accurate method to estimate the unstable regions to MHD instability, using both $N^2_{B,p}$ and $N^2_{m,p}$ frequencies.

\begin{figure}[htbp]
\begin{center}
\begin{tabular}{cc}
\includegraphics[width=0.5\linewidth]{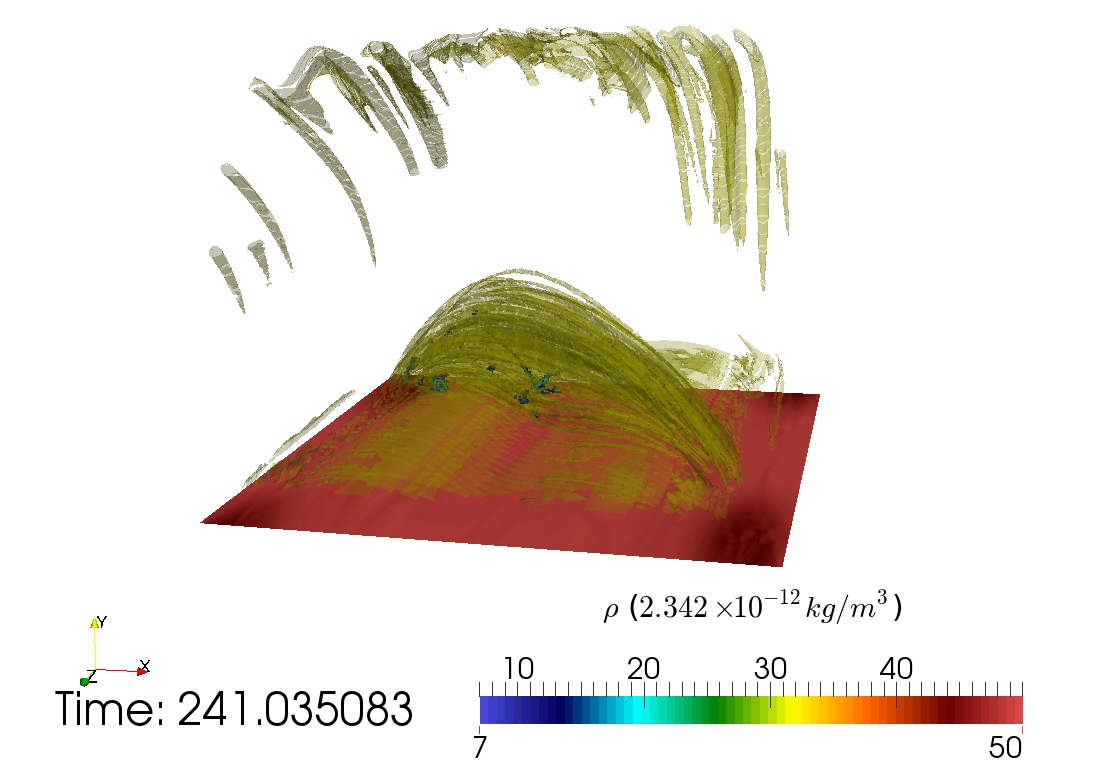}&
\includegraphics[width=0.5\linewidth]{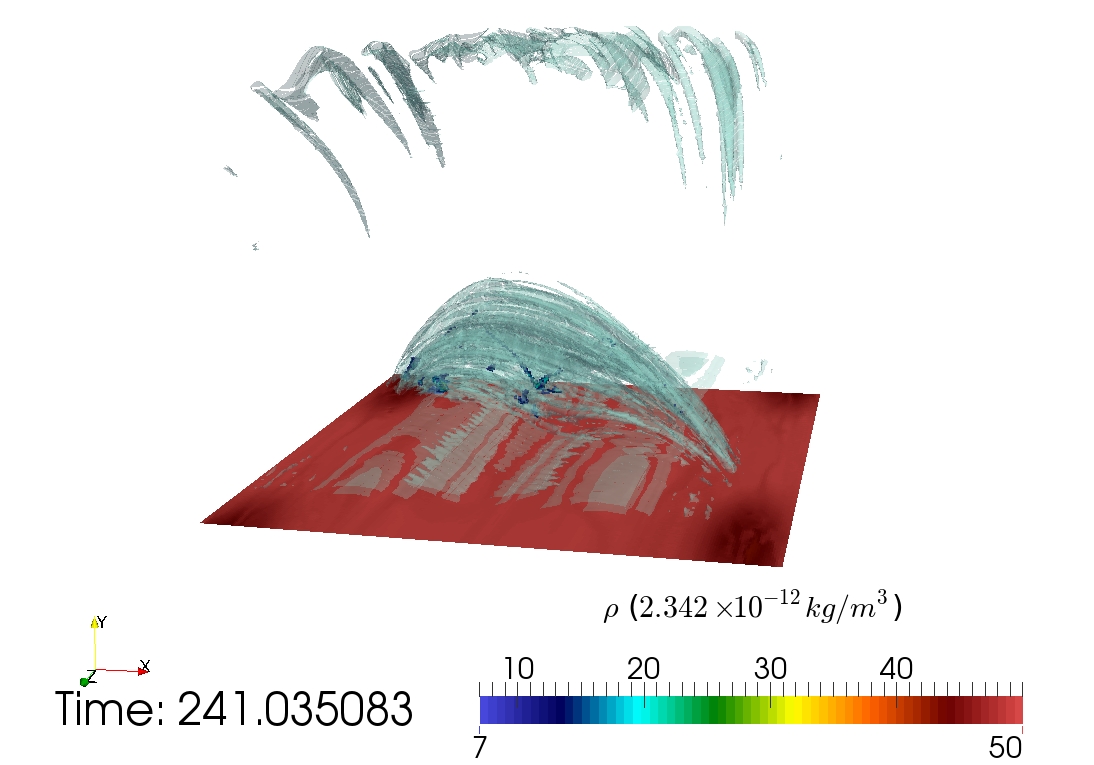}\\
\includegraphics[trim=0cm 0cm 0cm 8cm, clip=true,width=0.5\linewidth]{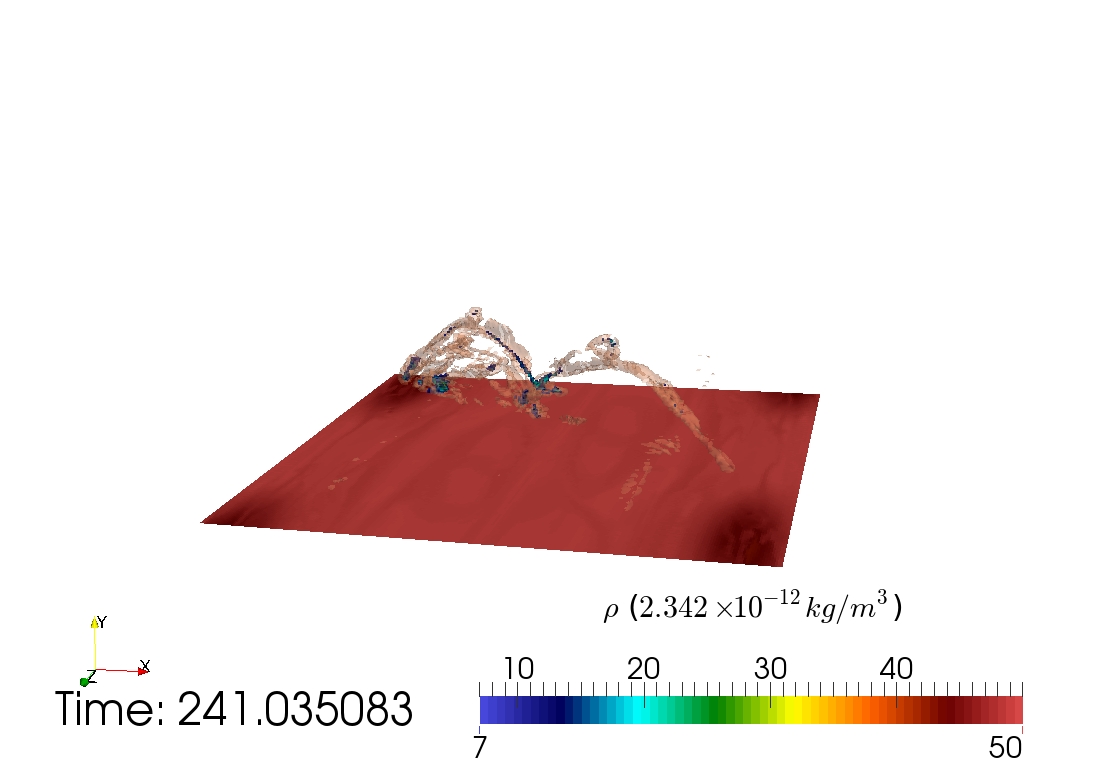}&
\includegraphics[trim=0cm 0cm 0cm 8cm, clip=true,width=0.5\linewidth]{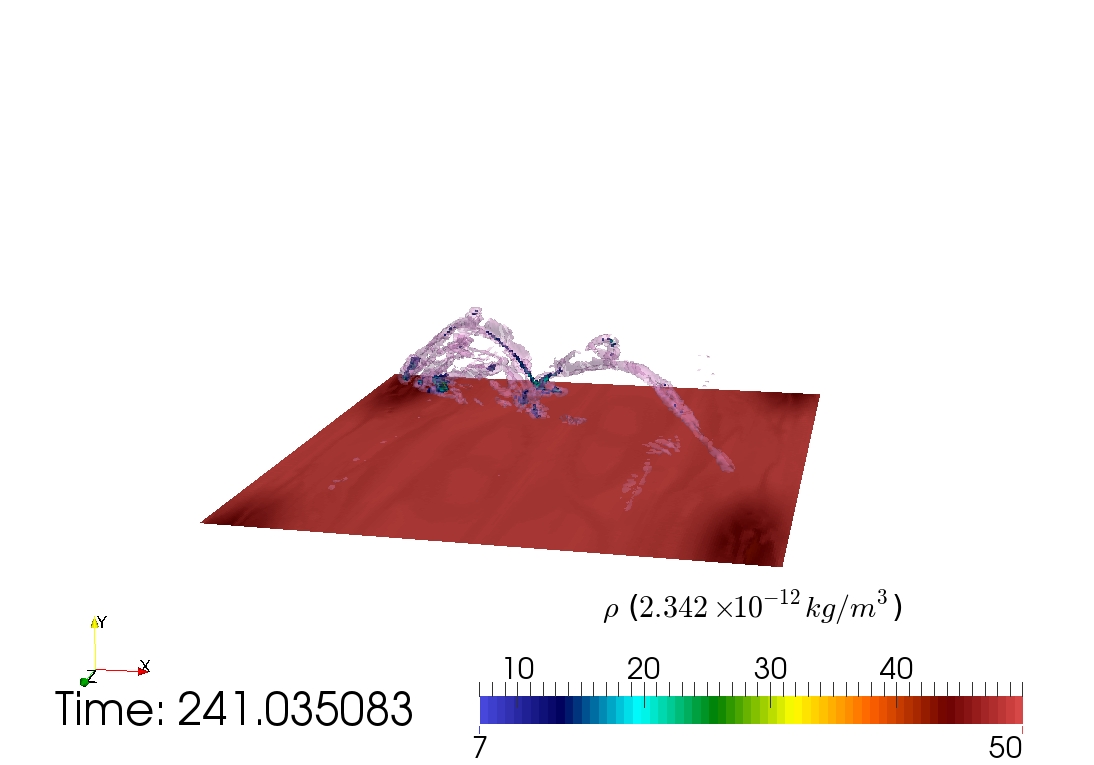}
\end{tabular}
\caption{In the four plots above we visualised for the snapshot at about 241min the results of four different criteria to estimate the unstable regions in our simulation box as they were presented in the text for both Brunt-V\"{a}is\"{a}l\"{a} frequencies, $N^2_{B}$ (left column) and $N^2_{m}$ (right column). We observe that the magnetohydrodynamic approach, i.e. the full-field projected Brunt-V\"{a}is\"{a}l\"{a} frequencies (bottom row) on the magnetic field lines, nicely correlates with the blob regions. The classic approach (top row) that takes into account only the vertical component of the gradient of the density and the pressure identifies also the (higher $\beta$) regions at the domain top, which also seem to show structures as in figure ~\ref{temp}.}
\label{compareBVprojBVclass}
\end{center}
\end{figure}

The field line projected Brunt-V\"{a}is\"{a}l\"{a} frequencies confirm that we have an overall stable configuration with well defined unstable regions that surround the blobs, throughout their circulation in the corona. 
They point out as the unstable region the top part of the internal loops that form the quadrupolar arcade system. 
We are led to conclude that the blob dynamics is driven by some variant of the CCI development. 

On the other hand, the classic way of estimating the Brunt-V\"{a}is\"{a}l\"{a} frequencies, taking into account only the vertical variation of pressure and density, rather than their three-dimensional gradients in the physical domain, estimates a larger unstable region, that extends in a large portion of our simulation box. 
This is shown in figure~\ref{compareBVprojBVclass} for a time $t=241\mathrm{min}$, where we contrast it with the full-field projected quantification in the bottom panels.
More specifically, using the classic formula two unstable regions emerge. One on the lower part of the box at the magnetic dip region under the locations of blob formation following also the blobs' evolution, as well as another region on top of the simulation box at the upper part of the loops of our magnetic topology. They correspond to the dynamics seen at the top of our arcade system, also seen earlier in Figure~\ref{temp}, where the plasma beta is higher.

Still, the most relevant criterion to estimate the unstable regions in low $\beta$ regions is the one using the three-dimensional gradients of density and pressure and the three-dimensional magnetic field, that captures quite accurately the region surrounding the blobs and highlighting the dynamically perturbed areas of the domain. 

\begin{figure}[htbp]
\begin{center}
\begin{tabular}{cc}
\includegraphics[width=0.5\linewidth]{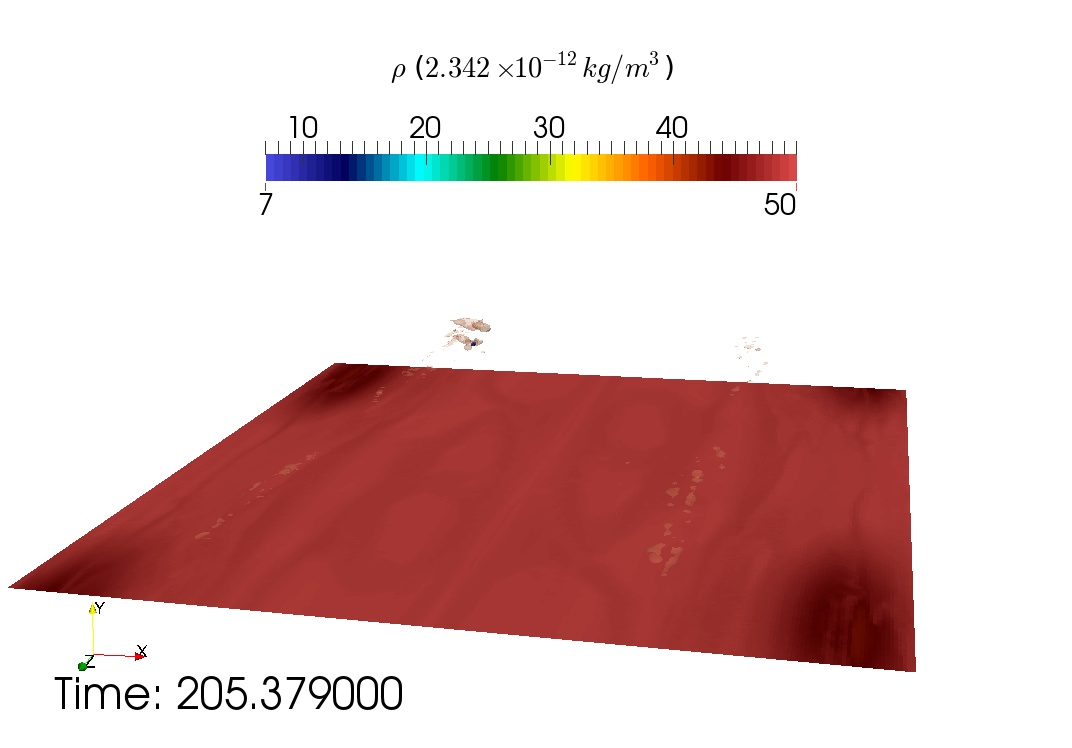}&
\includegraphics[width=0.5\linewidth]{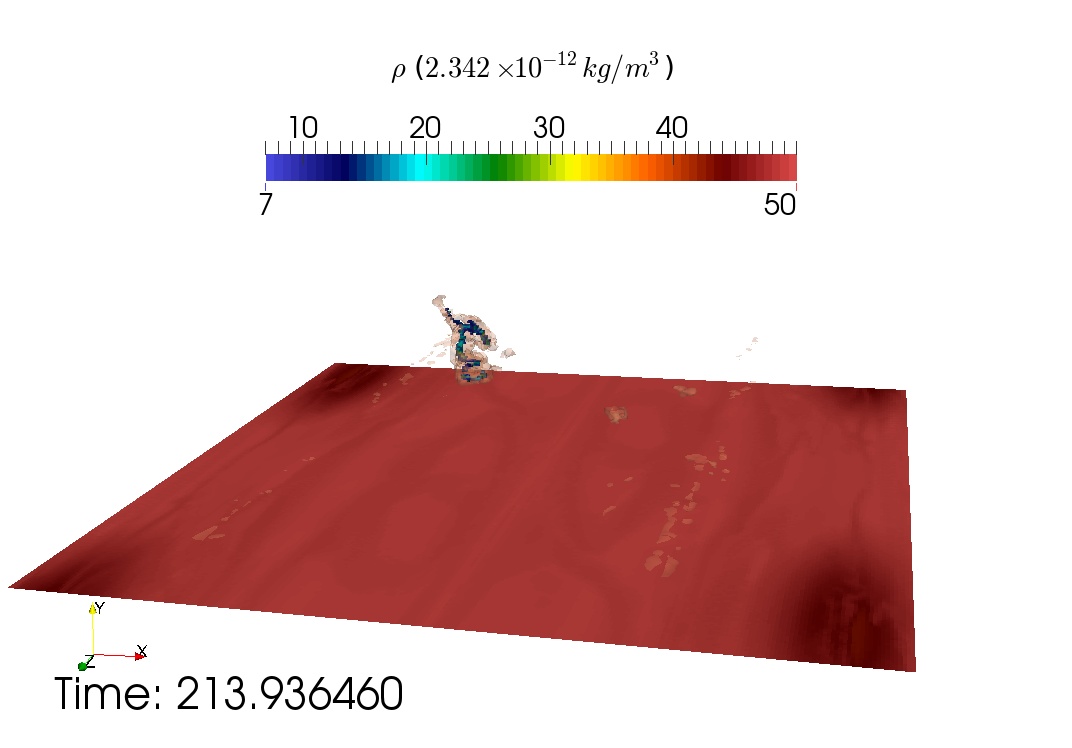}\\
\includegraphics[width=0.5\linewidth]{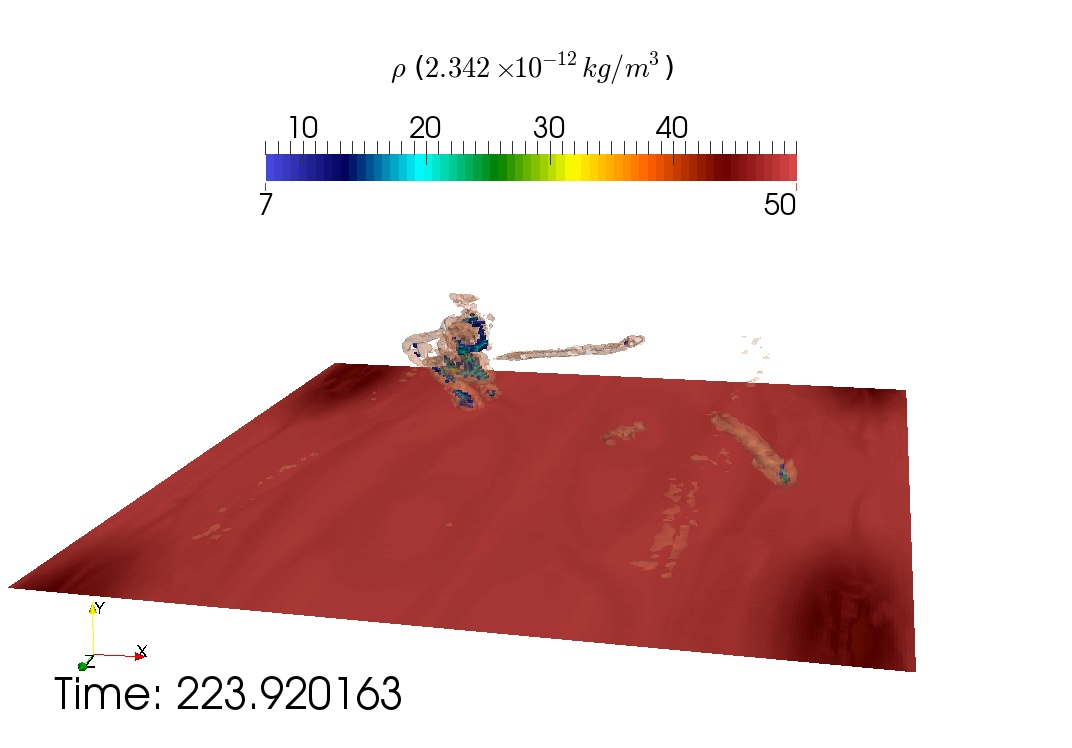}&
\includegraphics[width=0.5\linewidth]{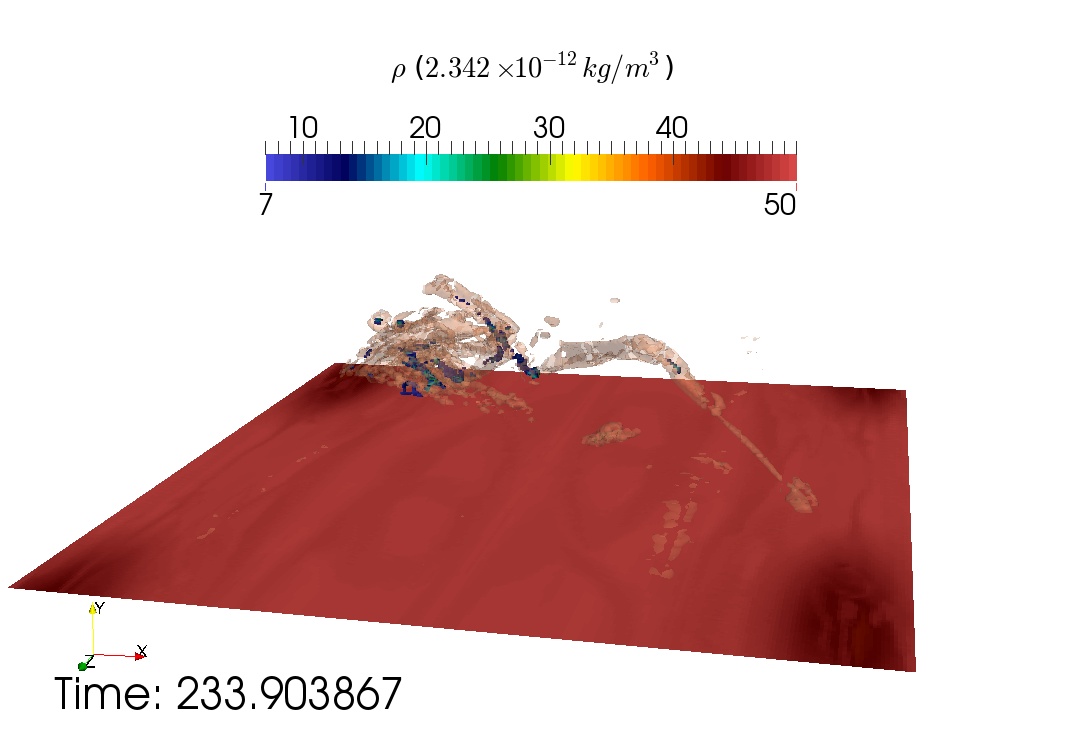}\\
\includegraphics[width=0.5\linewidth]{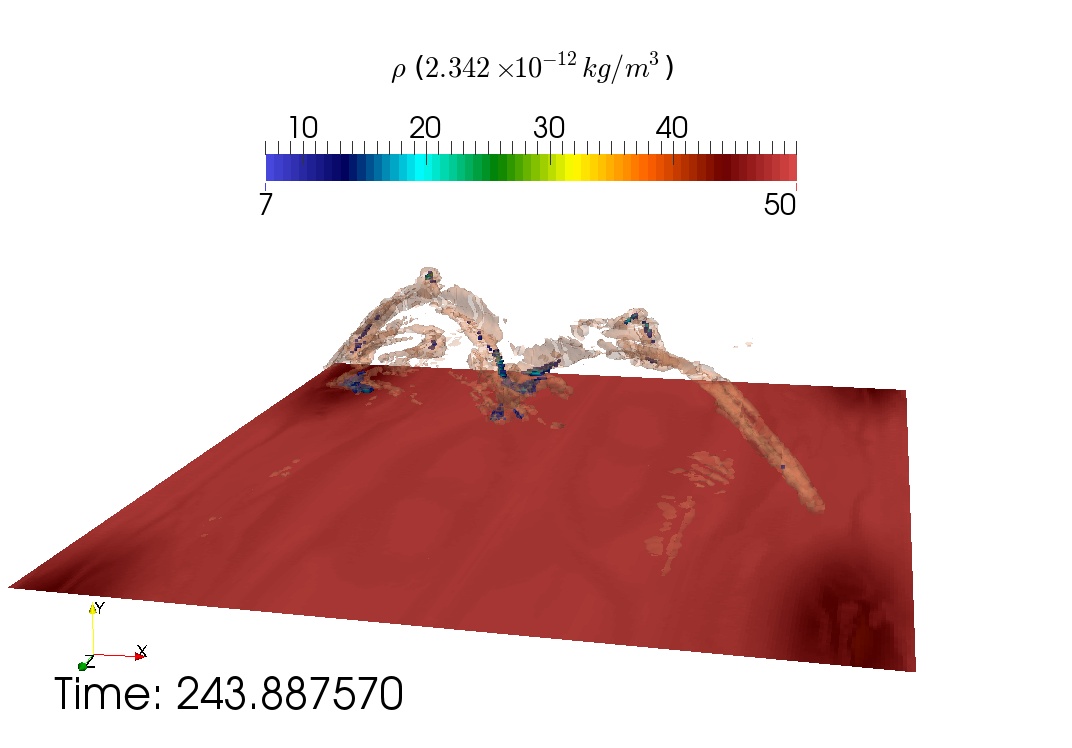}&
\includegraphics[width=0.5\linewidth]{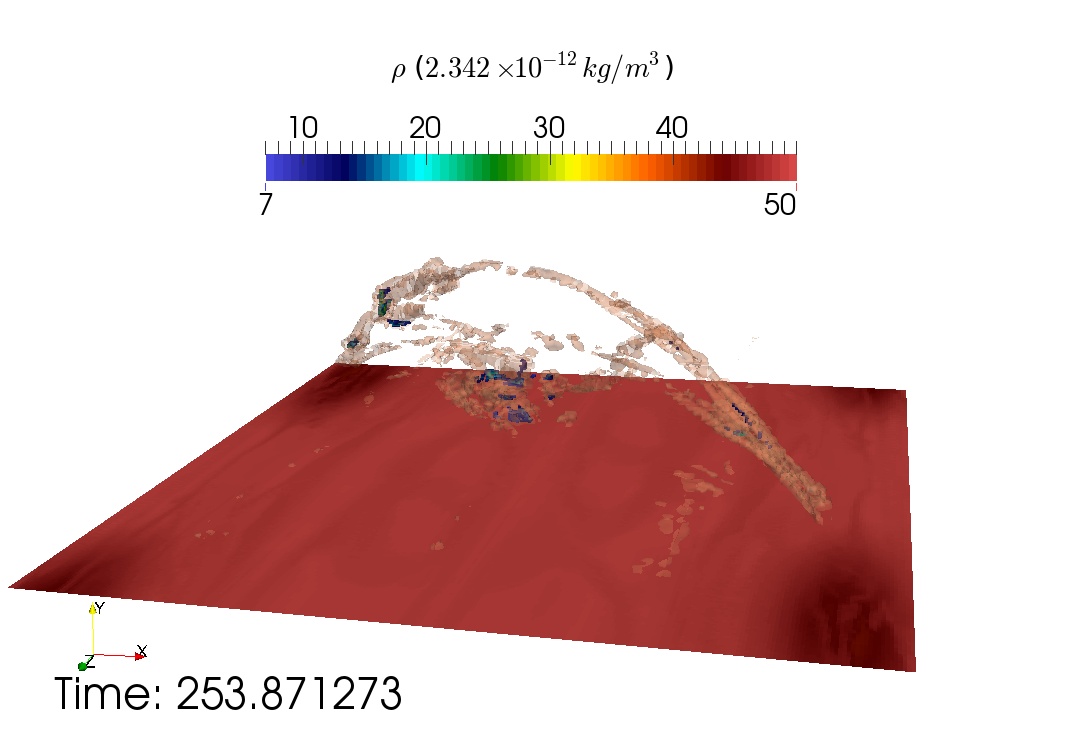}
\end{tabular}
\caption{We present six snapshots at different times for the unstable region as it is estimated by the full-field projected Brunt-V\"{a}is\"{a}l\"{a} frequency-criterion $N^2_{B,p}$. With brown colour and low opacity we can see the isosurface that corresponds to the unstable region and with saturated colour bar the density of the blobs. We conclude that there is a satisfying correspondence between the criterion and the position of the blobs at each time, as the unstable isosurface always surrounds the condensed matter.}
\label{BVseries}
\end{center}
\end{figure}

In figure~\ref{BVseries}, we demonstrate how the full-field projected Brunt-V\"{a}is\"{a}l\"{a} frequency-criterion $N^2_{B,p}$ captures the condensation regions throughout the simulation following the blobs as they circulate in the solar corona. We do so by using six snapshots corresponding to different physical times from the moment we observe the first blobs (about 205 min of physical time for the top left panel) till the final stages of the simulation (254 min for the bottom right panel of figure~\ref{BVseries}).

\subsection{Additional wave patterns}
\begin{figure}[htbp]
\begin{center}
\includegraphics[trim=2cm 2cm 2cm 0cm, clip=true,scale=0.4]{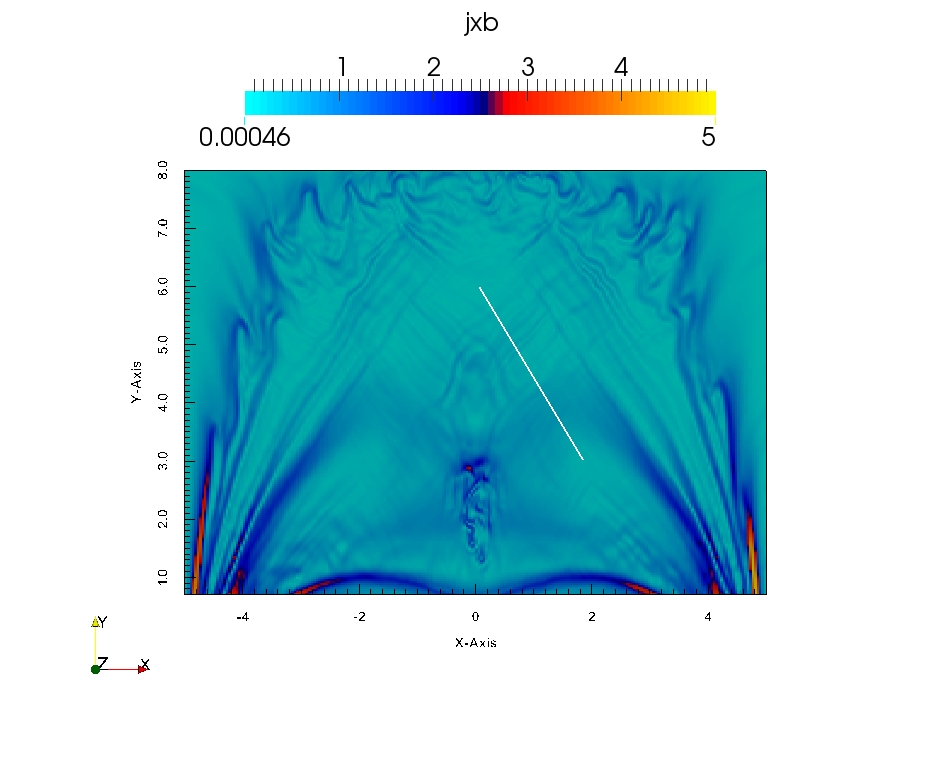}
\caption{The Lorentz force magnitude is demonstrated in the slice that corresponds to $x=z$ and cuts the domain diagonally at the snapshot that corresponds to about 205min of physical time. Several wave patterns appear throughout our simulation box. The white line represents the line along which we estimate the physical variations further on.}
\label{lorentz}
\end{center}
\end{figure}

There are several different physical phenomena and dynamic processes that take place inside our physical domain. 
As expected and in accordance with previous studies of prominence formation~\citep{Keppens14} we see several wave patterns all over the simulation box.
The representation of the Lorentz force magnitude indicates most evidently all the magnetohydrodynamic interactions, as demonstrated in figure~\ref{lorentz}. 
However, many of the observed wavy features are just an artefact of our (closed) boundary conditions, with repeated reflections at our boundaries as the simulation proceeds. 
Still, we are interested in localised waves that have an actual physical trigger at some point in the simulation and such features we will now try to describe.

From the plethora of wave features that are shown in the visualisation of the Lorentz force magnitude all over the simulation domain (figure~\ref{lorentz}), we noticed some clear wave patterns appearing in profiles of other quantities as well. 
One very characteristic example is the region on the top right part of the arcade in this plane $x=z$, as shown with the white line in figure~\ref{lorentz}.

\begin{figure}[htbp]
\begin{center}
\includegraphics[angle=-90,scale=0.4]{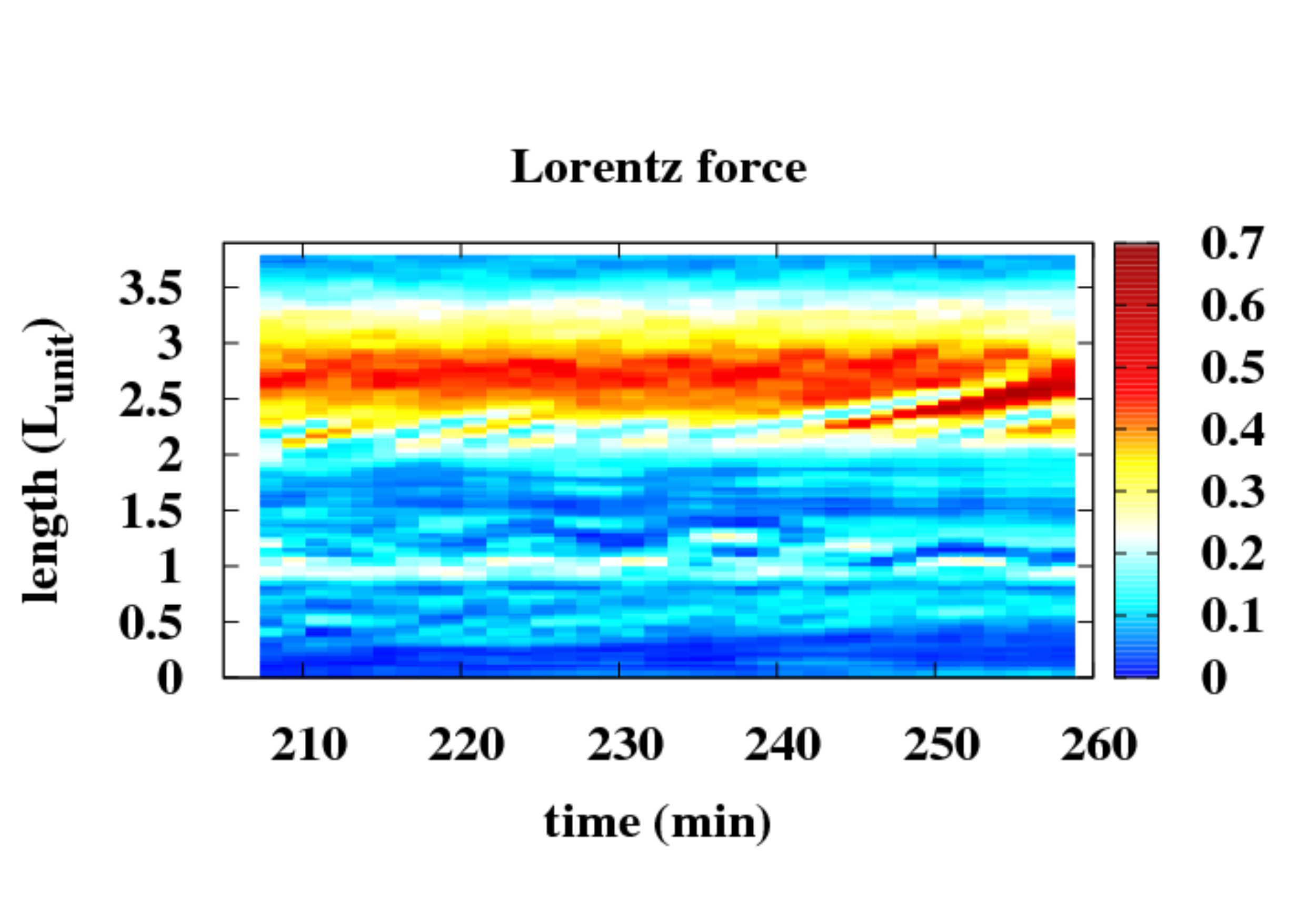}
\caption{The wave fronts identified by a Lorentz force magnitude view stacked in time along the line shown in figure~\ref{lorentz} to reveal their propagating speeds.}
\label{waves}
\end{center}
\end{figure}

There, a moving perturbation seemed to be localised, so we went on to quantify the evolution along this line. 
Our goal is to calculate the propagation speed of this wave and for that we quantify the values of the Lorentz force magnitude on the chosen line and stacked them over time, as depicted in the figure~\ref{waves}.

These appear at the middle part of the line length (total length is $\approx 4L_{unit}$ in figure~\ref{waves}) and from the slope we calculate the propagation speed of these waves.
We got an estimated speed for this position-time diagram of Lorentz force of about 6km/s. This apparent speed is about an order of magnitude smaller than the speed of the blobs while circulating around the lower parts of the corona and it is also about 2 orders of magnitude smaller than the local Alfv\'{e}n speed.
Animated views show clear indications of pressure perturbations moving in an oblique direction.

\begin{figure}[htbp]
\begin{center}
\includegraphics[trim=0cm 1cm 0cm 1cm,clip=true,width=0.7\linewidth,angle=-90]{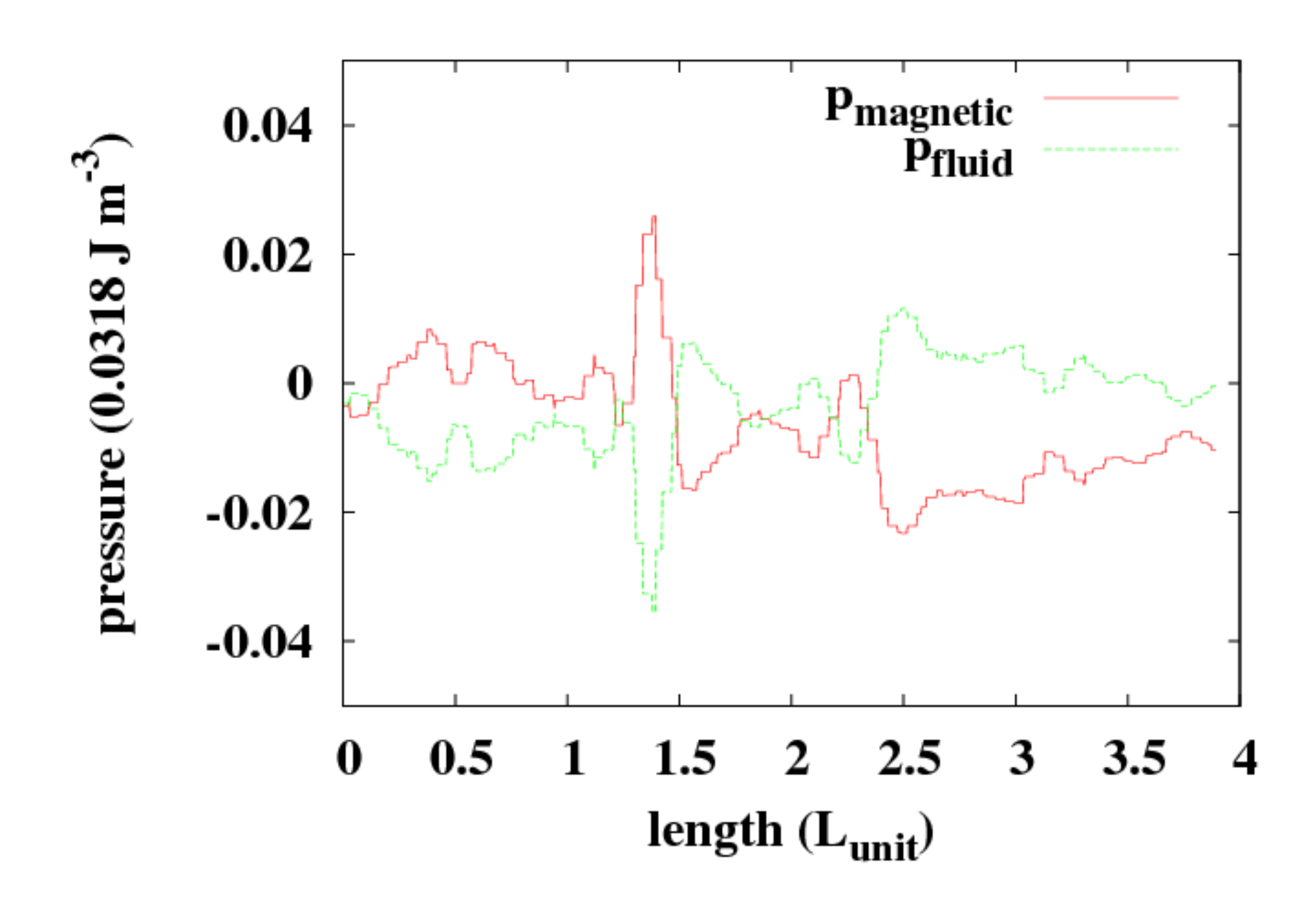}
\caption{Here we demonstrate the magnetic (red curve) and plasma (green curve) pressure deviations as they were quantified along the line shown in figure~\ref{lorentz}, at physical time of about $ 241\mathrm{min}$. The magnetic and fluid pressures are out of phase.}
\label{pressure}
\end{center}
\end{figure}

\begin{figure}[htbp]
\begin{center}
\begin{tabular}{cc}
\includegraphics[trim=0cm 1cm 0cm 1cm,clip=true,width=0.35\linewidth,angle=-90]{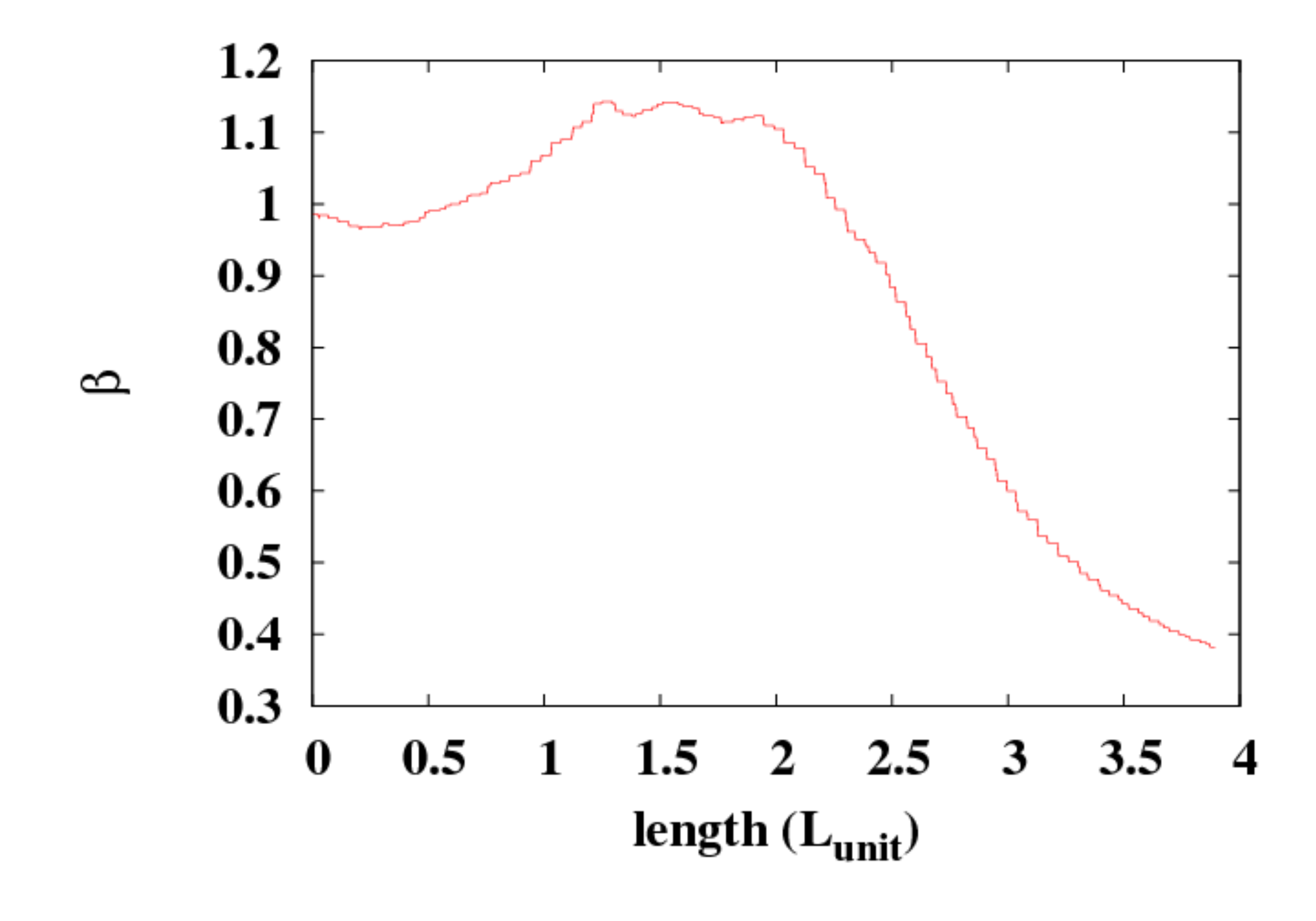}&
\includegraphics[trim=0cm 1cm 0cm 1cm,clip=true,width=0.35\linewidth,angle=-90]{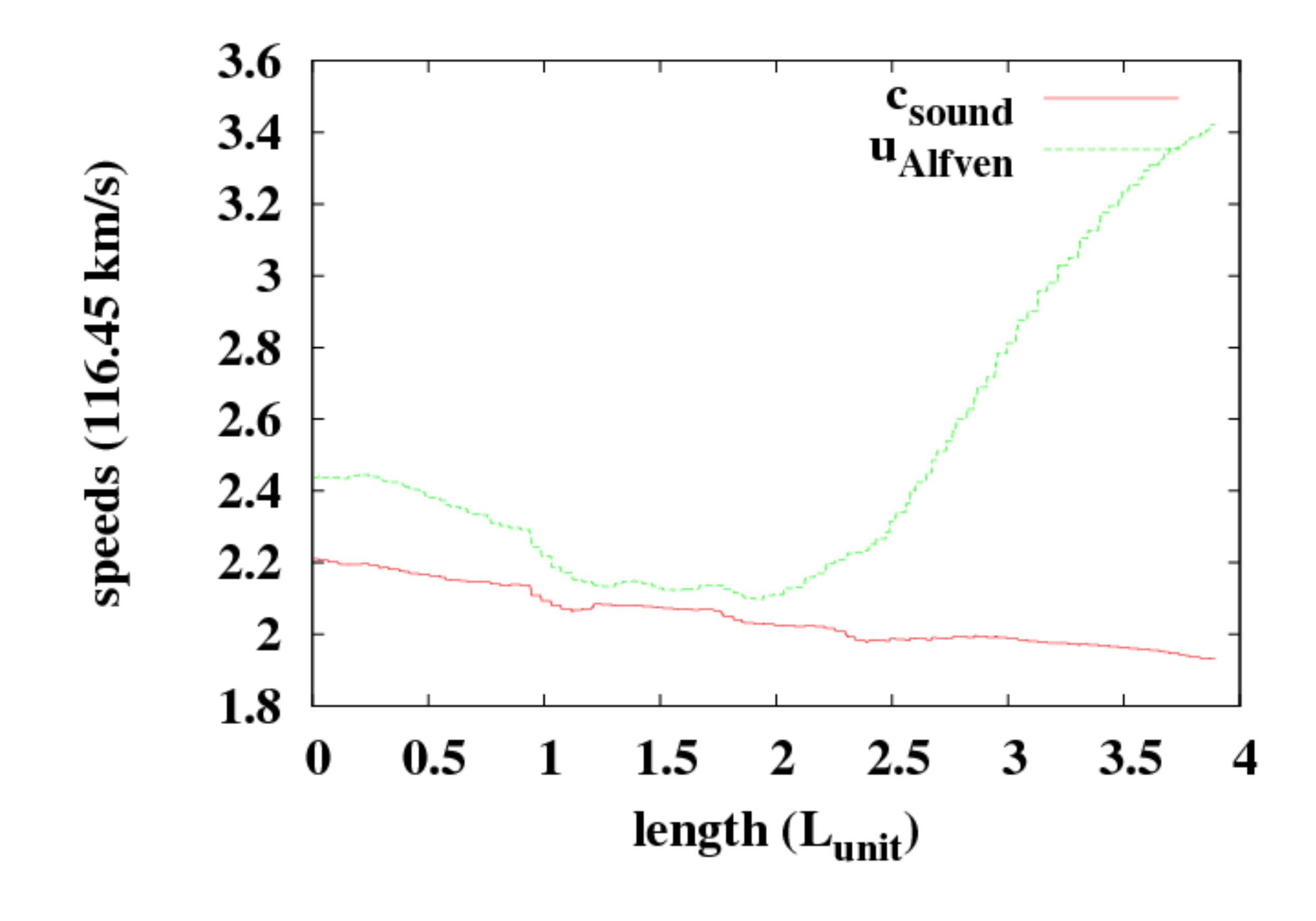}
\end{tabular}
\caption{Here we demonstrate the plasma $\beta$ (left panel) and the characteristic plasma speeds locally, i.e. sound speed (red curve in the right panel) and Alfv\'en speed (green curve in the right panel), as they were quantified along the line shown in figure~\ref{lorentz}, at physical time of about $ 241\mathrm{min}$.}
\label{betaspeeds}
\end{center}
\end{figure}

In the figure~\ref{pressure}, we show the magnetic and plasma pressure variations, $p_{fluid}=p(t)-p(t_{i})$ and $p_{magnetic}=(B^2(t)-B^2(t_{i}))/2$, along the line presented in figure~\ref{lorentz}, where as $t_{i}$ we use the moment when we first observe blobs in our simulation $t_{i}\approx 205\mathrm{min}$ and for $t$ we take $t=241 \mathrm{min}$.
Note that in this region, $\beta$ is in the range $[0.4,1.1]$, as shown in the left panel of figure~\ref{betaspeeds} (for a specific snapshot at 241min of physical time), so the local sound speed is in the range $[232.9,256.19] \mathrm{km/s}$, as shown in the right panel of figure~\ref{betaspeeds}.
From time series of similar figures of these spatial variations in pressure, as in figure~\ref{pressure}, we can estimate the order of magnitude of the wavelength of the oscillations as well as their period. 
We estimated a wavelength $\lambda$ of about $2.5\mathrm{Mm}$ and a period of about $257\mathrm{s}$, giving a phase speed of the order of $v_{phase}\approx 10 \mathrm{km/s}$.

We conclude that the magnetic pressure and the plasma pressure are out of phase.
This is a typical characteristic of slow magneto-acoustic waves in a uniform medium.
Even though, our medium is far from uniform, this result is a clear indication for the characterisation of the examined waves as slow magneto-sonic. 

\section{Conclusions}
We use a 3D magnetohydrodynamic simulation and include thermodynamic gain-loss terms in a solar quadrupolar arcade system with magnetic dip.
Due to evaporation from the chromosphere, eventually localised runaway condensation events occur and blobs of cool material are formed, as a result of thermal instability. We use the thermal instability criteria for isochoric as well as isobaric cases to show that the isochoric one is the most suitable and gives a satisfying estimation of the unstable regions where condensations create blobs. 
Indications of interchange instability are evident and a continuous coronal rain develops. The interchange instability takes over after the condensations make their appearance, as the configuration is gravitationally unstable with dense plasma on top of rarefied plasma. Our magnetic field dominates on the lower part of the simulation, as indicated by the plasma-$\beta$ parameter so that one must exploit field projected stability analysis. 
Blobs are heavily affected by the presence of the magnetic field and they seem to accumulate in the middle of our configuration, while splitting and merging to finally follow the magnetic field topology and fall back to the transition region. The latter occurs along the footpoints of the arcades with an accelerated speed. 
The estimated mass captured by the blobs is in agreement with the order of magnitude of observed mass loss rate due to coronal rain events, i.e. $5\times10^9\mathrm{g\ s^{-1}}$ according to \citet{Antolin12a}.
We examined commonly used quantifications of Brunt-V\"{a}is\"{a}l\"{a} frequencies, trying to improve the way the magnetic field topology is taken into account. They all help to identify the buoyant unstable regions with one of them taking into account the three dimensional magnetic field and projecting the pressure and density gradient along the field lines. When only taking into account the vertical component of pressure and density gradients we also find the (top) unstable low field regions in our domain, where no blobs are found but where other subtle velocity-temperature fluctuations do occur. 
We experimented with three different variations of the projection on the field lines and concluded that overall, the most trustworthy method for tracing the unstable regions is the projection with the full three dimensional magnetic field topology. This is also consistent with the CCI criterium for more idealised setups.

A plethora of waves becomes obvious throughout the simulation box when visualising the Lorentz force variation.
Oscillations along the magnetic field lines were also observed in intensity variations in \cite{OShea07} suggesting the occurrence of standing waves and propagating perturbations.
We found rather localised structures on the profiles of other physical quantities as well. 
We analysed a typical case and estimated its wave propagation speed to be about 6km/s and a phase speed of the order of 10km/s.
These speeds are of the order of magnitude as the redshifts of downflows observed in 1548\AA\ and 630\AA\ and the emission stops sharply after the condensed material has fallen back to the chromosphere \citep{Muller04}.
From the pressure variation we conclude that they are magneto-acoustic in nature.

The instability criteria that we used are only strictly applicable in the linear regime/phase of the instabilities. 
In our paper, we show how they can also serve to gain some insight on the physical processes, even when non-linear phenomena are the ones dominating.

Concluding, this simulation is one of the first ones that captures the mass-cycle in a coronal loop system initiating from footpoint-localized heating, causing chromospheric evaporation, triggering thermal instability that will result in plasma condensation and downflow, in full 3D setups. This physical sequence of processes is related to phenomena such as prominences, flares and coronal rain, that are overall closely intertwined, as indicated in previous studies \citep{Ahn14,Antolin12first,Antolin11,Karpen01,Kawaguchi70,Keppens14,Murawski11,Oliver14,Shimojo02}. A quadrupolar arcade system with a dip, a configuration usually met in prominences, was used in this experiment, as a natural continuation of previous simulations in coronal rain and prominence related phenomena \citep{Fang13,Keppens14}. Catastrophic cooling taking place at a height of about 25 Mm, in agreement with \citet{Antolin11}, leads to the formation of cool plasma condensations, that emit in EUV, as demonstrated in figures~\ref{sdo171},~\ref{sdo211}. Cool and dense plasma blobs are observed in cool chromospheric lines, such as H$\alpha$ and Ca II H, as they descend. A wide range of velocities was revealed from a few km to almost $100\ \mathrm{km\ s^{-1}}$, in agreement with previous studies \citep{Antolin12a,Antolin12b,Antolin10,Fang13,Muller04a, Oliver14,Shimojo02}. Temperatures with peak at 20,000 K and plasma-$\beta$ values of maximum 0.557, were found, both within ranges already reported \citep{Ahn14,Shimojo02},\citep{Antolin11}. Regarding our findings on the magneto-acoustic wave, \citet{Murawski11} also found that magneto-acoustic waves were linked to cool dense descending blobs, while \citet{Oliver14} supports that down falling blobs emit small amplitude sound waves of periods of about 100s. It must be stated that our magnetic field strength resembles more that of quiet Sun, rather than an active region case. Magnetic fields were estimated to an intensity in the range of 8-22$\mathrm{G}$ \citep{Antolin12first,Antolin11}, while at blob length scales, i.e. a few hundreds of kilometres, no perpendicular motion to the local magnetic field is expected according to \citet{Antolin12a}. Finally, the cool condensed plasma blobs appear as thin and elongated structures morphologically at the projected view of figure~\ref{beta} towards the end of the simulation, as observed in several cases \citep{Antolin10,Antolin11,Antolin12b}.

\section*{Acknowledgments}
This research was supported by projects GOA/2015-014 (KU Leuven, 2014-2018), and the Interuniversity Attraction Poles Programme initiated by the Belgian Science Policy Office (IAP P7/08 CHARM). The simulations used the VSC (flemish supercomputer center) funded by the Hercules foundation and the Flemish Government. CX acknowledges FWO Pegasus funding, SPM is aspirant FWO and acknowledges financial support by the Greek Foundation for Education and European Culture (IPEP). SPM gratefully acknowledges the ESPM meeting organizers, for awarding this work with the poster prize.






\end{document}